\documentclass[a4paper,11pt]{article}
\usepackage{jcappub}
\usepackage{lineno}
\usepackage{multirow}
\usepackage{graphicx}
\usepackage{amsmath}
\usepackage{amsfonts}
\usepackage{amssymb}
\usepackage{slashed}
\usepackage{color}
\usepackage{float}
\usepackage{hyperref}
\usepackage{units}
\usepackage{braket}
\usepackage{amsmath}

\title{Constraints on extended axion structures from the lensing of fast radio bursts}

\author[a]{Jan Tristram Acu\~{n}a}
\author[a]{, Kuan-Yen Chou}
\author[a,b]{and Po-Yan Tseng}

\affiliation[a]{Department of Physics, National Tsing Hua University, \\ No. 101 Kuang-Fu Rd., Hsinchu 300044, Taiwan R.O.C.}
\affiliation[b]{Physics Division, National Center for Theoretical Sciences, \\ Taipei 106319, Taiwan R.O.C.}

\emailAdd{jtacuna@gapp.nthu.edu.tw}
\emailAdd{kychou@gapp.nthu.edu.tw}
\emailAdd{pytseng@phys.nthu.edu.tw}
\date{\today}
\abstract{Axions are hypothetical pseudoscalar particles that have been regarded as promising dark matter (DM) candidates. On the other hand, extended compact objects such as axion stars, which are supported by gravity and axion self interactions, may have also been formed in the early Universe and comprise part of DM. In this work, we consider the lensing of electromagnetic signals from distant sources by axion stars, as a way to constrain the properties of axion stars and fundamental axion parameters. Accounting for the effect of the finite size of the axion star, we study the lensing effect induced by gravity, and by axion-photon interactions. The latter effect is frequency dependent, and is relevant in the low frequency band, which motivates the use of fast radio burst (FRB) signals as a probe. We calculate the predicted number of lensed FRB events by specifying the fundamental axion parameters, axion star radial profile, fraction of DM residing in axion stars, and imposing lensing criteria based on the flux ratio and time delay between the brightest images from lensing. Assuming an optimistic case of $10^4$ observed FRB events, and a timing resolution of $1\,\mu{\rm s}$, the lack of observed FRB lensing events in CHIME allows us to probe axion stars with mass $ \gtrsim 10^{-2} M_\odot$, corresponding to axion masses $\lesssim 10^{-10}\,{\rm eV}$ and for negligible axion-photon couplings. Even lighter axion stars up to $\sim 10^{-3} M_\odot$ can be probed, assuming axion-photon couplings of at least $10^{-6}\,{\rm GeV}^{-1}$. Our results indicate that while FRB lensing by axion stars lead to sensitivities that are competitive with conventional microlensing searches operating in the optical band, it remains a challenge to probe axion-photon induced lensing effects.}


\begin{document}
\maketitle
\flushbottom

\section{Introduction}
It is well-known that the QCD axion is the resulting pseudo Nambu-Goldstone boson of a spontaneously broken global U(1) Peccei-Quinn (PQ) symmetry. The U(1) Peccei-Quinn (PQ) symmetry has been proposed as a dynamical solution to the strong CP problem \cite{Peccei:1977hh,Peccei:1977ur,Weinberg:1977ma}, which is tied to the explanation of the smallness of the neutron electric dipole moment (EDM). The PQ symmetry breaking is taken to occur at some energy scale $f_a$, and the mass $m_a$ of the QCD axion is related to $f_a$. Apart from its utility to address the strong CP problem, the QCD axion may also serve as a viable dark matter (DM) candidate.

The axion does not only couple to quarks and gluons, but may also interact with other SM particles based on its pseudoscalar nature. For instance, the term $-g_{a\gamma\gamma}/4\, aF_{\mu \nu} \tilde{F}^{\mu \nu}$ allows for the tree level interaction between the axion and the photon, where $a$ represents the axion field, $F_{\mu \nu}$ and $\tilde{F}_{\mu \nu}$ are the electromagnetic field strength tensor and its dual, respectively~\cite{Adams:2022pbo}. A generalization of the QCD axion comes in the form of axionlike particles (ALPs), which still inherit the QCD axion interactions with SM particles. Since the $m_a$ and $f_a$ of ALPs are independent parameters, ALPs may not provide a solution to the strong CP problem but are less theoretically constrained. Conventional efforts to detect axions, such as in cavity haloscopes and light-shining-through-walls experiments (\textit{e.g.} \cite{Adair:2022rtw,ADMX:2020ote,HAYSTAC:2023cam,MADMAX:2019pub,Wei:2024fkf}), rely on the possibility of axion-photon conversion in the presence of a strong magnetic field. A comprehensive summary of axion parameter constraints can be found in \cite{AxionLimits}.

With regard to the production mechanisms (see, \textit{e.g.} \cite{Marsh:2015xka}, for a review), axions may be produced in the early Universe through the decay of cosmic strings \cite{Davis:1986xc,Harari:1987ht,Shellard:1998mi,Gorghetto:2018myk}, or alternatively through the vacuum misalignment mechanism \cite{Preskill:1982cy,Abbott:1982af,Dine:1982ah}. The latter produces a nonrelativistic and coherent population of axions. Meanwhile, axions radiated from cosmic strings initially produce a relativistic and incoherent population of axions. Through subsequent Hubble expansion, a portion of these axions become nonrelativistic and coherent. Furthermore, axions from cosmic strings can also become coherent through gravitational thermalization. This process occurs when overdense regions of the Universe, which correspond to patches with large axion field values, become gravitationally bound and form axion miniclusters \cite{Hogan:1988mp,Kolb:1993zz,Kolb:1993hw,Kolb:1994fi,Khlopov:1999tm}. Axions from both production mechanisms eventually form a Bose-Einstein condensate (BEC), when the occupation number becomes large \cite{Guth:2014hsa}. In the limit where the momentum of the axions is much smaller than $m_a$, which can be realized in the early Universe due to the fact that its momentum redshifts away with the scale factor, one can formulate a nonrelativistic effective field theory framework for the axion~\cite{Braaten:2015eeu,Hertzberg:2016tal,Eby:2018ufi,Zhang:2018slz,Braaten:2019knj}. The formation of BECs comprising of axions leads to various bound state configurations. These include axion stars (also known in the literature as axion clumps, \textit{e.g.}~\cite{Schiappacasse:2017ham}), objects that are composed of axions bound by gravity and attractive self interactions, which will be the main focus of this work.

Axion stars offer several signatures for indirect searches. For example, the collapse of the critical mass stars induced by the attractive self interaction of the ALPs emits relativistic axions~\cite{Levkov:2016rkk}, or emission of the resonance photons from the merging of axion clumps~\cite{Hertzberg:2018zte,Hertzberg:2020dbk}. Furthermore, microlensing searches provide stringent constraints on the abundance of compact objects with subsolar mass \cite{Fujikura:2021omw}. The characteristic signal for such observations is the amplification of the brightness of background stars from weak gravitational lensing, due to the passage of a compact object, such as an axion star, as it passes close to the line of sight from the observer to the star. In the presence of a nonzero axion-photon coupling, it is also possible to generate additional light bending as photons pass through an axion star. A study by \cite{Prabhu:2020pzm} accounted for this additional effect in the lensing of background stars, and the signal can possibly be probed by using future astrometric data from the Square Kilometer Array (SKA), to look for anomalies in the positions and proper motions of stars.

The effect of the axion-photon coupling in light bending is frequency dependent, and is more pronounced at low frequencies. This means that while the effect may be negligible in microlensing searches, which operate in the optical band, it can potentially be relevant for searches in the radio band. In this work, we take the lensing of fast radio bursts (FRBs) as a probe of axion stars, and, ultimately, the fundamental properties of axions. In obtaining the predicted lensing signal, we account for the finite size of the axion stars, and consider the contributions to the lensing signal induced by gravity and by the axion-photon interaction. We find that the axion star lenses must generally have a sufficiently large radius relative to their Einstein radii, in order for both novel lensing effects to appear. In addition, the axion-photon induced lensing effect will manifest only for axion-photon couplings that have been otherwise robustly excluded by axion haloscope experiments.

This paper is organized as follows. In section \ref{sec:axionstar}, we revisit the nonrelativistic effective field theory framework for axions, and obtain the relationship between the fundamental axion parameters and the macroscopic quantities describing the axion star. In section \ref{sec:revaxionbending}, we develop the formalism to track the trajectory of photons propagating in the presence of an axion medium, and work out the general expressions for the time delay and magnification of an image from lensing, through the use of lensing potentials. In section \ref{sec:frblensing} we lay down the procedure of calculating the number of lensing events, starting from: the lens equation, which determines the locations of images; the lensing cross section based on the lensing criteria; and the optical depth of light emission propagating through an extragalatic and Galactic population of axion stars. We formulate our lensing criteria based on the maximum threshold flux ratio between the two brightest images, and the minimum time delay between the images that can be measured by the CHIME radio telescope facility \cite{Leung:2022vcx,CHIMEFRB:2022xzl,CHIMEFRB:2021srp}. Finally, the absence of observed FRB lensing events in CHIME imposes sensitivities on the fundamental axion parameters and the fraction of axion stars $f_{\rm AS}$. We present our results in section \ref{sec:resultsanddiscussion}, where we show the main sensitivity forecasts in section \ref{sec:exclusion}. We conclude in section \ref{sec:conclusions}.
\section{Axion star properties}
\label{sec:axionstar}
The bosonic nature of axions, tied with a huge occupation number, allows the possibility of forming structures through Bose-Einstein condensation. Examples of structures that have been explored in the literature are axion droplets \cite{Sikivie:2009qn,Guth:2014hsa,Kolb:1993hw}, axion minihalos \cite{Xiao:2021nkb}, dilute and dense axion stars \cite{Colpi:1986ye,Braaten:2015eeu,Visinelli:2017ooc}. One may refer to \cite{Braaten:2019knj} and references therein for a review of axion stars, and these compact structures will be the focus of this work. The starting point is the following Lagrangian
\begin{eqnarray}
    \label{AxionLag}\mathcal{L} = \sqrt{-g}\left[\frac{1}{2}g^{\mu\nu} \partial_\mu a~\partial_\nu a - V(a) - g_{a\gamma\gamma}~a F_{\mu\nu}\tilde{F}^{\mu\nu}\right],
\end{eqnarray}
where $g_{a\gamma\gamma}$ is a dimensionful pseudoscalar coupling between the axion and the photon, $V(a)$ is the axion potential, which is amenable to an expansion up to quartic order
\begin{eqnarray}
    V(a) \approx \frac{m_a^2}{2}a^2 - \frac{\lambda}{4!}a^4,
\end{eqnarray}
and $g$ is the determinant of the spacetime metric induced by the energy momentum of the axion. Working in the weak field limit, the line element can be written as
\begin{eqnarray}
\label{WeakField}    ds^2 = \left(1 + 2\Phi\right)dt^2 - \left(1-2\Phi\right)d\vec{x}^2,
\end{eqnarray}
where $\Phi$ is the gravitational potential. For a specific axion field configuration in the nonrelativistic limit, we can perform the following decomposition
\begin{eqnarray}
    a(t,\vec{x}) = \frac{1}{\sqrt{2m_a}}\left[\exp(-i m_a t)\psi(t,\vec{x})+\exp(i m_a t) \psi^*(t,\vec{x})\right],
\end{eqnarray}
where we separate the rapidly oscillating term coming from the axion mass. In this case, $\vert \dot{\psi}(t,\vec{x}) \vert \ll m_a \vert \psi(t,\vec{x}) \vert$. Under the assumption of spherical symmetry, and taking the wavefunction to be real, we can write the axion field as
\begin{eqnarray}
    a(t,\vec{x}) = \sqrt{\frac{2}{m_a}} \cos\left(m_a t\right) \psi(r),\quad \psi(r) = \sqrt{\frac{m_a}{2}}a_0 f(r/R),
\end{eqnarray}
where an axion configuration can be specified by choosing the form of $f(x)$. In the presence of a global $U(1)$ symmetry, the axion number
\begin{eqnarray}
    N = \int d^3\vec{x}~\vert \psi\vert^2 
\end{eqnarray}
is conserved~\cite{Fujikura:2021omw}, so that we can write
\begin{eqnarray}
   a_0^2 = \frac{N}{2\pi m_a R^3} \left[\int dx~x^2 f^2(x)\right]^{-1}.
\end{eqnarray}
The classical Hamiltonian for a spherically symmetric axion field configuration can be shown to be
\begin{eqnarray}
    \nonumber H &=& \frac{2\pi}{m_a} \int dr~r^2 \left(\frac{d\psi}{dr}\right)^2 \frac{\lambda \pi}{4m_a^2}\int dr~r^2 \psi^4(r) \\
    &-& 8\pi^2 G_N m_a^2 \int dr~dr'~r^2 r'^2  \psi^2(r)\psi^2(r') \frac{1}{\text{max}\left\{r,r'\right\}},
\end{eqnarray}
or, in terms of $f$,
\begin{eqnarray}
    \label{Hamiltonian} H &=& \frac{Nm_a}{x^2}c_1 - \frac{\lambda N^2 m_a}{x^3}c_2 - \frac{G_N m_a^3 N^2}{x} c_3, \\
    c_1 &\equiv& \frac{1}{2}\frac{\int y^2 f'^2(y)~dy}{\int y^2 f^2(y)~dy}, \\
    c_2 &\equiv& \frac{1}{64\pi}\frac{\int y^2 f^4(y)~dy}{\left[\int y^2 f^2(y)~dy\right]^2}, \\
    c_3 &\equiv& \frac{1}{2}\frac{\int y^2 y'^2 f^2(y) f^2(y')\frac{1}{\text{max}\{y,y'\}}~dy~dy'}{\left[\int y^2 f^2(y)~dy\right]^2},
\end{eqnarray}
where $x=m_a R$. The values of these coefficients can be determined once the $f$ function is specified. Following \cite{Fujikura:2021omw}, we can adopt the following exponential-linear ansatz of the form
\begin{eqnarray}
    \psi_{\rm E}(r;R) = \sqrt{\frac{N}{7\pi R^3}}\left(1+\frac{r}{R}\right)\exp\left(-r/R\right),\quad f_{\rm E}(x) = (1+x) \exp(-x)
\end{eqnarray}
parameterized by $R$. We may also adopt the Gaussian ansatz \cite{Arvanitaki_2020,Chavanis:2011zi,Chavanis:2011zm}
\begin{eqnarray}
    \psi_{\rm G}(r; R) = \sqrt{\frac{N}{\pi^{3/2}R^3}}\exp\left(-\frac{r^2}{2R^2}\right),\quad f_{\rm G}(x) = \exp\left(-x^2/2\right).
\end{eqnarray}
The coefficients $c_1, c_2, c_3$, for each axion profile ansatz, are provided in Table \ref{tab:Coeffs}.
\begin{table}[t]
    \centering
    \begin{tabular}{|c|c|c|c|}\hline
        Profile & $c_1$ & $c_2$ & $c_3$ \tabularnewline\hline
        Exponential-linear & $\frac{3}{14} \simeq 0.21$ & $\frac{437}{200704\pi} \simeq 6.93 \times 10^{-4}$ & $\frac{5373}{25088} \simeq 0.21$ \tabularnewline\hline
        Gaussian & $\frac{3}{4} = 0.75$ & $\frac{1}{32\pi \sqrt{2\pi}} \simeq 3.97 \times 10^{-3}$ & $\frac{1}{\sqrt{2\pi}} \simeq 0.40$ \tabularnewline\hline
    \end{tabular}
    \caption{Coefficients $c_1, c_2, c_3$ in the Hamiltonian, for two axion profiles}
    \label{tab:Coeffs}
\end{table}
By extremizing the Hamiltonian in Eq. (\ref{Hamiltonian}) with respect to $x$ we find that the axion number is
\begin{eqnarray}
    N = \frac{2c_1 x}{c_3 G_N m_a^2 x^2 + 3c_2 \lambda}.
\end{eqnarray}
We expect two solutions for $x$, corresponding to the minimum or maximum of the Hamiltonian, which can be classified as stable or unstable, respectively.
At this point we can introduce the following dimensionless quantities
\begin{eqnarray}
    \tilde{N} = \frac{m_a^2 \sqrt{G_N}}{f_a'}N,\quad  \tilde{R} = m_a \sqrt{G_N} f_a' R,\quad \tilde{H} = \frac{m_a}{f_a'^3 \sqrt{G_N}}H,
\end{eqnarray}
where
\begin{eqnarray}
    f_a' \equiv \frac{f_a}{\gamma^{1/2}},\quad \gamma \equiv \frac{f_a^2}{m_a^2}\lambda.
\end{eqnarray}
Then we can write
\begin{eqnarray}
    \tilde{N} = \frac{2c_1 \tilde{R}}{c_3 \tilde{R}^2 + 3c_2}.
\end{eqnarray}
Following \cite{Fujikura:2021omw}, we refer to the benchmark value of $\gamma = 1 - 3 m_u m_d/(m_u+m_d)^2 \simeq 0.34$, which comes from the small field expansion of the QCD axion potential obtained in \cite{GrillidiCortona:2015jxo}. For completeness, we present both cases $\gamma > 0$ and $\gamma < 0$, which will lead to axion stars with slightly different qualitative properties. In both cases, there will exist a minimum rescaled axion star radius $\tilde{R}_{\rm min}$, corresponding to a maximum rescaled axion number $\tilde{N}_{\rm max}$, respectively given by
\begin{eqnarray}
    \tilde{R}_{\min} \equiv \sqrt{\frac{3 c_2 }{c_3}},\quad \tilde{N}_{\max} \equiv \frac{c_1}{\sqrt{3 c_2 c_3}}.
\end{eqnarray}
The allowed $\tilde{R}$ and $\tilde{N}$ values can be parameterized by $\zeta$, $\tilde{R}_{\rm min}$, and $\tilde{N}_{\rm max}$. For each case, we have
\begin{eqnarray}
    \gamma > 0&:& \quad \tilde{N}=\zeta\tilde{N}_{\max},\quad \tilde{R}=\frac{1+\sqrt{1-\zeta^2}}{\zeta}\tilde{R}_{\min}, \quad \zeta \in (0, 1]\\
    \gamma < 0&:& \quad \tilde{N} = \zeta\tilde{N}_{\max},\quad \tilde{R} = \frac{1 + \sqrt{1+\zeta^2}}{\zeta}\tilde{R}_{\min},\quad \zeta \in (0, \infty).
\end{eqnarray}
\begin{figure}[t]
    \centering
    \includegraphics[scale=0.5]{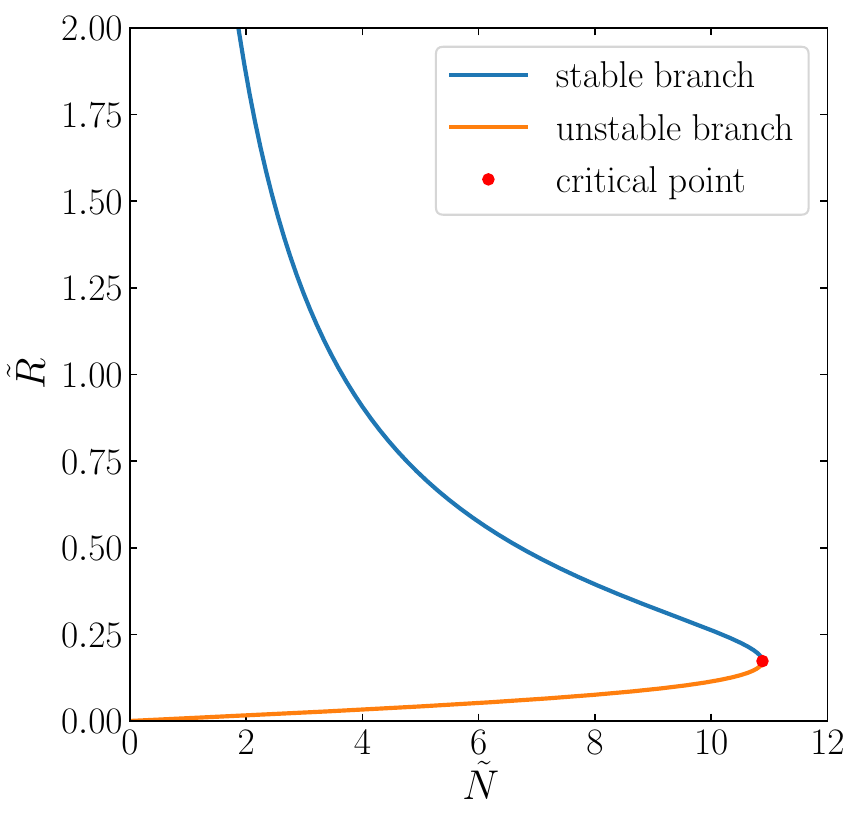}
    \caption{Two types of solution branches of axion star, in the specific case where we adopt the Gaussian profile, and $\gamma = 0.34$.}
    \label{fig: AS_N_R_parameters_gaussain_profile_config}
\end{figure}
In the specific case of the Gaussian profile, the number of axions, radius, and mass of an axion star in the stable branch are given respectively by
\begin{align}
    \label{NRelation}    N &\simeq 2.2787 \times 10^{60} \times \zeta \left( \frac{\unit[10^{-5}]{eV}}{m_a} \right)^2 \left( \frac{f_a}{\unit[10^{12}]{GeV}} \right) \left( \frac{0.34}{\vert \gamma\vert} \right)^{\frac{1}{2}}\\
    \label{RASRelation}    R &\simeq \unit[2.4350 \times 10^{4}]{m}\times \left( \frac{1+\sqrt{1\mp \zeta^2}}{\zeta} \right) \left( \frac{\unit[10^{-5}]{eV}}{m_a} \right) \left( \frac{\unit[10^{12}]{GeV}}{f_a} \right) \left( \frac{\vert\gamma \vert}{0.34} \right)^{\frac{1}{2}}\\
    \label{MASRelation}    M_{\rm AS} &=N m_a\simeq 2.0391 \times 10^{-11} M_{\odot} \times \zeta \left( \frac{\unit[10^{-5}]{eV}}{m_a} \right) \left( \frac{f_a}{\unit[10^{12}]{GeV}} \right) \left( \frac{0.34}{\vert \gamma\vert} \right)^{\frac{1}{2}}.
\end{align}
The upper (lower) sign in $R$ refers to the case $\gamma > 0$ ($\gamma < 0$). In Fig. \ref{fig: AS_N_R_parameters_gaussain_profile_config} we show the stable and unstable branches for axion star configurations following the Gaussian profile, with $\gamma = 0.34$. We take $\zeta=1$ in this work, which is indicated by the red dot in Fig. \ref{fig: AS_N_R_parameters_gaussain_profile_config}.

Before proceeding, we need to identify the conditions in which the above nonrelativistic effective field theory formalism, that we have adopted to arrive at the connection between axion star properties and fundamental axion parameters, remains valid. First, the polynomial expansion of the axion effective potential requires $a/f_a \ll 1$. In our case we have
\begin{eqnarray}
    \frac{a}{f_a} \simeq 6.97 \times 10^{-3}\left(\frac{m_a}{\unit[10^{-10}]{eV}}\right)^{3/2}\left(\frac{M_{\rm AS}}{M_\odot}\right)\left(\frac{0.34}{\vert \gamma \vert}\right)^{1/2} \frac{f(x)}{\sqrt{\int dx~x^2 f^2(x)}}.
\end{eqnarray}
Meanwhile, one way of checking the validity of the nonrelativistic approximation is by comparing the nonrelativistic kinetic term $T_1 = 1/(2m_a) (\nabla \psi)^2$ with the next-to-leading term in the fully relativistic kinetic term given by $T_2 = -1/(8m_a^3) (\nabla^2 \psi)^2$ \cite{Braaten:2019knj,Visinelli:2017ooc}. One can show that
\begin{eqnarray}
    \frac{T_2}{T_1} = -\frac{1}{4(m_a R)^2}\left[\frac{f''(x)}{f'(x)}+\frac{2}{x}\right]^2.
\end{eqnarray}
Note that the formation of the axion star requires that the Compton wavelength of an axion particle be comparable with the typical interparticle separation inside the axion star. This means that
\begin{eqnarray}
    \left(\frac{4\pi/3~R^3}{N}\right)^{1/3} \sim \frac{1}{m_a} \Rightarrow m_a R \sim N^{1/3},
\end{eqnarray}
so that $m_a R \gg 1$, assuming $N \gg 1$. Then the ratio $T_2/T_1$ can be used to check that the nonrelativistic approximation works for the Gaussian profile, in the limit of $x \gtrsim 1$, but may fail in regions close to the core of the axion star, as pointed out in \cite{Visinelli:2017ooc}. On the other hand, we will take the simplified approach where the failure of the nonrelativistic approximation in the core region can be neglected in subsequent calculations.
\section{Review: light propagation in axion electrodynamics}
\label{sec:revaxionbending}
A nonzero pseudoscalar axion-photon coupling $g_{a\gamma\gamma}$ definitely modifies the propagation of electromagnetic waves in the presence of an axion field. Most investigations of the impact of $g_{a\gamma\gamma}$ on light propagation consider a background axion field that permeates through spacetime, and which oscillates in a coherent manner (\textit{e.g.} \cite{Khmelnitsky:2013lxt}). In our work, we consider the effect induced by a compact object comprising of axions, on electromagnetic waves, emanating from a distant source, such as a fast radio burst (FRB). Earlier literature, \textit{e.g.} \cite{Laha:2018zav,Xiao:2022hkl,Gao:2023xbi,Kalita:2023eeq}, identified the utility of FRBs to probe astrophysical and cosmological phenomena, particularly in the context of searching for signals of extended dark matter structures. On the other hand, as we shall demonstrate in later sections, the specific case of FRB sources presents an ideal case to probe the effect of axion-photon interactions on lensing observables, since such an effect is pronounced at low frequencies, in contrast with lensing in the optical band. To quantitatively determine the axion induced effect on light propagation, we lay down the equations of motion describing the trajectory of a photon, working in the geometric optics limit. Given the frequency $f_0$ of the propagating EM wave, we need to have $2\pi f_0 \gg m_a$. For instance, a frequency of \unit[600]{MHz} translates to 
\begin{eqnarray}
    m_a \ll \unit[2.5 \times 10^{-6}]{eV}\left(\frac{f_0}{\unit[600]{MHz}}\right). 
\end{eqnarray}
Letting $\Vec{x}(t) = \Vec{\xi}(t)$ be the trajectory of such a photon, with wavevector $\Vec{k} = \Vec{\kappa}(t)$ and angular frequency $\omega = \epsilon(t)$, we have the following system of equations applicable in a dispersive medium:
\begin{eqnarray}
    \label{PropSystem1} \frac{d\Vec{\xi}}{dt} = \frac{\partial \omega}{\partial \Vec{k}}\Bigg\vert_{(\Vec{x},\Vec{k},\omega)=(\Vec{\xi},\Vec{\kappa},\epsilon)},\quad \frac{d\Vec{\kappa}}{dt} = -\frac{\partial \omega}{\partial \Vec{x}}\Bigg\vert_{(\Vec{x},\Vec{k},\omega)=(\Vec{\xi},\Vec{\kappa},\epsilon)},\quad \frac{d\epsilon}{dt} = \frac{\partial \omega}{\partial t}\Bigg\vert_{(\Vec{x},\Vec{k},\omega)=(\Vec{\xi},\Vec{\kappa},\epsilon)}.
\end{eqnarray}
The dispersion relation for photons propagating in an axion medium is given by \cite{McDonald:2019wou}
\begin{eqnarray}
    \label{FullDispersion} \omega_\eta^2 - \Vec{k}^2 = \eta g_{a\gamma\gamma}\left\{(\omega \dot{a}+\Vec{k}\cdot \Vec{\nabla}a)^2-(\omega^2-\Vec{k}^2)[\dot{a}^2-(\Vec{\nabla}a)^2]\right\}^{1/2},
\end{eqnarray}
where $g_{a\gamma\gamma}$ is the axion-photon coupling and $\eta = \pm 1$ is a parameter that corresponds to the helicity state of the photon: $+1$ for right handed helicity, and $-1$ for left handed helicity. We assume that the plasma frequency $\omega_p = 0$. Since we are usually interested in tracking to which order in $g_{a\gamma\gamma}$ a certain physical quantity appears, we can write $\omega_\eta$ as a perturbative series in $g_{a\gamma\gamma}$, \textit{i.e.}
\begin{eqnarray}
    \omega = k + g_{a\gamma\gamma} f_1 + g_{a\gamma\gamma}^2 f_2 + ...
\end{eqnarray}
Assuming that $\dot{a}+\hat{k}\cdot\Vec{\nabla}a \neq 0$, we can match terms in the LHS and RHS of Eq. (\ref{PropSystem1}), and find \cite{Mcdonald:2020hjm}
\begin{eqnarray}
\label{PertDispersion}    \omega_\eta = k + \frac{\eta g_{a\gamma\gamma}}{2}(\dot{a}+\hat{k}\cdot\Vec{\nabla}a)-\frac{g_{a\gamma\gamma}^2}{8k}[\dot{a}^2+(\hat{k}\cdot\Vec{\nabla}a)^2-2(\Vec{\nabla}a)^2] + \eta g_{a\gamma\gamma}^3~(\eta f_3)+...
\end{eqnarray}
Defining the projector
\begin{eqnarray}
    P_\perp(\hat{k})_{ij} \equiv \delta_{ij} - \hat{k}_i \hat{k}_j
\end{eqnarray}
orthogonal to the photon unit 3-vector, we can then show that
\begin{eqnarray}
    \nonumber   \frac{d\Vec{\xi}_\eta}{dt} &\approx& \hat{\kappa} + \frac{\eta g_{a\gamma\gamma}}{2\kappa}P_\perp(\hat{\kappa})\Vec{\nabla}a\Bigg\vert_{(\Vec{x},\Vec{k},\omega)=(\Vec{\xi},\Vec{\kappa},\epsilon)} \\
    \label{Approxdxdt} &+& \frac{g_{a\gamma\gamma}^2}{8\kappa^2}\left\{\hat{\kappa}[\dot{a}^2+(\hat{\kappa}\cdot\Vec{\nabla}a)^2-2(\Vec{\nabla}a)^2]-2(\hat{\kappa}\cdot\Vec{\nabla}a)P_\perp(\hat{\kappa})\Vec{\nabla}a]\right\}\Bigg\vert_{(\Vec{x},\Vec{k},\omega)=(\Vec{\xi},\Vec{\kappa},\epsilon)}+... \\
    \nonumber   \frac{d\Vec{\kappa}_\eta}{dt} &\approx& -\frac{\eta g_{a\gamma\gamma}}{2}[\Vec{\nabla}\dot{a}+(\hat{\kappa}\cdot \Vec{\nabla})\Vec{\nabla}a]\Bigg\vert_{(\Vec{x},\Vec{k},\omega)=(\Vec{\xi},\Vec{\kappa},\epsilon)} \\
    \label{Approxdkdt}  &+&\frac{g_{a\gamma\gamma}^2}{8\kappa}[2\dot{a}\Vec{\nabla}\dot{a}+2(\hat{\kappa}\cdot \Vec{\nabla}a)(\hat{\kappa}\cdot \Vec{\nabla})\Vec{\nabla}a-4(\Vec{\nabla}a \cdot \Vec{\nabla})\Vec{\nabla}a]\Bigg\vert_{(\Vec{x},\Vec{k},\omega)=(\Vec{\xi},\Vec{\kappa},\epsilon)}+...
\end{eqnarray}
It is important to note that in obtaining the equation of motion for $\Vec{\kappa}$, we need to perform the gradient operation on $\omega$ \textit{before} evaluating expressions at the photon trajectory. Then, for instance, the term
\begin{eqnarray}
    \Vec{\nabla}(\hat{k}\cdot \Vec{\nabla}a)\Bigg\vert_{(\Vec{x},\Vec{k},\omega)=(\Vec{\xi},\Vec{\kappa},\epsilon)} = (\hat{k}\cdot \Vec{\nabla})(\Vec{\nabla}a)\Bigg\vert_{(\Vec{x},\Vec{k},\omega)=(\Vec{\xi},\Vec{\kappa},\epsilon)}=(\hat{\kappa}\cdot \Vec{\nabla})(\Vec{\nabla}a)\Bigg\vert_{(\Vec{x},\Vec{k},\omega)=(\Vec{\xi},\Vec{\kappa},\epsilon)}.
\end{eqnarray}
Note that Eqs. (\ref{Approxdxdt}) and (\ref{Approxdkdt}) do not yet include the gravitational contribution. From the line element in Eq. (\ref{WeakField}) we can work out the geodesic equations, and show that
\begin{eqnarray}
    \label{GravityTrajectory}  \left(\frac{d\vec{\xi}_\eta}{dt}\right)_{\rm gravity} \approx \left(1 + 2\Phi\right)\hat{\kappa}_\eta\Big\vert_{\Vec{x}=\Vec{\xi}_\eta},\quad \left(\frac{d\vec{\kappa}_\eta}{dt}\right)_{\rm gravity} \approx -2\kappa \left[\vec{\nabla}\Phi - \hat{\kappa} (\hat{\kappa}\cdot\vec{\nabla}\Phi) \right]\Big\vert_{\Vec{x}=\Vec{\xi}_\eta}.
\end{eqnarray}
We are assuming that $\vert\dot{\Phi}\vert \ll \vert \vec{\nabla}\Phi\vert$, so that the gravitational potential is static.

\subsection{Astrophysical observables: bending angle, time delay, and magnification}
\begin{figure}[t]
    \centering
    \includegraphics[width=15cm]{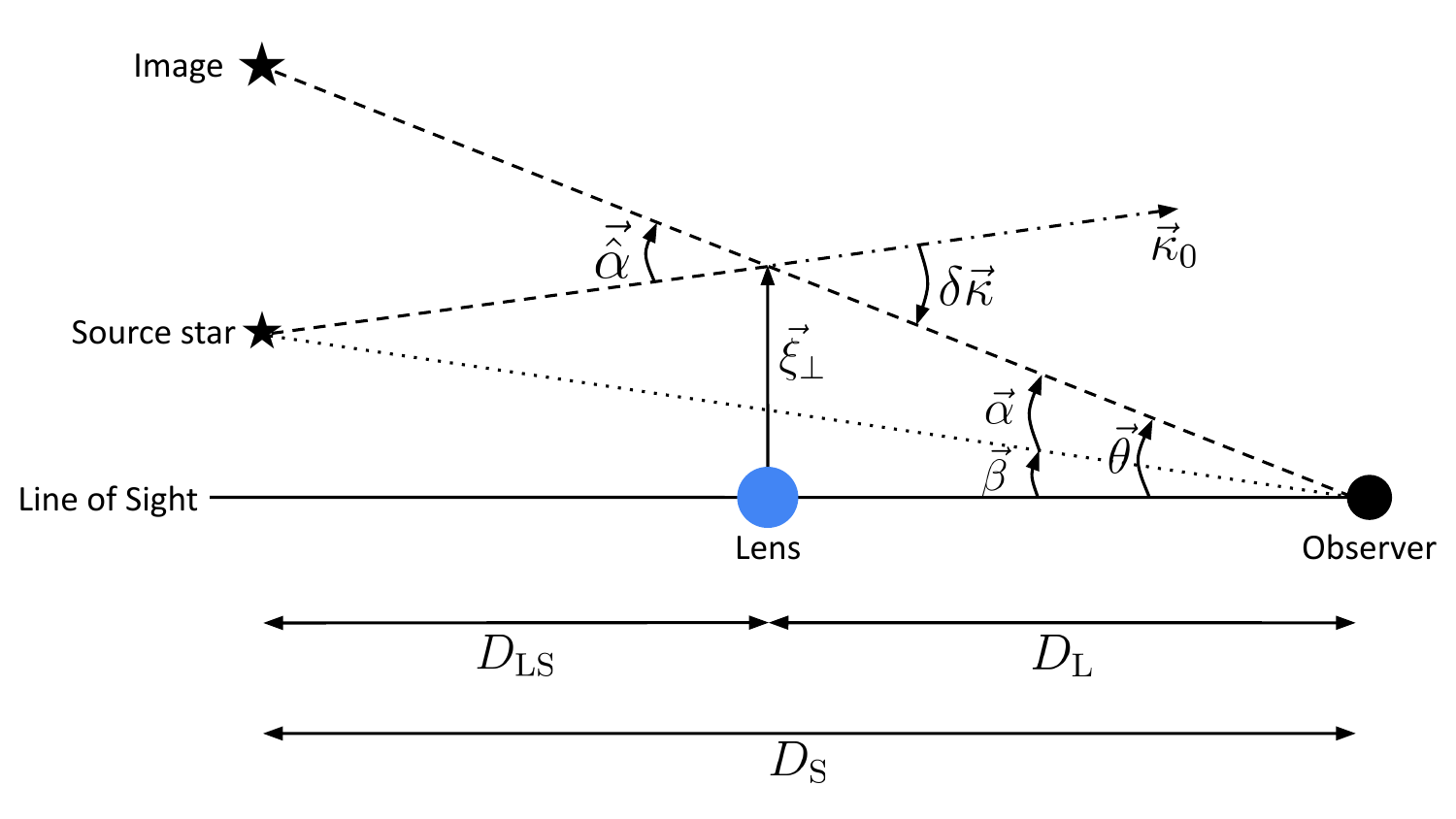}
    \caption{Geometrical setup of the lensing configuration.}
    \label{fig:GeometricalSetupGravitationalLensing}
\end{figure}

The axion-photon induced change in momentum of the propagating photon, in the presence of an axion configuration is given by 
\begin{eqnarray}
    \left(\frac{\delta \vec{\kappa}}{\kappa_0}\right)_{\rm axion} \approx \frac{g_{a\gamma\gamma}^2}{8\kappa_0^2}\int_{-\infty}^\infty dt'~\vec{\nabla}_\perp \left(\partial a\right)^2\Big\vert_{t = t', \vec{x} = \hat{\kappa}_0 t' - \vec{\xi}_*},\quad \vec{\xi}_*(t) = \vec{r}_0 + \vec{v}_* t.
\end{eqnarray}
By assuming that the spherically symmetric axion field configuration oscillates very rapidly in time, \textit{i.e.}~the axion field oscillates undergoes several cycles of oscillation, within the period where light traverses the axion star, we can show that 
\begin{eqnarray}
    \left(\frac{\delta \vec{\kappa}}{\kappa_0}\right)_{\rm axion} \approx \frac{(g_{a\gamma\gamma} a_0)^2}{8(\kappa_0 R)^2}~\frac{1}{R^2}\int_{-\infty}^\infty dt'~P_\perp(\hat{\kappa}_0)\vec{r}~F(r/R;m_aR),
\end{eqnarray}
where
\begin{eqnarray}
   \label{DefFrR} F(r/R; m_a R) \equiv \frac{(m_a R)^2 f(r/R) f'(r/R) - f'(r/R) f''(r/R)}{r/R}.
\end{eqnarray}
As shown in Fig. \ref{fig:GeometricalSetupGravitationalLensing}, $P_\perp(\hat{\kappa}_0)\vec{r} = \vec{\xi}_\perp$ is just the vector pointing from the axion star to the point of closest approach of the photon trajectory, and this is a constant. Noting that
\begin{eqnarray}
    r^2 = \xi_\perp^2 + t'^2,
\end{eqnarray}
we can write the amount of deflection as an Abel integral given by
\begin{eqnarray}
   \left(\frac{\delta \vec{\kappa}}{\kappa_0}\right)_{\rm axion} \approx \frac{\vec{\xi}_\perp}{R}\frac{(g_{a\gamma\gamma} a_0)^2}{8(\kappa_0 R)^2}~\mathcal{F}(\xi_\perp/R;m_a R), 
\end{eqnarray}
where $\mathcal{F}$ is the Abel transform of $F$, \textit{i.e.}
\begin{eqnarray}
    \mathcal{F}(\xi_\perp/R;m_a R) \equiv 2\int_{\xi_\perp/R}^\infty \frac{y~F(y; m_a R)}{\sqrt{y^2 - (\xi_\perp/R)^2}}dy,
\end{eqnarray}
and is exactly known for both the exponential and Gaussian profiles. Note that $\mathcal{F} < 0$ since the photon deflection is always opposite $\vec{\xi}_\perp$. Meanwhile, the case of the gravity induced bending simply follows from Eq. (\ref{GravityTrajectory}), so that
\begin{eqnarray}
    \left(\frac{\delta \vec{\kappa}}{\kappa_0}\right)_{\rm gravity} = -2\int_{-\infty}^\infty dz~\vec{\nabla}_\perp \Phi.
\end{eqnarray}
We can rewrite the above expression by taking the divergence of both sides along the direction perpendicular to the direction of propagation of the photon, and noting that the Poisson equation gives
\begin{eqnarray}
    \nabla_\perp^2 \Phi + \partial_z^2 \Phi = 4\pi G_N \rho.
\end{eqnarray}
Then taking the case of a spherically symmetric potential, we have
\begin{eqnarray}
    \vec{\nabla}_\perp \cdot \left(\frac{\delta \vec{\kappa}}{\kappa_0}\right)_{\rm gravity} = -(4G_N)~2\pi \Sigma(s),\quad \Sigma(s) \equiv \int_{-\infty}^\infty dz~\rho(s,z),
\end{eqnarray}
where $s$ is the cylindrical radial coordinate. Implicitly we assumed that $\partial_z \Phi\vert_{z \rightarrow \pm \infty} = 0$, which is equivalent to the statement that the size of the lens along the direction of light propagation is much less than the distance between the source and the observer. Then
\begin{eqnarray}
    \left(\frac{\delta \vec{\kappa}}{\kappa_0}\right)_{\rm gravity} \approx -\vec{\xi}_\perp\frac{4 G_N \mathcal{M}(\xi_\perp)}{\xi_\perp^2},\quad \mathcal{M}(\xi_\perp) \equiv 2\pi \int_0^{\xi_\perp} \Sigma(s')~s' ds'.
\end{eqnarray}
~\\
We may then write the total deflection angle $\vec{\hat{\alpha}}$, as seen by an observer in the lens plane, which includes both the gravity and axion-induced contributions, as
\begin{eqnarray}
    \label{hatAlpha}\vec{\hat{\alpha}}\left(\vec{\xi}_\perp\right) = -\frac{\vec{\xi}_\perp}{R}\frac{(g_{a\gamma\gamma} a_0)^2}{8(\kappa_0 R)^2}~\mathcal{F}(\xi_\perp/R;m_a R) +\vec{\xi}_\perp\frac{4 G_N \mathcal{M}(\xi_\perp)}{\xi_\perp^2},
\end{eqnarray}
where $\vec{\hat{\alpha}}$ is defined in Fig. \ref{fig:GeometricalSetupGravitationalLensing}. 
As a side comment, one may wonder about plasma effects in light propagation in an axion medium. It has been shown by \cite{McDonald:2019wou} that the axion-induced bending angle, in the presence of an optically dense medium for photons, is first order in the axion photon coupling, and is given by
\begin{eqnarray}
    \vec{\hat{\alpha}} = \pm \frac{g_{a\gamma\gamma}}{2\kappa_0}\int dt~\left[n_0 \nabla_\perp \dot{a} + \left(\hat{\kappa}_0\cdot \nabla\right) \nabla_\perp a\right];
\end{eqnarray}
 the different signs indicate that left and right circularly polarized photons are bent by the same amount, but in opposite directions. In the limit $n_0 = 1$, the above contribution is zero because the integrand becomes a total derivative \cite{Blas:2019qqp}. On the other hand, the plasma contribution to $n_0$ gives  
\begin{eqnarray}
   \label{FirstOrder} \vec{\hat{\alpha}} = \pm \frac{g_{a\gamma\gamma}}{2\kappa_0}\int dt~\left(-\frac{\omega_p^2}{2\kappa_0^2}\right) \nabla_\perp \dot{a},
\end{eqnarray}
where $\omega_p$ is the plasma frequency, in the presence of free electrons. The integration over $t$ is understood to be a line of sight integral. However, since the axion field oscillates rapidly, relative to the light travel time through the axion star, the time average of $\dot{a}$ gives zero, so that we may neglect the plasma contribution \cite{Prabhu:2020pzm}.

Note that the deflection angle is a vectorial quantity defined on the \textit{lens plane}, which is perpendicular to the line of sight containing the lens. We may establish a relationship between the source angular position $\vec{\beta}$ and image angular position $\vec{\theta}$ using
\begin{eqnarray}
    \label{LensEquation} \vec{\beta} = \vec{\theta} - \vec{\hat{\alpha}}\left(D_{\rm L} \vec{\theta}\right)\frac{D_{\rm LS}}{D_{\rm S}},
\end{eqnarray}
which follows from simple geometrical arguments. Eq. (\ref{LensEquation}) will be referred to as the \textit{lens equation} in the following contains. One may observe that the bending angle $\vec{\hat{\alpha}}$ is parallel to $\vec{\xi}_\perp = D_{\rm L}/R~\vec{\theta}$, and thus it can be written as a gradient. Following \cite{schneider1985new,Narayan:1996ba},  one can define a \textit{lensing potential} $\Psi$ as
\begin{eqnarray}
    \vec{\nabla}_{\vec{\theta}}\Psi = \vec{\alpha}  = \frac{D_{\rm LS}}{D_{\rm S}}\vec{\hat{\alpha}},
\end{eqnarray}
and the lens equation Eq. (\ref{LensEquation}) can be rewritten as
\begin{eqnarray}
    \vec{0} = \vec{\nabla}_{\vec{\theta}}t(\vec{\theta}),\quad t(\vec{\theta}) \equiv (1+z_{\rm L})\frac{D_{\rm L} D_{\rm S}}{D_{\rm LS}}\left[\frac{1}{2}\left(\vec{\theta}-\vec{\beta}\right)^2 - \Psi\right].
\end{eqnarray}
As shown in \cite{schneider1985new}, the function $t(\vec{\theta})$ has the interpretation of being a time delay function; when we evaluate the time delay function for two images and take the difference, the result is the actual difference in arrival times associated with the two images. The first term is the \textit{geometric} time delay contribution, owing to the fact that the geodesic trajectory taken by the photon from the source to the observer acquires an extra path length relative to the straight-line path from the source to the observer; on the other hand, the term associated with $\Psi$ is the \textit{Shapiro} time delay contribution, which comes from both gravitational and axion-photon interactions. Explicitly, the general expression for the lensing potential is given by
\begin{eqnarray}
    \Psi &=& \Psi_{\rm gravity} + \Psi_{\rm axion},\\
    \Psi_{\rm gravity} \equiv \frac{2D_{\rm LS}}{D_{\rm S}  D_{\rm L} } \int_{-\infty}^\infty dt'\Phi &,& \quad \Psi_{\rm axion} \equiv -\frac{D_{\rm LS}}{D_{\rm S}  D_{\rm L} }\frac{g_{a\gamma\gamma}^2}{8\kappa_0^2}\int_{-\infty}^\infty dt'\left(\partial a\right)^2.
\end{eqnarray}
Finally, the bending of the trajectory of a light bundle through lensing will affect the luminosity of an astrophysical object. Treating the lens equation as a mapping of source to image positions, the determinant of the Jacobian matrix
\begin{eqnarray}
    \mathcal{A}_{ij} \equiv \frac{\partial \beta_j}{\partial \theta_i} = \delta_{ij} - \partial_{\theta_i}\partial_{\theta_j}\Psi
\end{eqnarray}
relates the size of the area element on the source to the area element on the image. Then the magnification of the image is simply given by the inverse of the determinant of the Jacobian. In this regard, the inverse of the Jacobian matrix is just the \textit{magnification tensor} $\mathcal{M}$. In the case where the lens is circularly symmetric, the magnification tensor takes the following simplified expression:
\begin{eqnarray}
    \mathcal{M}_{ij} = \left(\delta_{ij} - \hat{\theta}_i \hat{\theta}_j\right)\frac{1}{1 - \Psi'(\theta)/\theta} + \hat{\theta}_i \hat{\theta}_j~\frac{1}{1 - \Psi''(\theta)},
\end{eqnarray}
where $\hat{\theta}$ is the unit vector in the direction of the image at $\vec{\theta}$. The magnification of the image is then
\begin{eqnarray}
    \mu(\theta) = \frac{1}{\left[1 - \Psi'(\theta)/\theta\right]\left[1 - \Psi''(\theta)\right]} = \frac{\theta}{\beta}\frac{d\theta}{d\beta},
\end{eqnarray}
where the last equality follows from the lens equation.
\section{Estimating the number of lensing events}
\label{sec:frblensing}
\subsection{Writing the lens equation}
In the case of weak gravitational lensing induced by a pointlike object with mass $M$, the lens equation takes the form
\begin{eqnarray}
    \beta = \theta - \frac{\theta_{\rm E}^2}{\theta},\quad \theta_{\rm E}^2 \equiv 4G_N M\frac{D_{\text{LS}}}{D_{\rm L} D_{\rm S}},
\end{eqnarray}
where the \textit{Einstein angle} $\theta_{\rm E}$ sets the angular scale within which gravitational effects can significantly distort the image of some source object. To see this, we examine the image positions of a source located at $\beta$, which are given by 
\begin{eqnarray}
    \theta_\pm = \theta_{\rm E} \left[\frac{\beta}{2\theta_{\rm E}} \pm \sqrt{1 + \left(\frac{\beta}{2\theta_{\rm E}}\right)^2}\right].
\end{eqnarray}
In the limit $\beta \ll \theta_{\rm E}$, we have $\theta_\pm \approx \pm\theta_{\rm E}$, while the opposite limit $\beta \gg \theta_{\rm E}$ gives $\theta_+ \approx \beta$, $\theta_- \approx 0$. In other words, the source object must be within the Einstein angle of the pointlike lens in order for gravitational lensing to be noticeable. With respect to an incident circular bundle of light rays, whose axis passes through the pointlike object, the effective cross section of the lens is $\pi D_{\rm L}^2 \theta_{\rm E}^2$. In general, if we have a collection of lenses, with comoving number density $n_{\rm L}$, and effective cross section $\sigma(z_{\rm L})$, along the line of sight between the observer and the source at redshift $z_{\rm S}$, the probability that lensing will occur is given by $\exp(-\tau)$, where (\textit{e.g.}~\cite{Munoz:2016tmg, Oguri:2022fir, Kalita:2023eeq})
\begin{eqnarray}
    \tau(z_{\rm S}) = \int_{\chi(0)}^{\chi(z_{\rm S})} d\chi(z_{\rm L})~(1+z_{\rm L})^2 n_{\rm L} \sigma(z_{\rm L})
\end{eqnarray}
is the \textit{optical depth} of the medium. Note that the integration is performed with respect to the comoving distance to the lens, which is dependent on the redshift of the lens $z_{\rm L}$. 

\begin{figure}[t]
    \centering
    \includegraphics[scale=0.4]{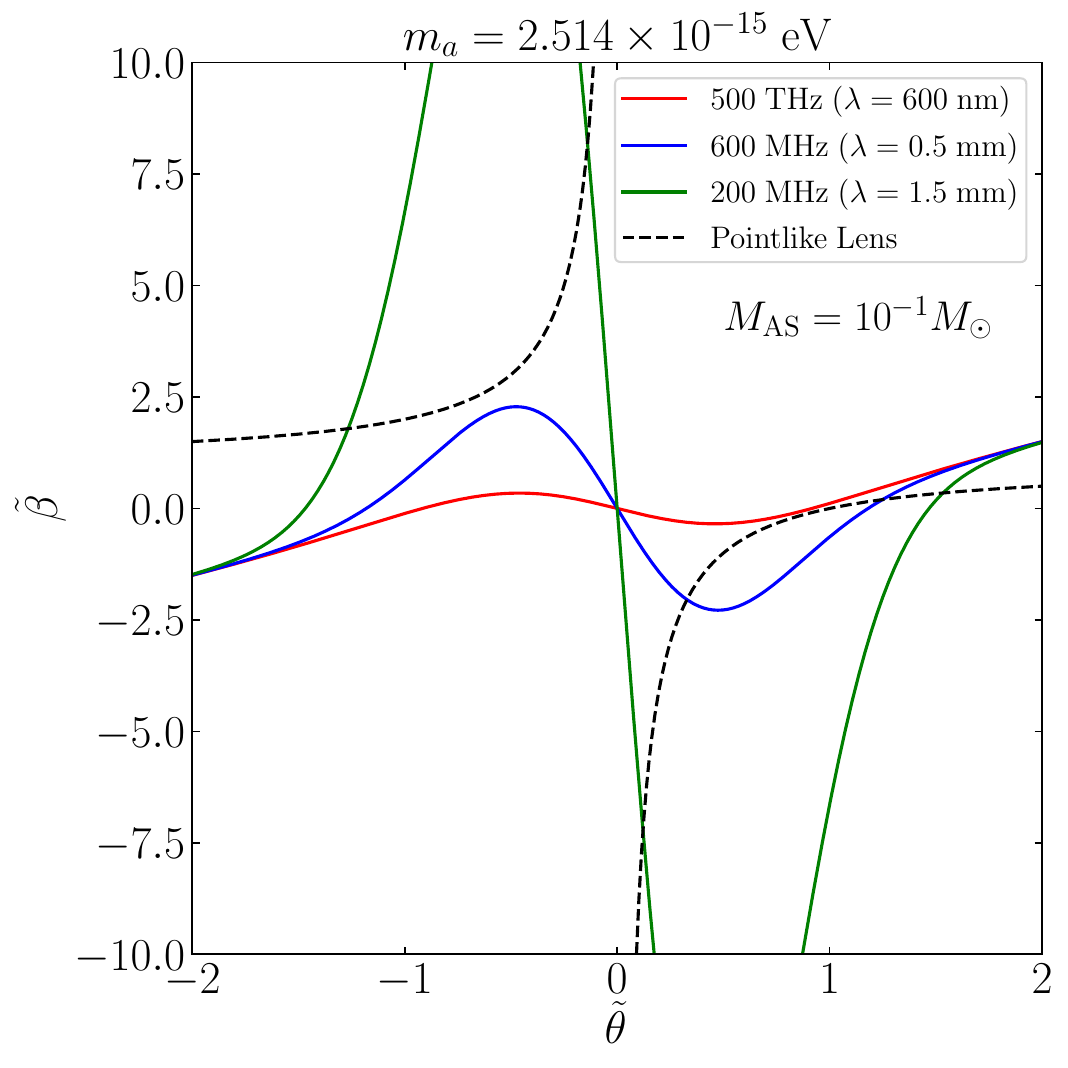}\includegraphics[scale=0.4]{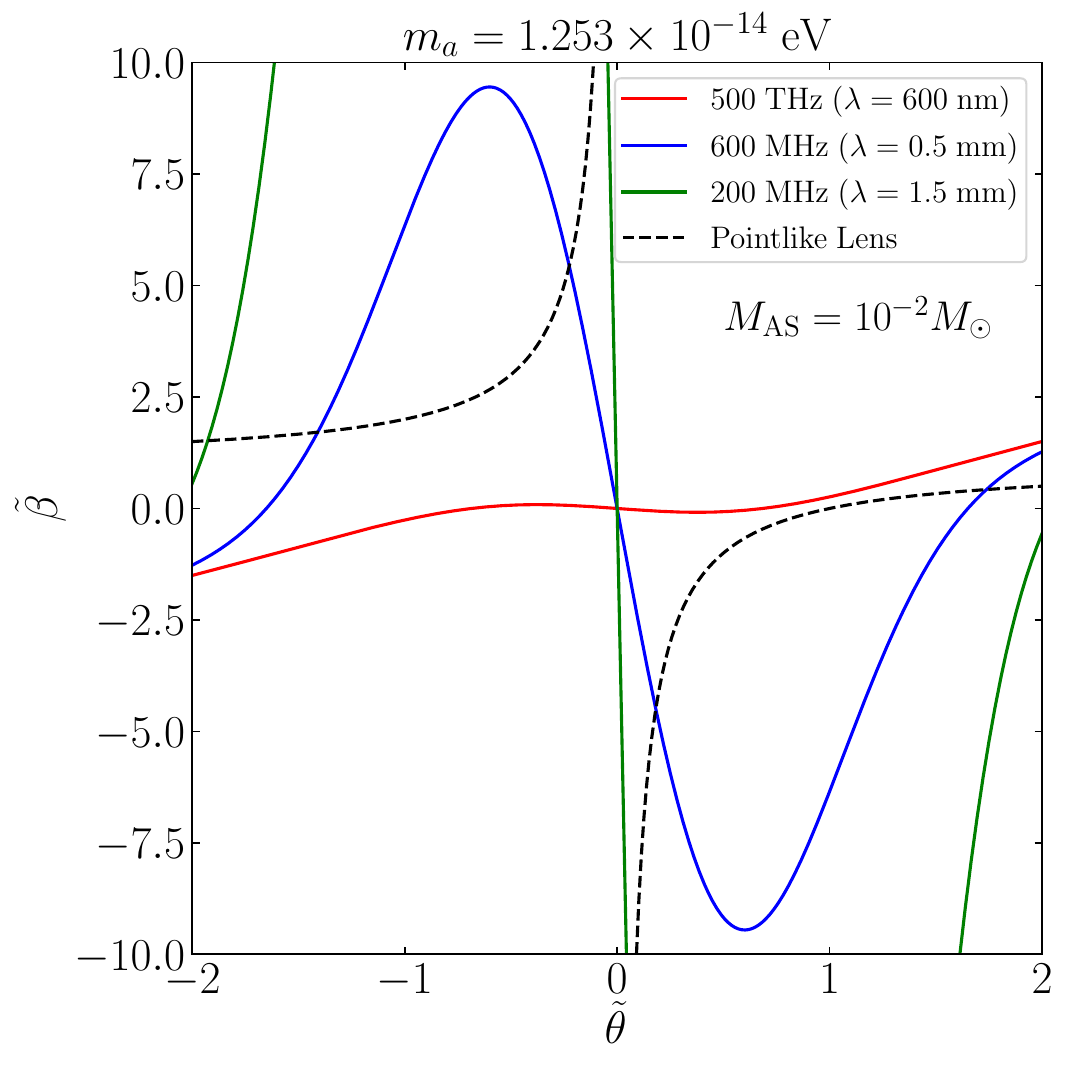}
    \caption{Plots of the right hand side of the reduced lens equation given in Eq. (\ref{ReducedLens}), for different axion star configurations and three different FRB frequencies. In both panels, $g_{a\gamma\gamma}=\unit[10^{-3}]{GeV^{-1}}$, the lens and source redshifts are $z_{\rm L} = 0.4$ and $z_{\rm S} = 1.5$, respectively.}
    \label{fig:LensEquation}
\end{figure}

For the case of lensing by axion stars, the situation slightly differs from the case of typical pointlike gravitational lensing, since the bending angle $\hat{\alpha}$ is nonsingular at $\theta = 0$. For $\theta \rightarrow 0$, 
\begin{eqnarray}
    \hat{\alpha} \approx \theta~\frac{D_{\rm L}}{R}\left\{\frac{4G_N M_{\rm AS}}{R} + \frac{g_{a\gamma\gamma}^2 N}{8\pi R^2 (\kappa_0 R)^2 (m_a R)}\left[2(m_a R)^2+1\right]\right\}.
\end{eqnarray}
From the lens equation, it can be shown that for a given source position $\beta > 0$, there can be \textit{three} images for $\beta < \beta_c$, \textit{two} images for $\beta = \beta_c$, and \textit{one} image for $\beta > \beta_c$. The critical value $\beta_c$ corresponds to the case where the magnification is infinite, \textit{i.e.}
\begin{eqnarray}
    \frac{1}{\mu(\theta_c)} = 0,\quad \beta_c = \theta_c - \Psi'(\theta_c).
\end{eqnarray}
We illustrate this qualitative behavior in Fig. \ref{fig:LensEquation} by plotting the right hand side of the lens equation, where the angular scales are rescaled with respect to $\theta_{\rm E}$, \textit{i.e.} we implicitly introduced
\begin{eqnarray}
    \tilde{\beta} \equiv \frac{\beta}{\theta_{\rm E}},\quad \tilde{\theta} \equiv \frac{\theta}{\theta_{\rm E}},
\end{eqnarray}
so that the lens equation becomes
\begin{eqnarray}
   \tilde{\beta} = \tilde{\theta} - \frac{1}{\theta_{\rm E}}\Psi'\left(\tilde{\theta}\theta_{\rm E}\right).
\end{eqnarray}
Explicitly for the case of the Gaussian profile we have
\begin{eqnarray}
 \label{ReducedLens}   \tilde{\beta} = \tilde{\theta} - \frac{1-\exp(-w_{\rm E}^2 \tilde{\theta}^2)}{\tilde{\theta}} - A w_{\rm E}^2 \tilde{\theta} \left[1 + \frac{1}{2(m_a R)^2} - \frac{w_{\rm E}^2}{(m_a R)^2} \tilde{\theta}^2\right]\exp\left(-w_{\rm E}^2 \tilde{\theta}^2\right),
\end{eqnarray}
where
\begin{eqnarray}
    A \equiv \frac{g_{a\gamma\gamma}^2}{16 \pi G_N (\kappa_0 R)^2},\quad w_{\rm E} \equiv \frac{D_{\rm L} \theta_{\rm E}}{R}.
\end{eqnarray}
The parameter $w_{\rm E}$ is the size of the Einstein radius relative to the axion star radius. It is a quantity that characterizes the degree in which the axion star can be regarded as an object with a finite size; when $w_{\rm E} \rightarrow \infty$, the axion star is essentially pointlike, while in the limit $w_{\rm E} \sim O(1)$, the axion star is effectively an extended object. Meanwhile, the dimensionless parameter $A$ controls the relative contribution of the axion-photon to the bending angle. Note that the criterion that ensures the presence of three images implies that there is a minimum $w_{\rm E}$ value set by $A$ and $m_a R$, so that
\begin{eqnarray}
\label{wEThreshold}    w_{\rm E} > A',\quad A'^2 \equiv \frac{1}{1 + A\left[1+(m_a R)^{-2}\right]} \simeq \frac{1}{1+A},
\end{eqnarray}
where the last equality holds when $m_a R \gg 1$.

In the event that finite size effects cannot be ignored, the amount of lensing by axion stars is dependent on the wavelength of the propagating electromagnetic radiation, and the frequency dependence enters in the lens equation through the $A$ parameter in Eq. (\ref{ReducedLens}). In probing axion-photon interactions, we note that the size of the effect is proportional to $g_{a\gamma\gamma}^2/\kappa_0^2$, so that the axion-induced lensing effect is potentially enhanced in the presence of a source that emits low frequency electromagnetic waves. This is the main motivation to consider \textit{fast radio bursts} (FRBs), lying in the radio band, \textit{i.e.} around $\sim \unit[O(100)]{MHz}$, as the astrophysical probe that we consider in this work. To illustrate this claim, we take the case of two axion star configurations composed of ALPs, with $(m_a, M_{\rm AS}) = (\unit[2.5 \times 10^{-15}]{eV}, 10^{-1} M_\odot)$ and $(\unit[1.25 \times 10^{-14}]{eV}, 10^{-2} M_\odot)$. The axion-photon coupling is fixed at $g_{a\gamma\gamma} = \unit[10^{-3}]{GeV^{-1}}$, and the lens and source redshifts are set at $z_{\rm L} = 0.4$ and $z_{\rm S} = 1.5$, respectively. One can easily check that the respective $w_{\rm E}$ values for the two axion star configurations are $w_{\rm E} \simeq 1.134$ and $w_{\rm E} \simeq 1.435$, so we are in the regime where the finite size effect is active. In Fig. \ref{fig:LensEquation} we present the right hand side of the reduced lens equation, Eq. (\ref{ReducedLens}), as solid curves, for three sample frequencies (wavelengths): $f_0 = \unit[500]{THz}, \unit[600]{MHz}, \unit[200]{MHz}$ ($\lambda \simeq \unit[600]{nm}, \unit[0.5]{mm}, \unit[1.5]{mm}$). As shown in Fig. \ref{fig:LensEquation}, each solid curve exhibits the presence of two extrema and an inflection point, with the peak becoming more pronounced for longer wavelengths. This peculiar characteristic of lensing by axion stars can then be attributed to two effects that are not present in typical pointlike lensing, shown as dashed curves, namely: \textit{finite-size} effect, and the \textit{axion-induced} lensing effect. In what follows we shall restrict our attention to cases where $\beta \leq \beta_c$. For $\beta < \beta_c$, we choose the image locations $\theta_\pm = \theta_{\rm E} \tilde{\theta}_\pm$  which give the two largest absolute image magnifications. We assign $\theta_+$ to be the image with the larger absolute magnification than $\theta_-$. We note here that the reduced lens equation in the form given by Eq. (\ref{ReducedLens}) is a nonlinear equation in $\tilde{\theta}$, which can only be solved numerically. 

\subsection{Conditions for lensing}
We now formulate our lensing criteria, which are necessary to obtain the lensing cross section, and eventually the optical depth of FRBs propagating through a population of axion stars. First we take the flux ratio
\begin{eqnarray}
    R_f \equiv \frac{\vert \mu(\theta_+)\vert}{\vert \mu(\theta_-)\vert}
\end{eqnarray}
to be below some critical value, to ensure that the image at $\theta_-$ is not overwhelmed by the brightness of the image at $\theta_+$; in our analysis, we choose this value to be \cite{Munoz:2016tmg}
\begin{eqnarray}
    R_{f,\max}=5.
\end{eqnarray}
We also require that the absolute time delay between the images
\begin{eqnarray}
    \Delta t = \left\vert t(\theta_+) - t(\theta_-)\right\vert
\end{eqnarray}
is at least equal to the timing resolution of the observational facility, but below the maximum observation time. In the case of CHIME \cite{Leung:2022vcx, CHIMEFRB:2022xzl}, a radio telescope that is suitable to observe FRBs from the sky, it is possible to achieve a timing resolution of as low as in the order of nanoseconds. On the other hand, an FRB signal has a typical duration of the order of milliseconds. In practice, looking at the flux signal alone may pose a challenge to discriminate between the peculiar features contained in the FRB signal, and the expected double peak signal from lensing. A technique was developed in \cite{CHIMEFRB:2021srp}, which overcomes this limitation. The method relies on the fact that gravitational lensing will produce multiple images that are coherent and with distinct brightness, that will manifest as interference fringes in the wave domain. For any voltage signal captured by the telescope, one can construct an autocorrelation function, from which one should be able to eventually extract information about the flux ratio and the time delay.

Thus, as a benchmark, we choose
\begin{eqnarray}
    \Delta t_{\min} = \unit[1]{\mu s}.
\end{eqnarray}
On the other hand, the same telescope scans over the whole sky as the Earth rotates, so that a certain area of the sky is only observable for a limited period per day. This observing time window, which depends on the latitude and longitude of the target area relative to CHIME, sets the maximum time delay that the facility can observe. From the technical specifications of CHIME \cite{CHIMEFRB:2021srp}, sky areas that are close to the equator can be observed up to a maximum of only a few minutes per day; on the other hand, sky areas near the North Celestial Pole correspond to a maximum observing time window of 24 hours in a day.
\begin{figure}[t]
    \centering
    \begin{tabular}{cc}
        \includegraphics[scale=0.36]{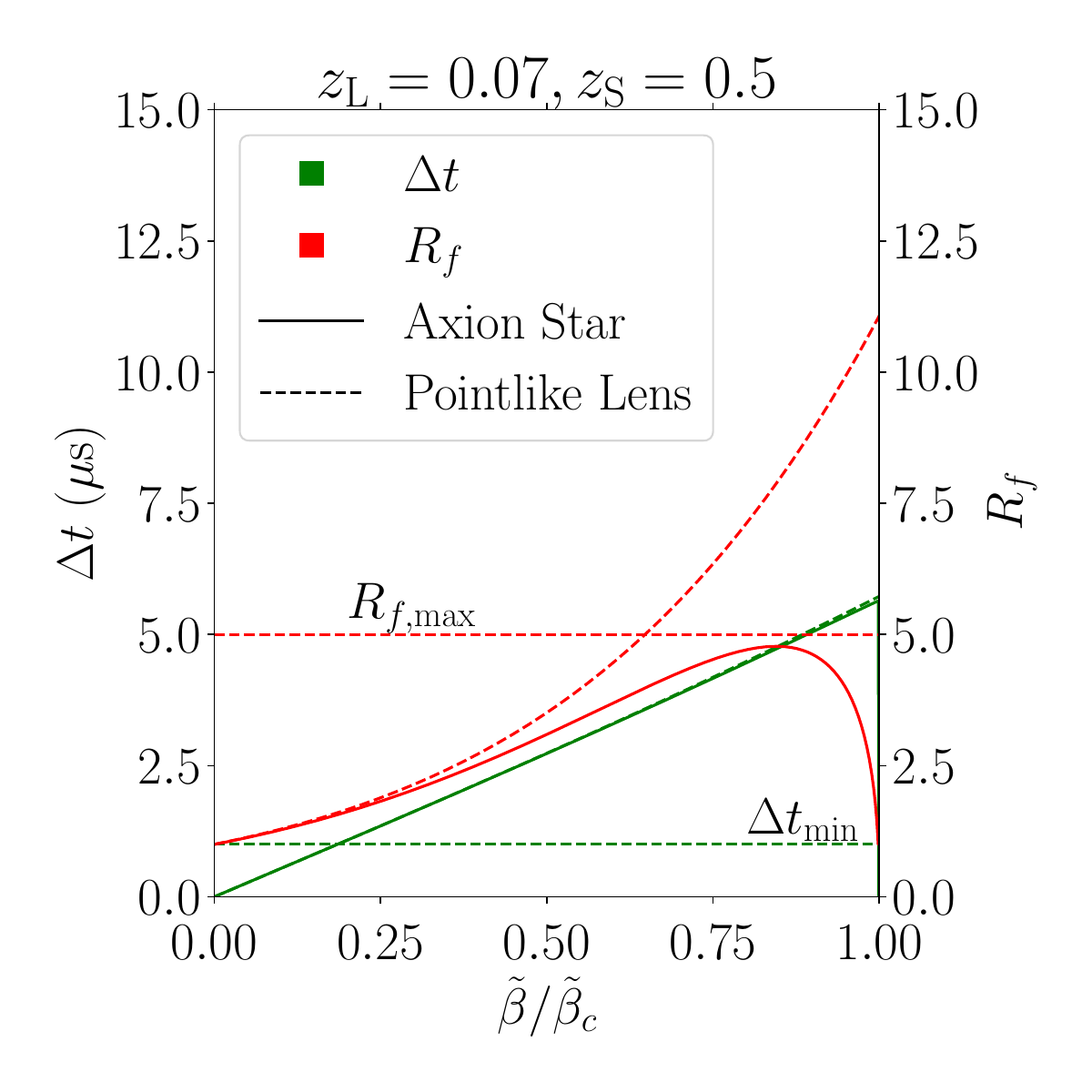}&\includegraphics[scale=0.36]{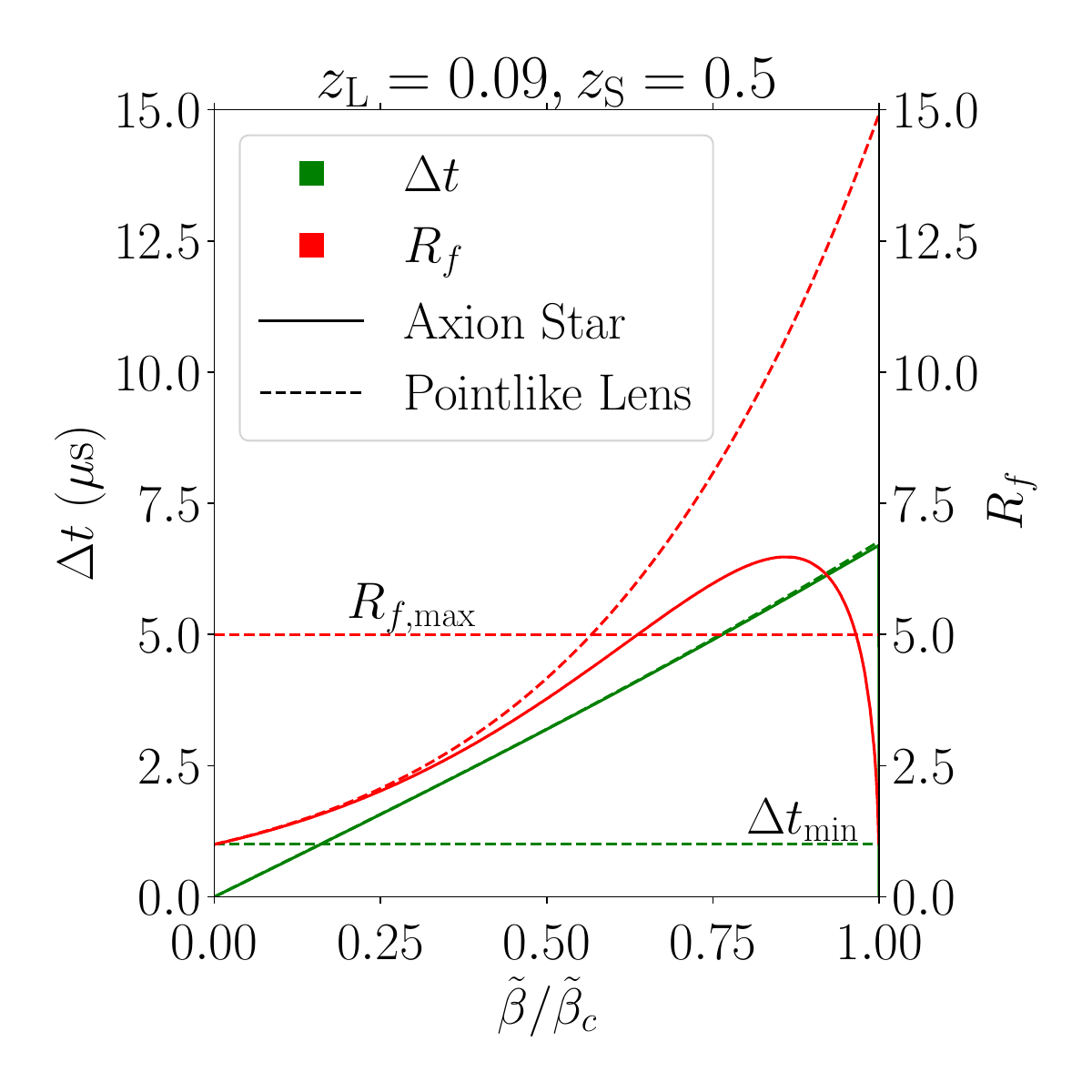}
    \end{tabular}
    \caption{Plot of $R_f$ and $\Delta t$ vs. $\tilde{\beta}/\tilde{\beta_c}$ for $m_a = \unit[4.4661 \times 10^{-15}]{eV}$, $M_{\rm AS} = 10^{-1} M_\odot$, and $g_{a\gamma\gamma} = \unit[10^{-6}]{GeV^{-1}}$. The two panels refer to $z_{\rm L} = 0.07$ (left) and $0.09$ (right), and the source redshift is $z_{\rm S} = 0.5$. We set $R_{f, \max}=5$ and $\Delta t_{\min}=1 \mu$s.}
    \label{fig:Dt_Rf_curve}
\end{figure}

To obtain a general idea on how the absolute time delay and flux ratio depend on the source position, which are necessary to implement the lensing criteria in practice, we show plots of the absolute time delay between the $\theta_\pm$ images and the flux ratio in Fig. \ref{fig:Dt_Rf_curve}, assuming the presence of axion stars with mass $M_{\rm AS} = 10^{-1} M_\odot$, composed of ALPs with mass $m_a = \unit[4.4661 \times 10^{-15}]{eV}$. The left and right panels correspond respectively to two different lens positions $z_{\rm L} = 0.07$ and $z_{\rm L} = 0.09$, for an FRB source at $z_{\rm S} = 0.5$. For comparison, we included the absolute time delay and flux ratio in the \textit{pointlike lens limit}, as dashed curves in both panels; in this limit the time delay and flux ratio are both monotonically increasing functions of the impact parameter $\beta$, where the expressions are reproduced below from \cite{Munoz:2016tmg}:
\begin{eqnarray}
\label{PointLikeDt} \Delta t &=& 4G_N M_{\rm AS} \left(1 + z_{\rm L}\right)\left[\frac{\tilde{\beta}}{2}\sqrt{\tilde{\beta}^2+4}+\ln\left(\frac{\sqrt{\tilde{\beta}^2+4}+\tilde{\beta}}{\sqrt{\tilde{\beta}^2+4}-\tilde{\beta}}\right)\right],\\
\label{PointLikeRf} R_f &=& \frac{\tilde{\beta}^2+2+\tilde{\beta}\sqrt{\tilde{\beta}^2+4}}{\tilde{\beta}^2+2-\tilde{\beta}\sqrt{\tilde{\beta}^2+4}}.
\end{eqnarray}
For reference, note that
\begin{eqnarray}
    4G_N M_{\rm AS} \simeq \unit[19.7]{\mu s}\left(\frac{M_{\rm AS}}{M_\odot}\right).
\end{eqnarray}
In contrast, for the case of lensing by an axion star, we observe in Fig. \ref{fig:Dt_Rf_curve} that the flux ratio decreases for a certain range of source positions. This can easily be seen by looking at two extreme limits for the source position: $\tilde{\beta} \rightarrow 0$ and $\tilde{\beta} \rightarrow \tilde{\beta}_c$. In the former case, we find that
\begin{eqnarray}
    \frac{1}{\mu_\pm} \approx \pm\tilde{\beta}\left[\frac{2F'(\tilde{\theta}_E^2)}{\tilde{\theta}_E}\right],\quad F(x) \equiv x - 1 + (1 - A w_E^2 x) \exp(-w_E^2 x),
\end{eqnarray}
and $\tilde{\theta}_E$ is the position of the image, when the source is along the line of sight of the observer, so that $R_f \rightarrow 1$. In the latter case, we have
\begin{eqnarray}
    \frac{1}{\mu_\pm} \approx 2\tilde{\beta}_c \left[2\tilde{\theta}_c F''(\tilde{\theta}_c^2) + \frac{\tilde{\beta}_c}{2\tilde{\theta}_c^2} \right]\left(\frac{\tilde{\theta}_\pm-\tilde{\theta}_c}{\tilde{\theta}}\right),
\end{eqnarray}
where $\tilde{\theta}_c$ is the image position at the maximum source position which still leads to multiple lens images. In the limit $\tilde{\beta} \rightarrow \tilde{\beta}_c$, the image positions of the two brightest images both approach $\tilde{\theta}_c$, and hence $R_f \rightarrow 1$. Meanwhile, the time delay is a monotonically increasing function of the source position, except in the limit $\tilde{\beta} \rightarrow \tilde{\beta}_c$ where the two brightest images are almost coincident.

\begin{figure}[t!]
    \centering
    \begin{tabular}{c|c|c}
    $M_{\rm AS} = 10^{-1} M_\odot$&$M_{\rm AS} = 10^{-2} M_\odot$&$M_{\rm AS} = 10^{-3} M_\odot$\\ 
    \hline  \hline 
    \includegraphics[scale=0.3]{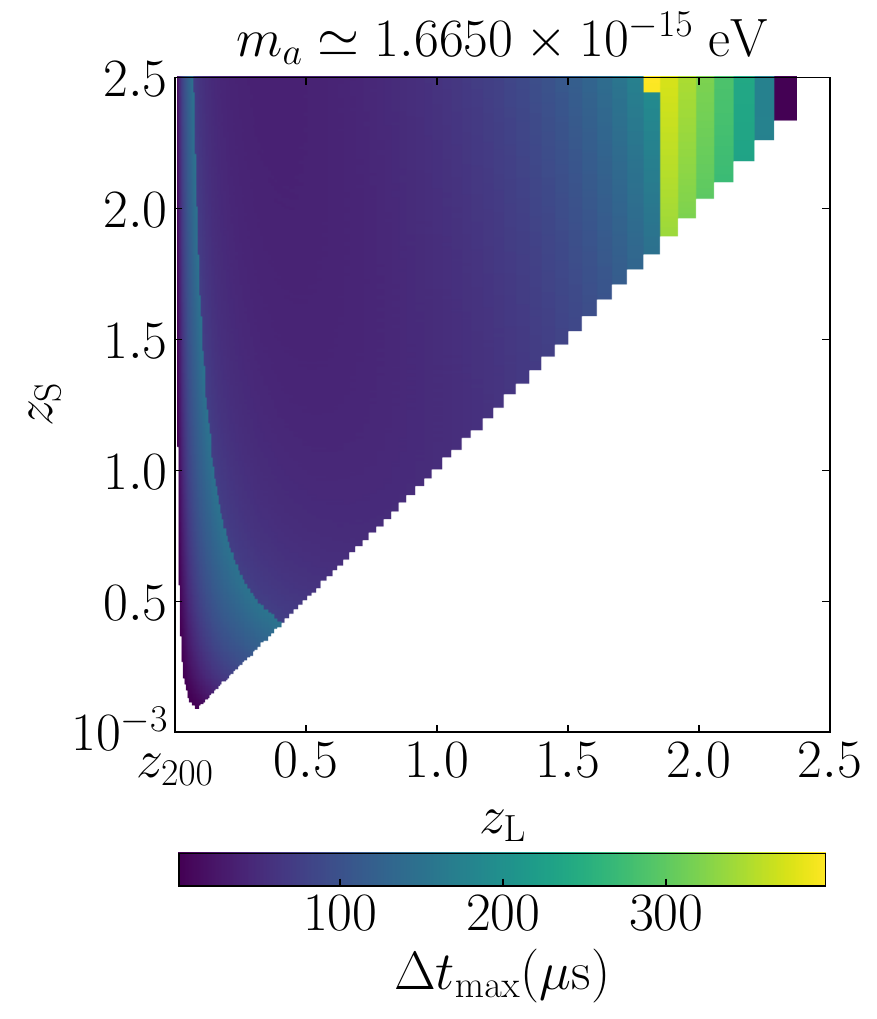} &  \includegraphics[scale=0.3]{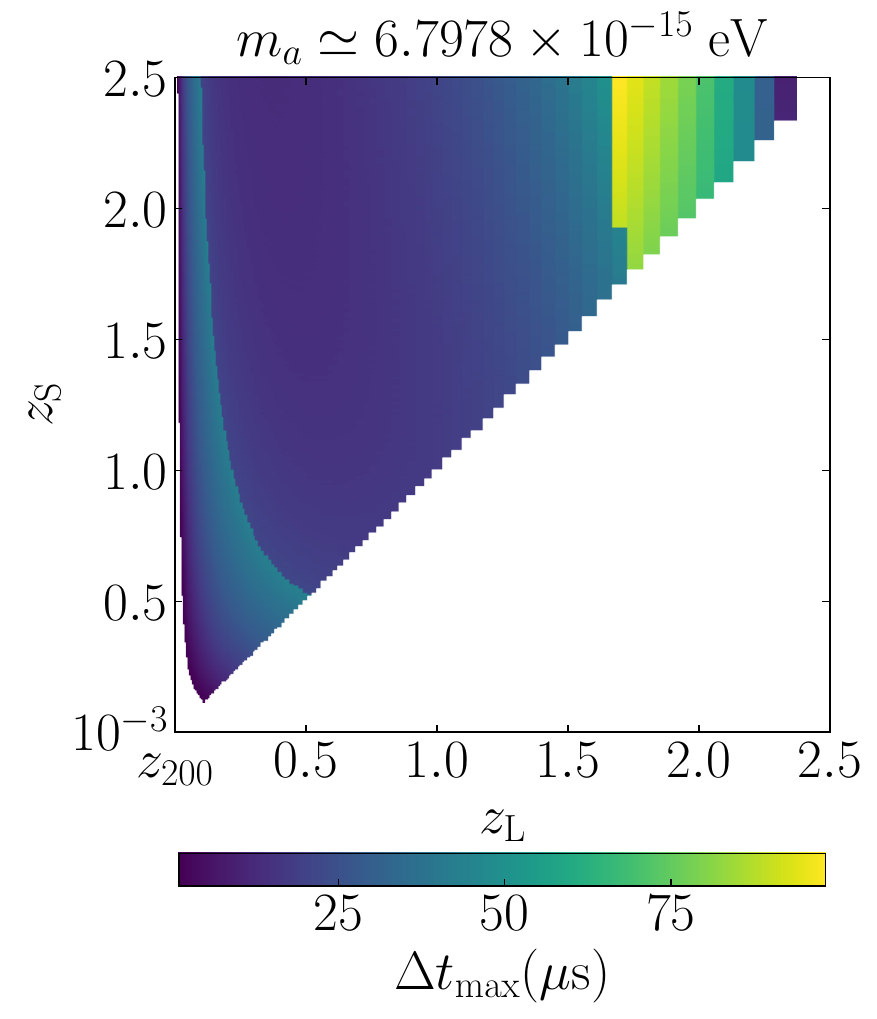} & \includegraphics[scale=0.3]{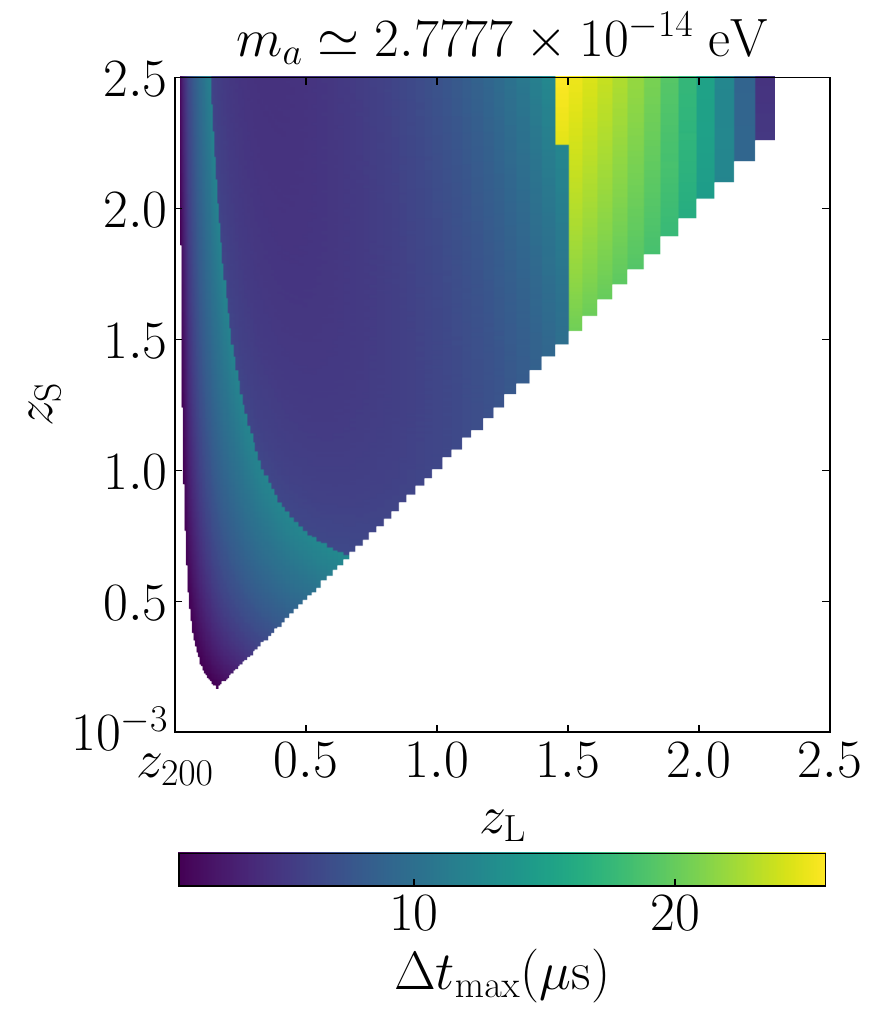} \\
    (a) & (b) & (c) \\
    \includegraphics[scale=0.3]{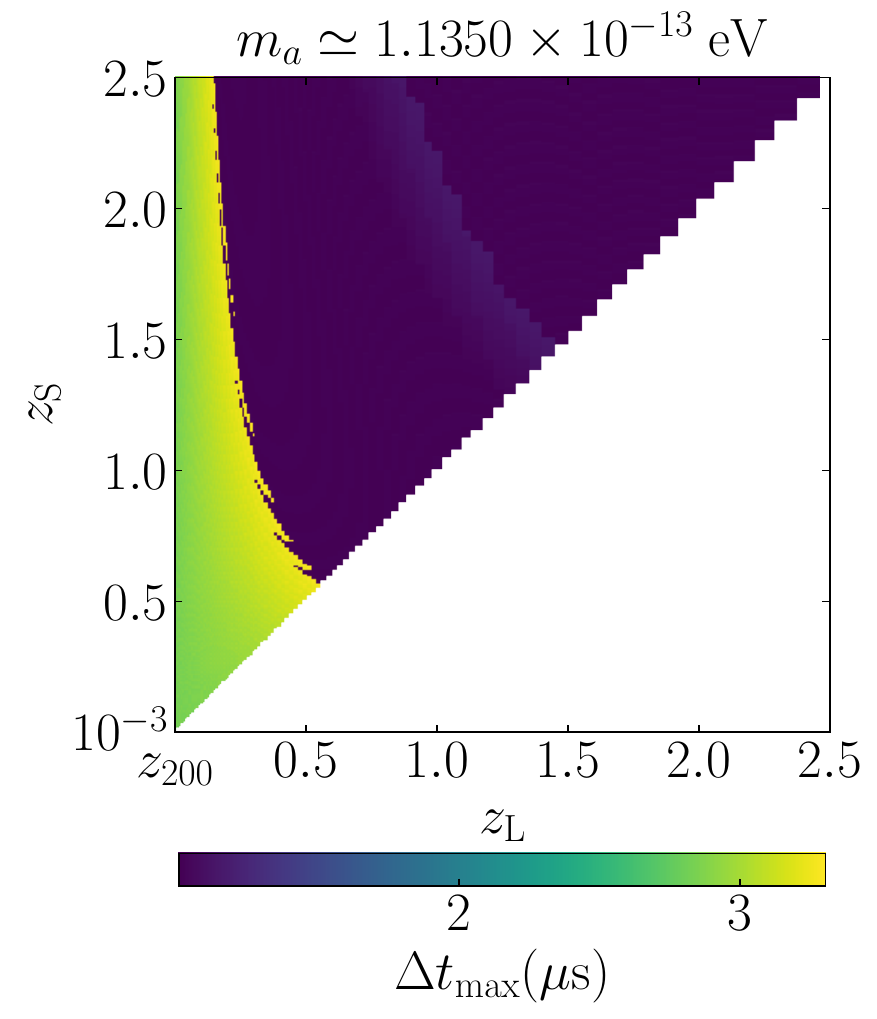} & \includegraphics[scale=0.3]{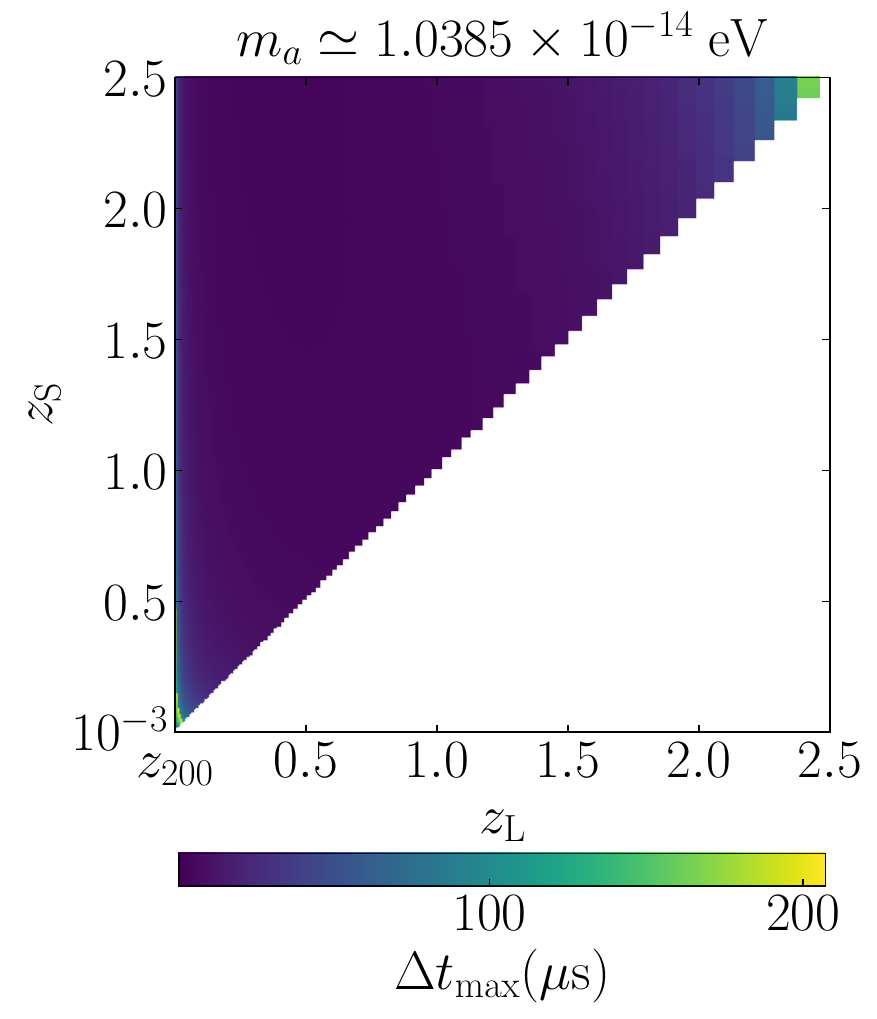} & \includegraphics[scale=0.3]{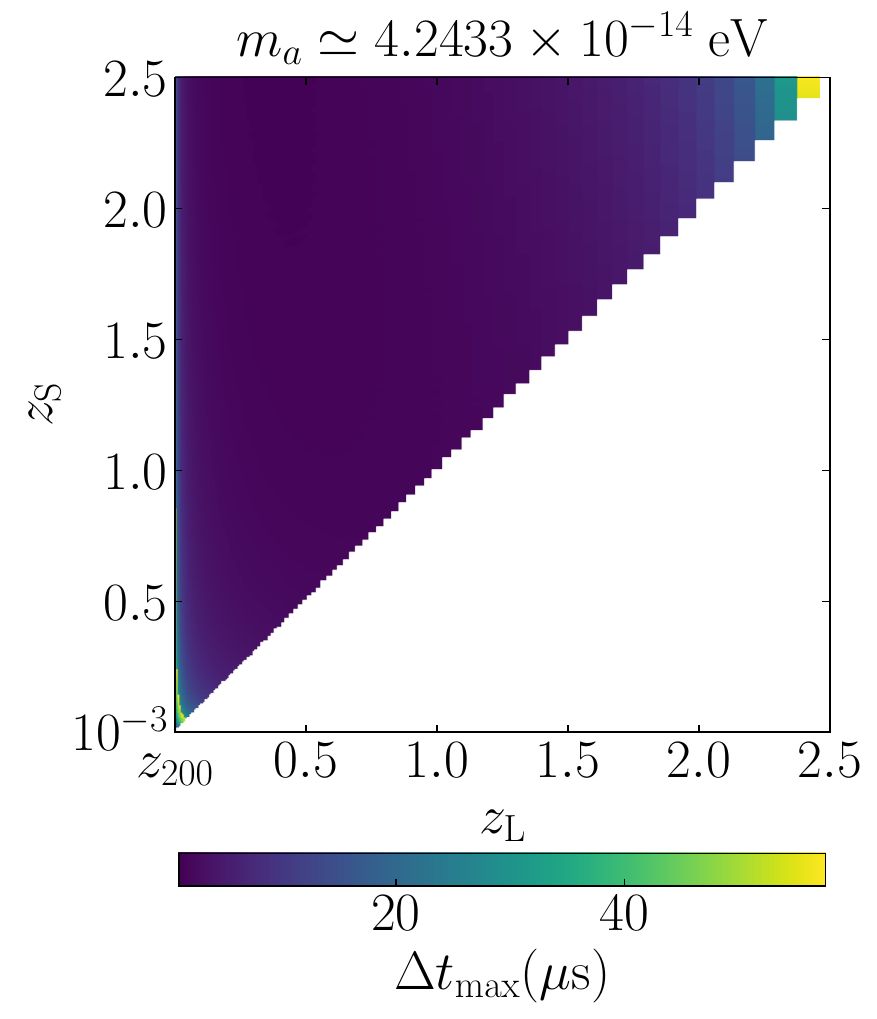} \\
    (d) & (e) & (f) \\
    \end{tabular}
    \caption{Scans of $\Delta t_{\rm max}$ over the $z_{\rm S}$-$z_{\rm L}$ plane, for a selection of ALP axion stars with masses $10^{-1} M_\odot, 10^{-2} M_\odot$, and $10^{-3} M_\odot$. We set $g_{a\gamma\gamma}=10^{-2}\;{\rm GeV}^{-1}$.}
    \label{fig:Dtmax_zSzL}
\end{figure}

Notice also that the maximum time delays in each of the sample cases shown in Fig. \ref{fig:Dt_Rf_curve} are way below the lower bound for the maximum observation time $\sim \unit[O(1)]{min}$, which provide indications that we can safely ignore the criterion set by the maximum observable time delay. More generally, we can gain more confidence in this claim by looking at Fig. \ref{fig:Dtmax_zSzL}, which correspond to scans in the lens-source redshift plane of $\Delta t_{\rm max}$, which we define to be the maximum time delay in the range of source positions where $R_f \leq R_{f, \rm max}$ and $\Delta t \geq \Delta t_{\min}$. Taking $g_{a\gamma\gamma} = \unit[10^{-2}]{GeV^{-1}}$, each column corresponds to a choice for the axion star mass, where each panel demonstrates cases where $\Delta t_{\rm max}$ is sizable enough to potentially reach the bottom threshold of $\sim O(1)$ minute associated with the maximum observation time. We have checked that the lensing criterion imposed by the maximum observation time can be safely ignored for most of the parameter space of interest, and we only show a selection in Fig. \ref{fig:Dtmax_zSzL}.

\subsection{Lensing cross section, optical depth, and number of lensing events}
\begin{figure}[t!]
    \centering
    \begin{tabular}{cc}
    \includegraphics[scale=0.3]{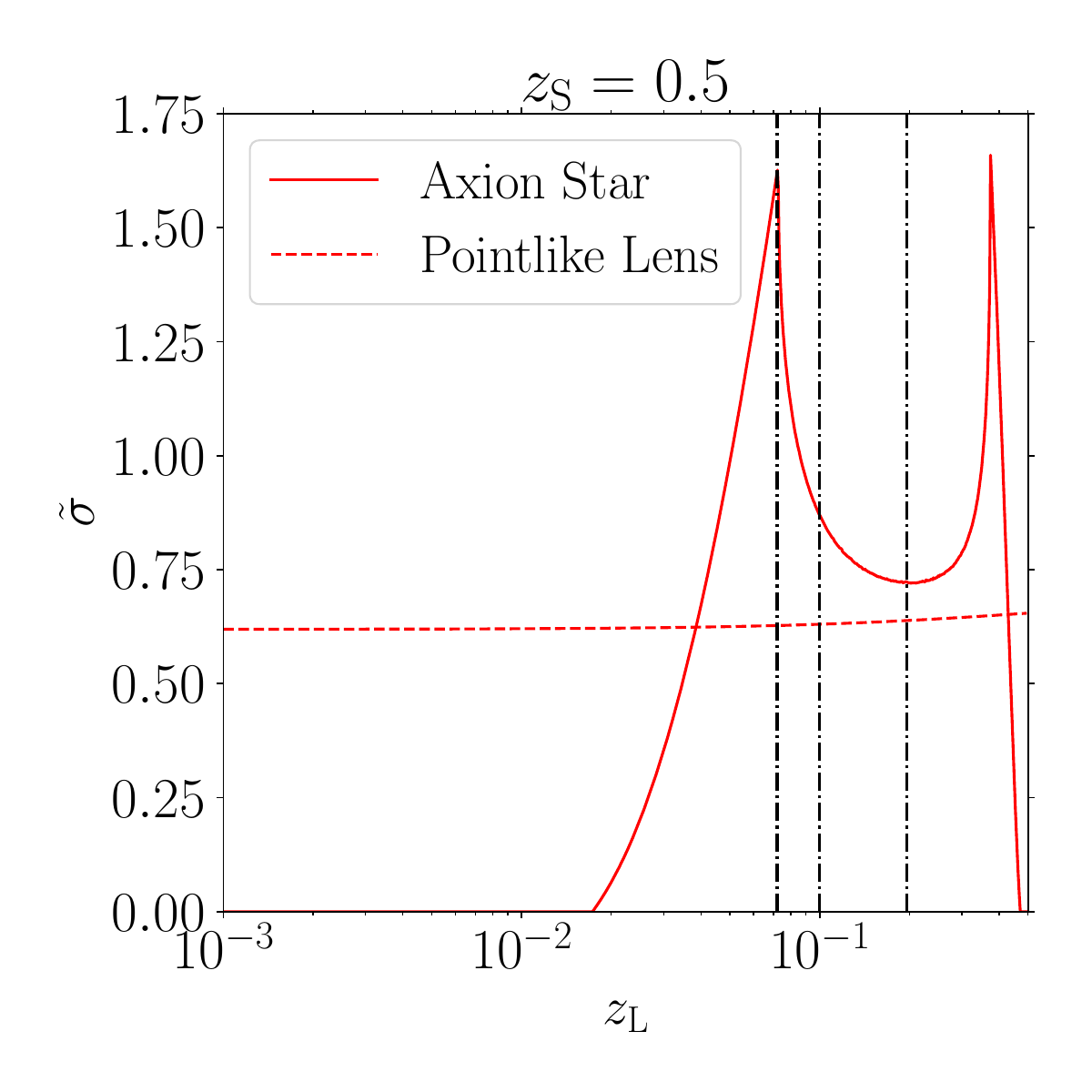} & \includegraphics[scale=0.3]{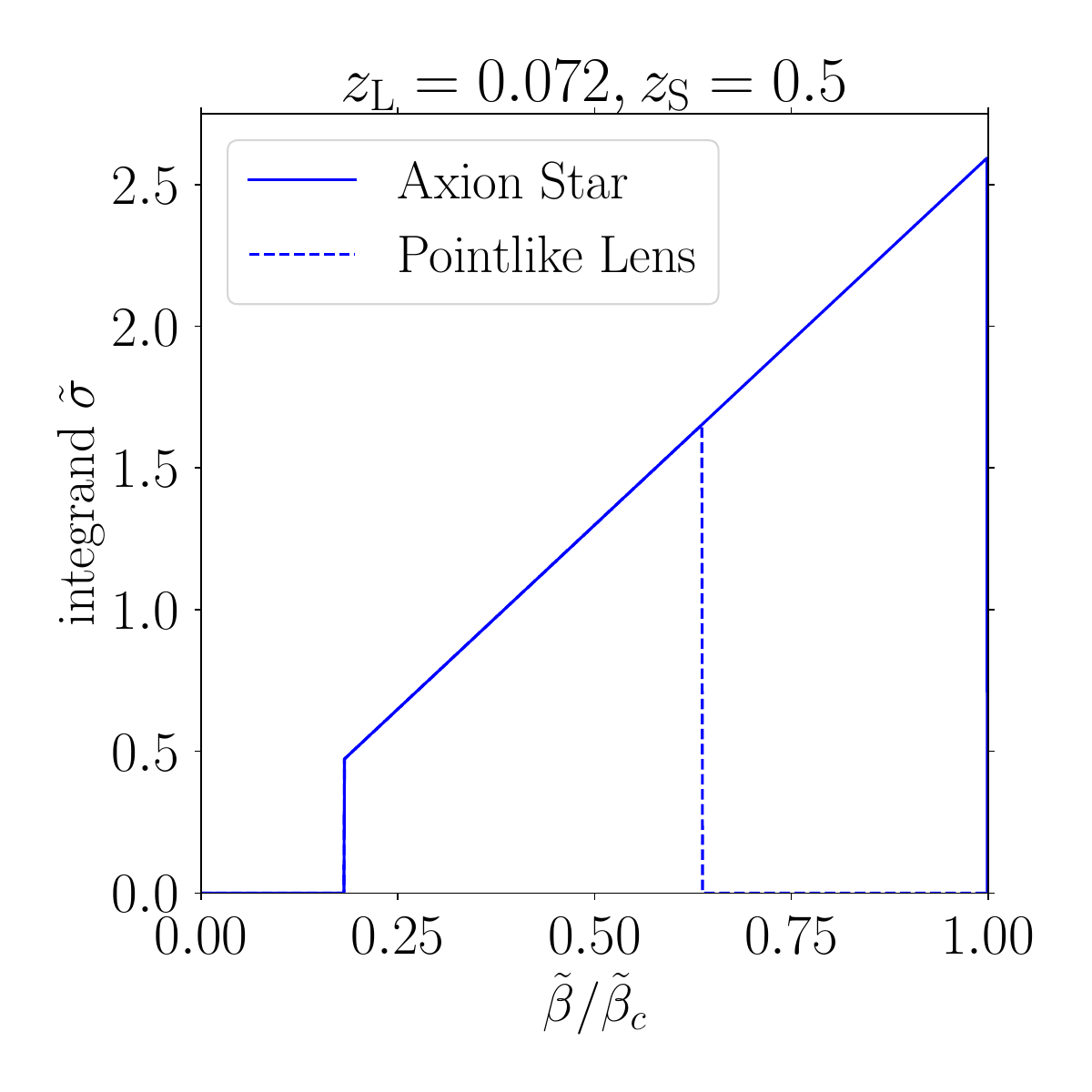}\\
    (a)&(b)\\
    \includegraphics[scale=0.3]{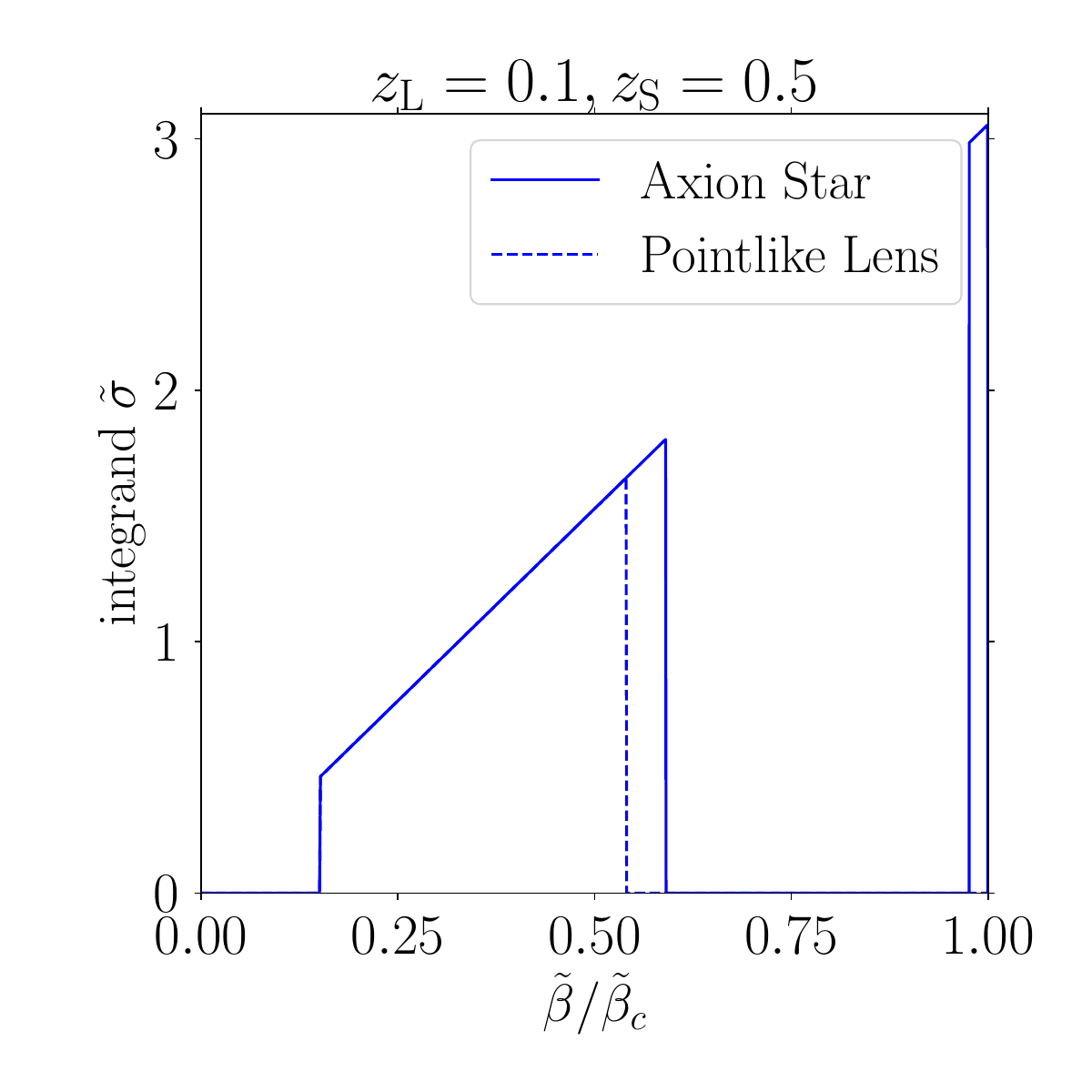} & \includegraphics[scale=0.3]{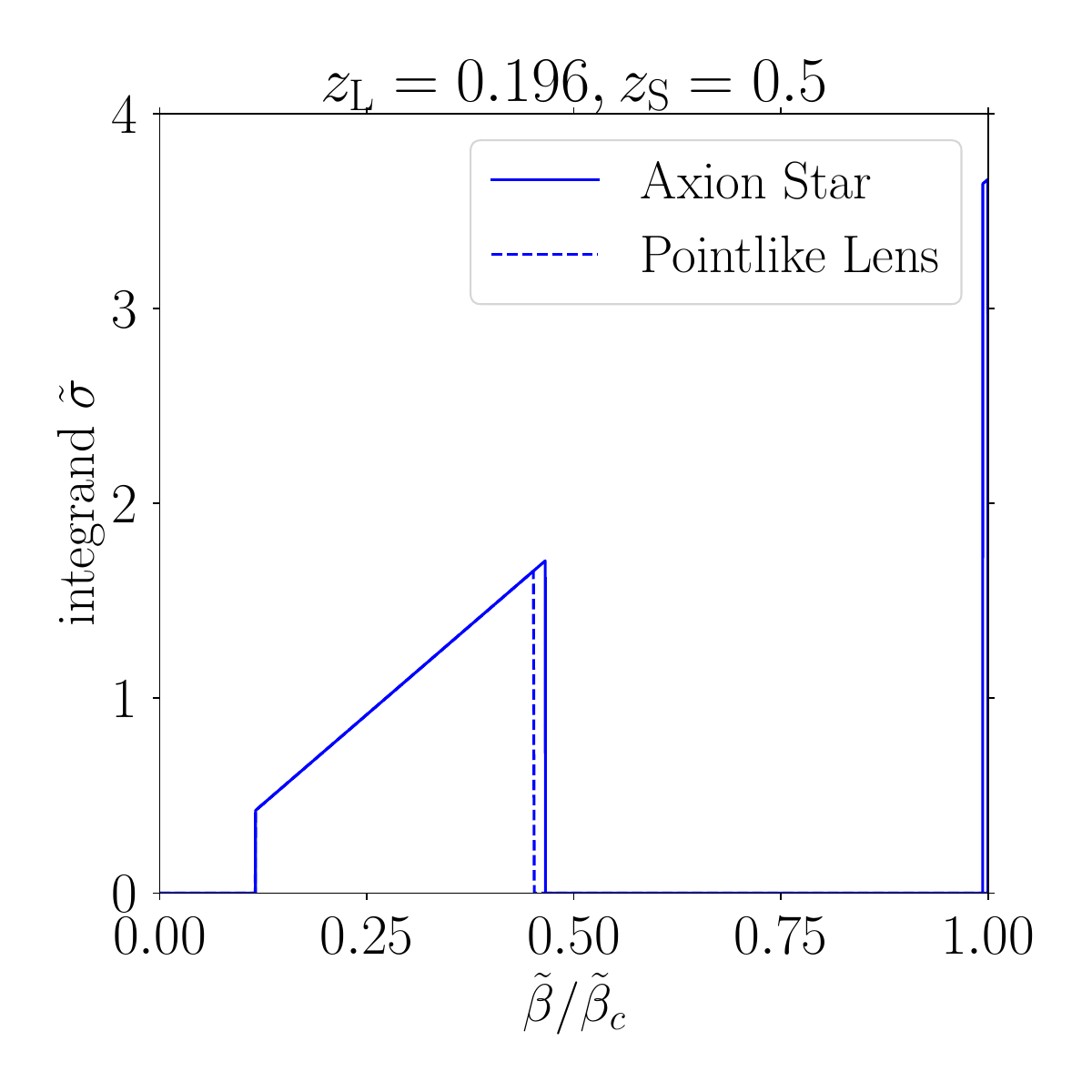}\\
    (c)&(d)
    \end{tabular}
    \caption{Taking the same axion star configuration in Fig. \ref{fig:Dt_Rf_curve}, we show the plot of $\tilde{\sigma}$ versus $z_{\rm L}$ (Panel a), and the corresponding integrand of $\tilde{\sigma}$ for $z_{\rm L} = 0.072, \; 0.1, \; 0.196$.}
    \label{fig:xsec}
\end{figure}

Calculating the lensing probability requires us to know the effective lensing cross section of an axion star. In the pointlike case, the effective lensing cross section can be less than the estimate $\pi D_{\rm L}^2 \theta_{\rm E}^2$ which just uses the Einstein radius; if we incorporate the lensing criteria imposed by the maximum flux ratio and the timing resolution of the radio telescope facility, then the cross section is just the area of the annular region set by $\beta_{\rm min}$ and $\beta_{\rm max}$, \textit{i.e.} 
\begin{eqnarray}
    \sigma = \pi D_{\rm L}^2 \left(\beta_{\rm max}^2 - \beta_{\rm min}^2\right) &=& 4\pi G_N M\frac{D_{\text{LS}} D_{\rm L}}{D_{\rm S}}\left(\tilde{\beta}_{\rm max}^2 - \tilde{\beta}_{\rm min}^2\right),\\
    \beta_{\max} = \left[\frac{1+R_{f, \max}}{\sqrt{R_{f, \max}}}-2\right]^{1/2}&,&\quad \Delta t(\beta_{\min}) = \Delta t_{\min}.
    \label{eq:sigma}
\end{eqnarray}
However, in our current setup, there are situations where $R_f$ may increase or decrease with $\beta$, so that the range of impact parameter values where lensing can occur is no longer restricted to lie in a single interval. This motivates the need to generalize the notion of the lensing cross section: within an infinitesimal range of impact parameters $[\beta, \beta+d\beta]$ which satisfy the lensing criteria, the corresponding infinitesimal cross section is $d\sigma = 2\pi (D_{\rm L} \beta) (D_{\rm L} d\beta)$. Then 
\begin{eqnarray}
   \label{ASXSec} \sigma = 4\pi G_N M\frac{D_{\text{LS}} D_{\rm L}}{D_{\rm S}} \tilde{\sigma}(\tilde{\beta_c}; D_{\rm L}, D_{\rm S}),
\end{eqnarray}
where
\begin{eqnarray}
    \label{DefnSigmaTilde} \tilde{\sigma}(\tilde{\beta}_c; D_{\rm L}, D_{\rm S}) &\equiv& 2\int_0^{\tilde{\beta}_c} \Theta\left(R_{f, \text{max}}-R_f(\tilde{\beta})\right)\Theta\left(\Delta t(\tilde{\beta}) - \Delta t_{\min}\right)\tilde{\beta}~d\tilde{\beta},\\
    \Theta(x) &\equiv& \begin{cases}
        1,\quad x > 0\\
        ~\\
        0,\quad x \leq 0.
    \end{cases}
\end{eqnarray}
We shall denote $I_\beta$ to be the range of source angular positions that satisfy the lensing criteria, \textit{i.e.} those $\beta$ where the product of the Heaviside functions in Eq. (\ref{DefnSigmaTilde}) is 1. In Panels b, c, d of Fig. \ref{fig:xsec} we take the same benchmark axion star configuration as in Fig. \ref{fig:Dt_Rf_curve}, with the same FRB source redshift, and plotted the integrand of Eq. (\ref{DefnSigmaTilde}) for a selection of lens redshifts $z_{\rm L} = 0.07, 0.09, 0.2$. The dotted lines assume a pointlike lens, while the solid lines correspond to an axion star with finite size, and with axion-photon interactions taken into account. Indeed we see that $I_\beta$ in the pointlike case covers a single interval, with endpoints corresponding to $\beta_{\rm min}$ and $\beta_{\rm max}$; meanwhile, $I_\beta$ in the case of axion star lensing consists of disjoint intervals in $\beta$. In Panel a, we show $\tilde{\sigma}$ versus lens redshift, for the same $z_{\rm S} = 0.5$, for the case of axion star lensing (solid) and the case of pointlike lensing (dashed). We observe that $\tilde{\sigma}$ for axion star lensing, compared to the pointlike lensing case, has more features; in particular, $\tilde{\sigma}$ is no longer a monotonic function of $z_{\rm L}$, and we notice the presence of peaks at certain redshift values.

Note that the cross section in Eq. (\ref{ASXSec}) applies for a single axion star, for a given lensing configuration where the lens and source positions are located respectively at $D_{\rm L}$ and $D_{\rm S}$. Given a population of axion stars through which the signal propagates from an FRB source at $D_{\rm S}$, the infinitesimal \textit{optical depth} is \cite{Munoz:2016tmg}
\begin{eqnarray}
    d\tau(D_{\rm S}) = n(D_{\rm L})\sigma(D_{\rm L}, D_{\rm S}) \left(1+z_{\rm L}\right)^2 d\chi(z_{\rm L}),\quad n(D_{\rm L}) \equiv \frac{\rho_{\rm AS}(D_{\rm L})}{M_{\rm AS}},
\end{eqnarray}
where $\chi(z_{\rm L})$ is the comoving distance at redshift $z_{\rm L}$, $n(D_{\rm L})$ is the number density of axion stars at $D_{\rm L}$. In the last line, we assumed that the population of axion stars possess the same mass $M_{\rm AS}$. Furthermore, the axion stars may reside both in extragalactic space, and in the Milky Way Galaxy. For the extragalactic population, the comoving number density is determined by the energy density of the dark matter (DM) component in the Universe's energy budget, and the fraction of DM, $f_{\rm AS}$, residing in axion stars, \textit{i.e.}
\begin{eqnarray}
    n_{\rm egal} = \frac{f_{\rm AS} \rho_{\rm DM}}{M_{\rm AS}} = \frac{f_{\rm AS}}{M_{\rm AS}}~\frac{3H_0^2 \Omega_{\rm DM}}{8\pi G_N},
\end{eqnarray}
where the cosmological parameters are obtained from \cite{Planck:2018vyg}. Meanwhile, the Galactic component is assumed to have the same fraction $f_{\rm DM}$ of the DM energy density in the Milky Way. Adopting the NFW profile, we have
\begin{eqnarray}
   n_{\rm gal}(r_{\rm L}) =  \frac{1}{M_{\rm AS}}\frac{\rho_0}{(r_{\rm L}/R_s)\left(1+r_{\rm L}/R_s\right)^2},
\end{eqnarray}
where $r_{\rm L}$ is the distance of the axion star from the Galactic center. The best fit values for the parameters in the NFW profile, $\rho_0$ and $R_s$, are taken from \cite{McMillan:2016jtx}. If we adhere to a coordinate system centered about the Sun's location, the distance between the observer and the axion star in terms of the redshift of the lens $z_{\rm L} \ll 1$ is $D_{\rm L} \approx z_{\rm L}/H_0$. Then
\begin{eqnarray}
    r_{\rm L} \simeq \sqrt{R_\odot^2 + \frac{z_{\rm L}^2}{H_0^2} - 2R_\odot \frac{z_{\rm L}}{H_0} x},\quad R_{\odot} = \unit[8.21]{kpc},
\end{eqnarray}
where $x$ is the direction cosine of the vector pointing towards the axion star from the observer, projected along the axis joining the observer and the Galactic center. To simplify our analysis, we assume that the sources are uniformly distributed over solid angle. Then we average the density distribution over solid angle so that
\begin{eqnarray}
\langle \rho_{\rm NFW}(z_{\rm L}) \rangle \equiv \int \frac{d\Omega}{4\pi} \rho_{\rm NFW}(r_{\rm L}(z_{\rm L},x)) = \frac{1}{2}\int_{-1}^1 dx~\rho_{\rm NFW}(r_{\rm L}(z_{\rm L},x)).
\end{eqnarray}
Now that we have established the number densities for the two axion star populations, we can now write down the different contributions to the optical depth: the \textit{extragalactic contribution} to the optical depth is
\begin{eqnarray}
    \tau_{\rm egal}(M_{\rm AS}, z_{\rm S}) &=& \frac{3}{2}f_{\rm AS}\Omega_{\rm DM}\int_0^{z_{\rm S}} dz_{\rm L}~\frac{D_{\text{LS}} D_{\rm L}}{D_{\rm S}}\frac{H_0^2}{H(z_{\rm L})}\left(1+z_{\rm L}\right)^2~\tilde{\sigma}(\tilde{\beta}_c),\\
    H(z) &=& H_0 \sqrt{\Omega_{\rm m}(1+z)^3 + \Omega_\Lambda}.
\end{eqnarray}
Meanwhile the \textit{Galactic contribution} is
\begin{eqnarray}
    \tau_{\text{gal}}(M_{\rm AS}, z_{\rm S}) \simeq 4\pi G_N f_{\rm AS}\int_0^{z_{200}} \frac{z_{\rm L}~dz_{\rm L}}{H_0^2}~\langle \rho_{\rm NFW}(z_{\rm L}) \rangle~\tilde{\sigma}(\tilde{\beta}_c).
\end{eqnarray}
In the last line, we used the approximation $D_{\text{LS}} \approx D_{\rm S}$, $\chi(z) \approx z/H(z)$, and $H(z) \approx H_0$ for $z \ll 1$. For the Galactic contribution, we have taken the boundary of the Galactic halo to be at the virial radius $r_{200}$ within which the average mass density is 200 times the critical density at present time. In terms of the concentration parameter $c=15.4$ \cite{McMillan:2016jtx}, we have $r_{200}=cR_s$ \cite{Navarro:1995iw}. The corresponding redshift of the virial radius is $z_{200} \simeq 6.78 \times 10^{-5}$. Given a collection of FRB sources, we take the following source redshift distribution for constant comoving number density \cite{Munoz:2016tmg}
\begin{eqnarray}
    \label{RedshiftDist}    N(z)=\mathcal{N}\frac{\chi^2(z)}{H(z)(1+z)}\exp\left[-\frac{d^2_{\rm L}(z)}{2d^2_{\rm L}(z_{\text{cut}})}\right],
\end{eqnarray}
where $z_{\rm cut} = 0.5$, and the normalization constant $\mathcal{N}$ is set by taking the integral of the source redshift distribution, over all source redshifts, to unity. In Eq. (\ref{RedshiftDist}), $d_{\rm L}(z)$ is the luminosity distance of the FRB source at redshift $z$. Taking all FRB sources into consideration, we can perform a convolution of the source redshift distribution, with the optical depth of both Galactic and extragalactic space for a source located at $z_{\rm S}$, which will give us the \textit{integrated optical depth}
\begin{eqnarray}
    \Bar{\tau}(M_{\rm AS}) = \bar{\tau}_{\rm egal} + \bar{\tau}_{\rm gal},\quad \bar{\tau}_i \equiv \int dz_{\rm S}~\tau_i(M_{\rm AS},z_{\rm S})N(z_{\rm S}),
\end{eqnarray}
where $i = \rm egal,\; \rm gal$. The integrated optical depth provides us with a prediction for the number of lensed FRB sources in the optically thin regime, given by
\begin{eqnarray}
\label{NLens}    N_{\text{lensed}}=(1-e^{-\Bar{\tau}})N_{\text{obs}}.
\end{eqnarray}
At present, no FRB lensing event has been observed by radio telescope observations such as CHIME \cite{Leung:2022vcx, CHIMEFRB:2022xzl}, and this can be used as a criterion to place constraints on the integrated optical depth, and eventually, the fundamental axion parameters and the fraction $f_{\rm AS}$, which is assumed to be the same for the Galactic and extragalactic populations.
\section{Results and discussion}
\label{sec:resultsanddiscussion}
We consider two distinct scenarios in which the components of axion stars are made up of: QCD axions, in which the relationship between the axion mass $m_a$ and the axion decay constant $f_a$ is fixed by
\begin{eqnarray}
    m_a = \unit[10^{-5}]{eV}\left(\frac{\unit[5.691 \times 10^{11}]{GeV}}{f_a}\right);
\end{eqnarray}
or axionlike particles (ALPs) in which $m_a$ and $f_a$ are treated as free parameters. For the latter case, the relationship between $m_a$ and $f_a$ can be fixed by setting the axion star mass $M_{\rm AS}$ and then use the relationship between $M_{\rm AS}$, $m_a$, and $f_a$ in Eq. (\ref{MASRelation}).
\begin{figure}[t!]
    \centering
    \begin{tabular}{cc}
    \includegraphics[scale=0.4]{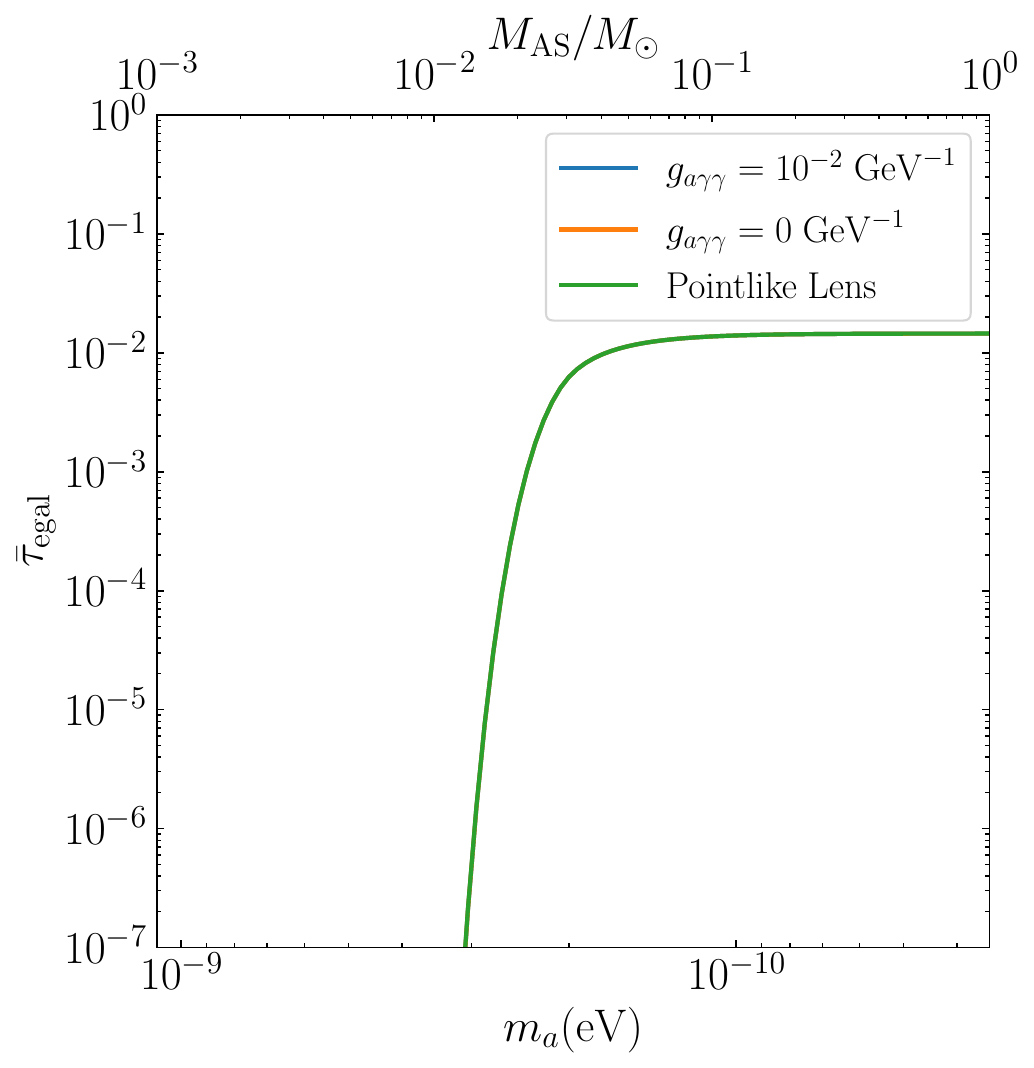}&\includegraphics[scale=0.4]{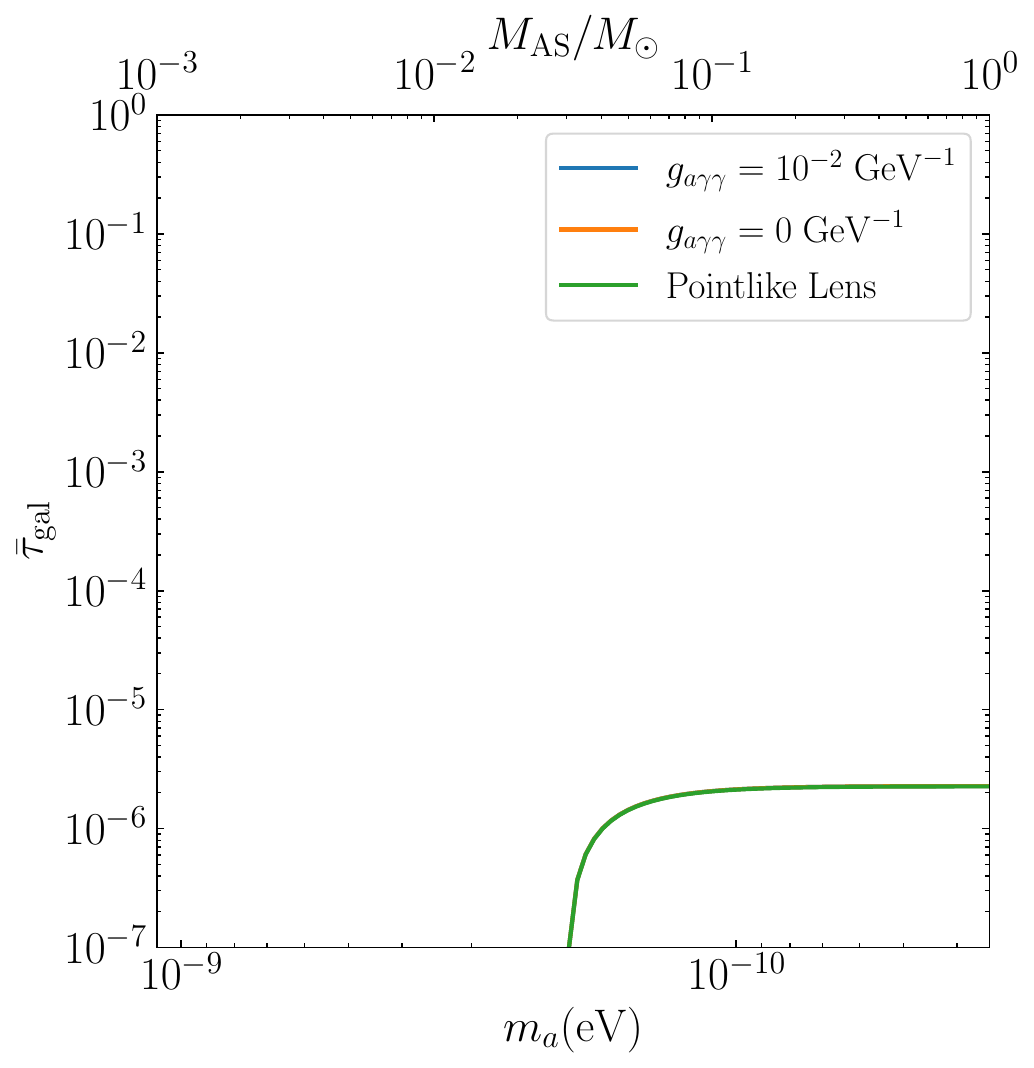}\\
    (a)&(b)
    \end{tabular}
    \caption{Plots of the integrated optical depth versus axion star mass, for the extragalactic and Galactic contributions, are shown for the QCD axion case. In both panels, the curves for $g_{a\gamma\gamma} = 0$ and $\unit[10^{-2}]{GeV^{-1}}$ converge with the pointlike limit.
    }
    \label{fig:QCDAxionCase}
\end{figure}

\subsection{Axion stars: QCD axion case}
We first discuss the case of stars made up of QCD axions, where we take the axion star population to saturate the DM energy density, \textit{i.e.} $f_{\rm AS} = 1$. In Fig. \ref{fig:QCDAxionCase}, we show the results for the integrated optical depth versus the axion star mass, in the range $10^{-3} M_\odot \leq M_{\rm AS} \leq M_\odot$, for the extragalactic (Panel a) and Galactic contributions (Panel b). We have included the sensitivity curve assuming a pointlike lens, as well as cases where the axion-photon coupling is $g_{a\gamma\gamma} = \unit[10^{-2}]{GeV^{-1}}$ and $\unit[0]{GeV^{-1}}$. We verified that, as we vary the axion-photon coupling from $g_{a\gamma\gamma} = \unit[10^{-10}]{GeV^{-1}}$ to $\unit[1]{GeV^{-1}}$, the trend in the integrated optical depth versus axion star mass remains unchanged, relative to both cases where the axion-photon interactions are turned off, and in the case of lensing by a pointlike object. We can infer that the finite size and the axion-induced lensing effects are suppressed; by examining the reduced lens equation Eq. (\ref{ReducedLens}), the suppression may come from the exponential term, which could be the result of a large $w_{\rm E}$. We find that $w_{\rm E}$ is large enough for $10^{-3} M_\odot \leq M_{\rm AS} \leq M_\odot$ with the extragalactic range of redshifts from $10^{-5} \leq z_{\rm L} \leq 10^{-2}$, as well as the Galactic range $10^{-20} \leq z_{\rm L} \leq 10^{-5}$ with an FRB source at redshift $z_{\rm S} = 0.5$. The large values of $w_{\rm E}$ make the axion star effectively pointlike, except only in the range of extremely small redshifts below $\sim 10^{-20}$. In fact, in the QCD axion star case, we have 
\begin{eqnarray}
    w_{\rm E}^2 \simeq 3.42 \times 10^{22} h^{-1}\left(\frac{M_{\rm AS}}{M_\odot}\right) \left(D_{\rm S} H_0\right)\left(\frac{D_{\rm L}}{D_{\rm S}}\right)\left(1-\frac{D_{\rm L}}{D_{\rm S}}\right).
\end{eqnarray}
On the other hand, one can understand the behavior of the integrated optical depth versus axion star mass by adopting the time delay and flux ratio expressions in eqs. (\ref{PointLikeDt}) and (\ref{PointLikeRf}), applicable in the pointlike limit. In the limit of lower axion star masses, where we see a steep decrease in the integrated optical depth, Eq. (\ref{PointLikeDt}) tells us that the prefactor in $\Delta t$ becomes smaller. Then within $0 \leq \beta \leq \beta_c$, it is more likely that the time delay is mostly below the threshold $\Delta t_{\min}$. Since the flux ratio $R_f$ is unaffected by $M_{\rm AS}$, the set $I_\beta$ can become empty and the lensing cross section will be zero. In the opposite limit, the integrated optical depth tends to a fixed value for sufficiently large $M_{\rm AS}$ (small $m_a$). Here the set $I_\beta$ can be nonempty because the prefactor in $\Delta t$ is sufficiently large. Then the optical depth in the pointlike case is \cite{Narayan:1996ba}
\begin{eqnarray}
    \label{TauEstimatePointLike}
    \tau(z_{\rm S}) &\simeq& \int_0^{D_{\rm S}}\left[n_{\rm L, egal} + n_{\rm L, gal} (D_{\rm L})\right]~\pi D_{\rm L}^2 \theta_{\rm E}^2~d(D_{\rm L})\\
    &\simeq& \frac{2\pi}{3}G_N \left(\rho_{\rm AS, egal} D_{\rm S}^2 + \rho_{\rm AS, ave, gal} R_{\rm gal}^2\right).
\end{eqnarray}
The estimation above can be used to understand the relative sizes of the extragalactic and Galactic contributions to the integrated optical depth. First, note that the cosmological DM density $\rho_{\rm DM} \sim \unit[10^{-6}]{GeV/cm^3}$, while the local DM density is $\sim \unit[10^{-1}]{GeV/cm^3}$; however, the possible enhancement in the integrated optical depth due to the larger number density of axion stars in the Galactic halo, can be suppressed by the short distance covered by light within the Galaxy, compared with the extragalactic distance between the source and the observer. Performing a convolution of Eq. (\ref{TauEstimatePointLike}) with the source redshift distribution Eq. (\ref{RedshiftDist}), that is peaked at $z_* \sim 0.5$, will further suppress the Galactic contribution to the integrated optical depth. 

\subsection{Axion stars: ALP case}
\begin{figure}[t!]
    \centering
    \begin{tabular}{cc}
    \includegraphics[scale=0.4]{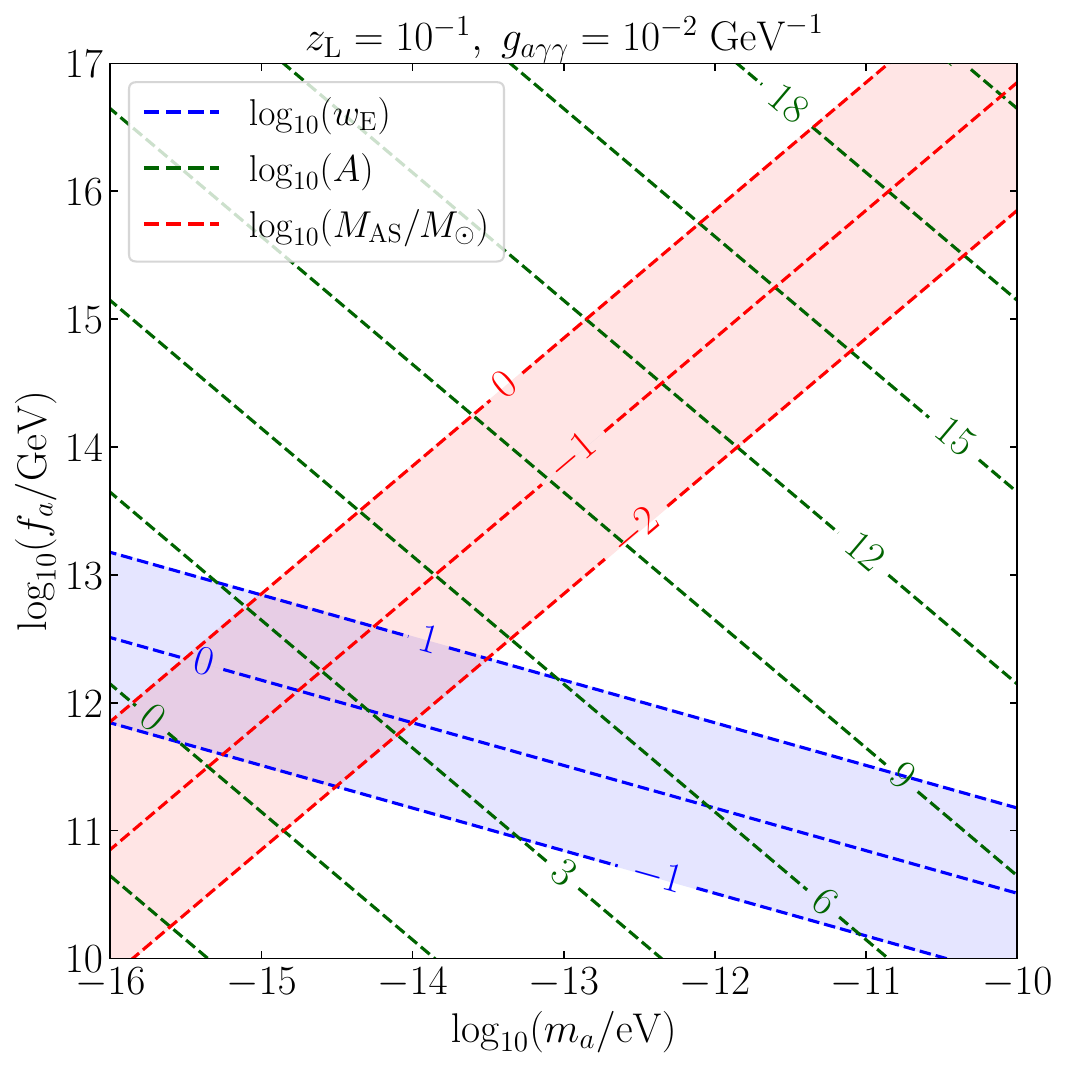}&\includegraphics[scale=0.4]{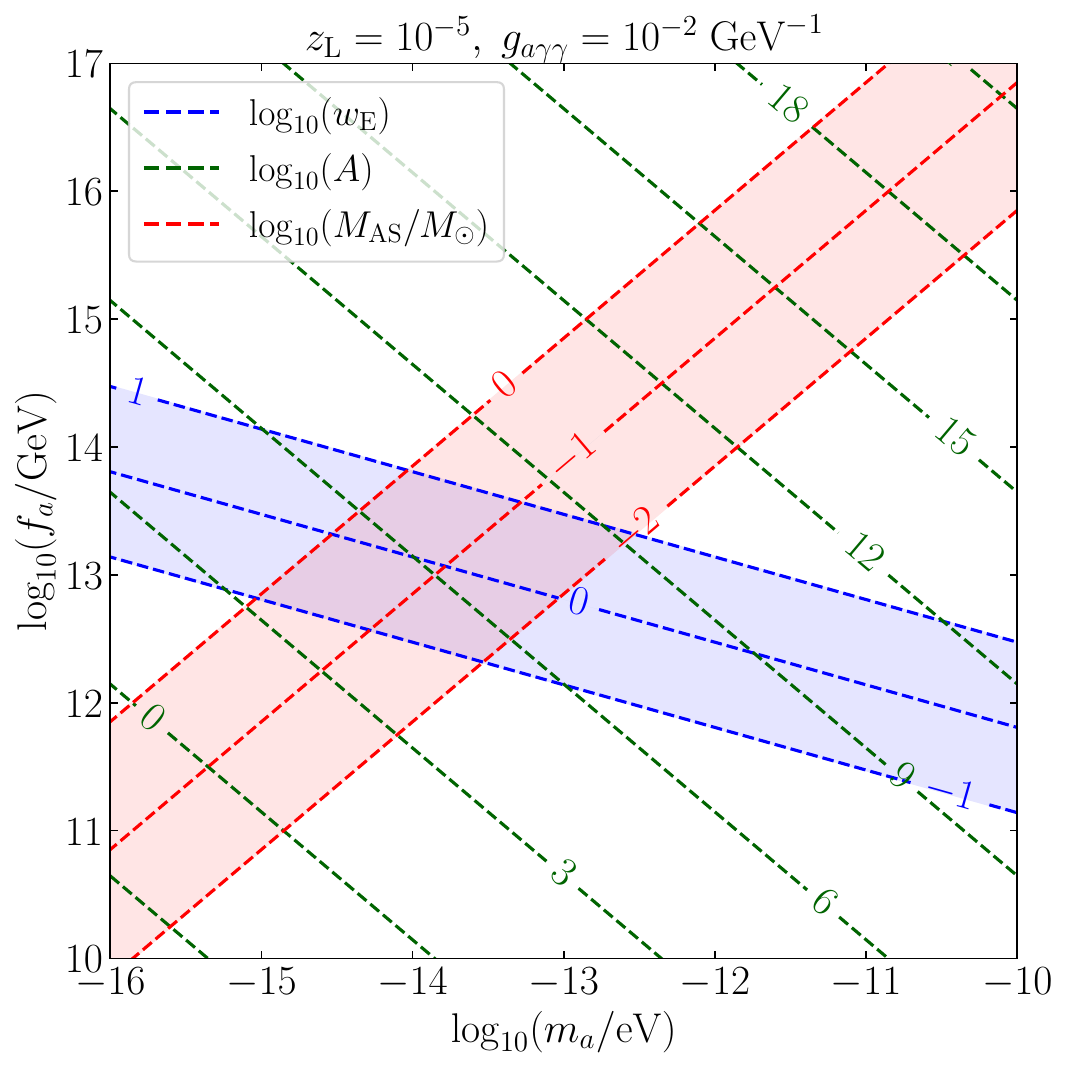}
    \end{tabular}
    \caption{``Checkerboard" plots indicating the relevant regions of interest in the $m_a$-$f_a$ plane, for $z_{\rm L}=10^{-1}$ (left) and $z_{\rm L}=10^{-5}$ (right) and fixing $g_{a\gamma\gamma} = \unit[10^{-2}]{GeV^{-1}}$. Please refer to the text for more details.}
    \label{fig:ALPCheckerboard}
\end{figure}

The case of ALPs presents a more interesting scenario: since the parameters $m_a$ and $f_a$ are essentially free parameters, this can potentially open up a viable parameter region where the novel effects, namely the finite size of the axion star and the axion-induced lensing effect, can be sizable. To obtain some guidance in identifying the relevant region in the fundamental axion parameter space $(m_a, f_a, g_{a\gamma\gamma})$, we locate those parameters that will lead to: (i) axion star masses within $10^{-2}M_\odot$ to $M_\odot$; (ii) $w_{\rm E}$ values between $0.1$ and $10$; and (iii) sufficiently large $A$ values. The rationale behind condition (i) comes from the expectation that the finite size and axion-induced lensing effects on the integrated optical depth are subleading compared to the pointlike case, and we have seen in Fig. \ref{fig:QCDAxionCase} for the QCD case that the relevant axion star mass range is between $10^{-2}M_\odot$ to $M_\odot$. Conditions (ii) and (iii) are necessary criteria to enhance the finite-size and axion-induced lensing effects. Taking $g_{a\gamma\gamma} = \unit[10^{-2}]{GeV^{-1}}$ as a benchmark value, we show the ``checkerboard" plots in Fig. \ref{fig:ALPCheckerboard}, where we project contours of constant $w_{\rm E}$ (dashed blue lines), $A$ (dashed green lines), and $M_{\rm AS}$ (dashed red lines), in the $m_a$-$f_a$ plane. Notice that since only $w_{\rm E}$ depends on the lens redshift, which we have fixed to $z_{\rm L} = 10^{-1}$ and $10^{-5}$ in the panels of Fig. \ref{fig:ALPCheckerboard}, only the positions of the blue contour lines change in each panel. Imposing the guiding criteria (i)-(iii) will put us in the scan range $\unit[10^{-16}]{eV} \leq m_a \leq \unit[10^{-10}]{eV}$ and $\unit[10^{10}]{GeV} \leq f_a \leq \unit[10^{17}]{GeV}$. Note that changing $g_{a\gamma\gamma}$ amounts to a simple rescaling of the labels of the dashed green contours. Thus, in order to see the lensing effects due to axion-photon interactions, we are forced to be in the region of the parameter space where $g_{a\gamma\gamma}$ is unusually large relative to $1/f_a$. While there are proposed mechanisms that can generate an exponential hierarchy between $g_{a\gamma\gamma}$ and $1/f_a$ \cite{Choi:2015fiu,Kaplan:2015fuy,Farina:2016tgd}, other experimental probes such as axion haloscopes robustly rule out a huge $g_{a\gamma\gamma}$. On the other hand, scaling down $g_{a\gamma\gamma}$ to a level that evades such robust constraints would imply that the finite size effect will dominate the lensing signal.
\begin{figure}[t!]
    \centering
    \begin{tabular}{ccc}
    \includegraphics[scale=0.27]{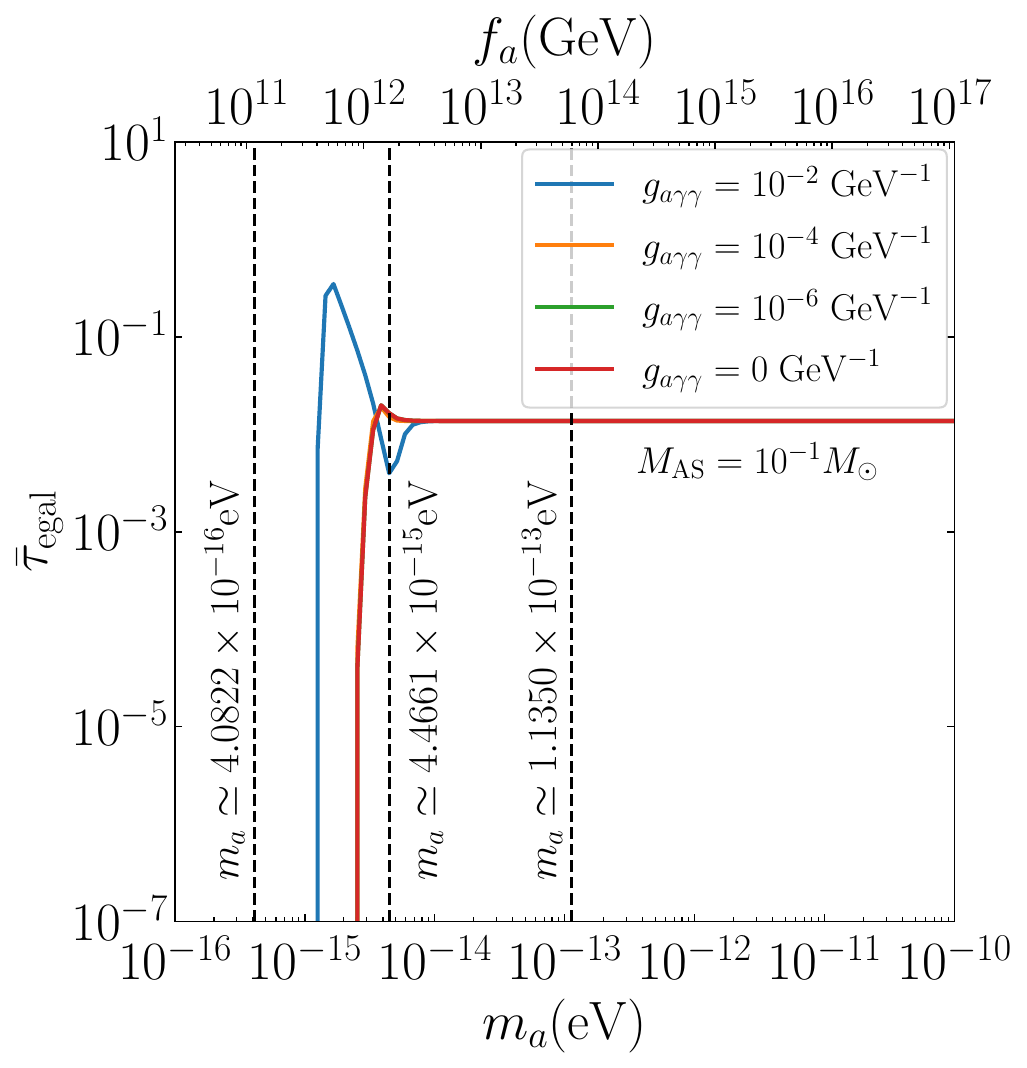}&\includegraphics[scale=0.27]{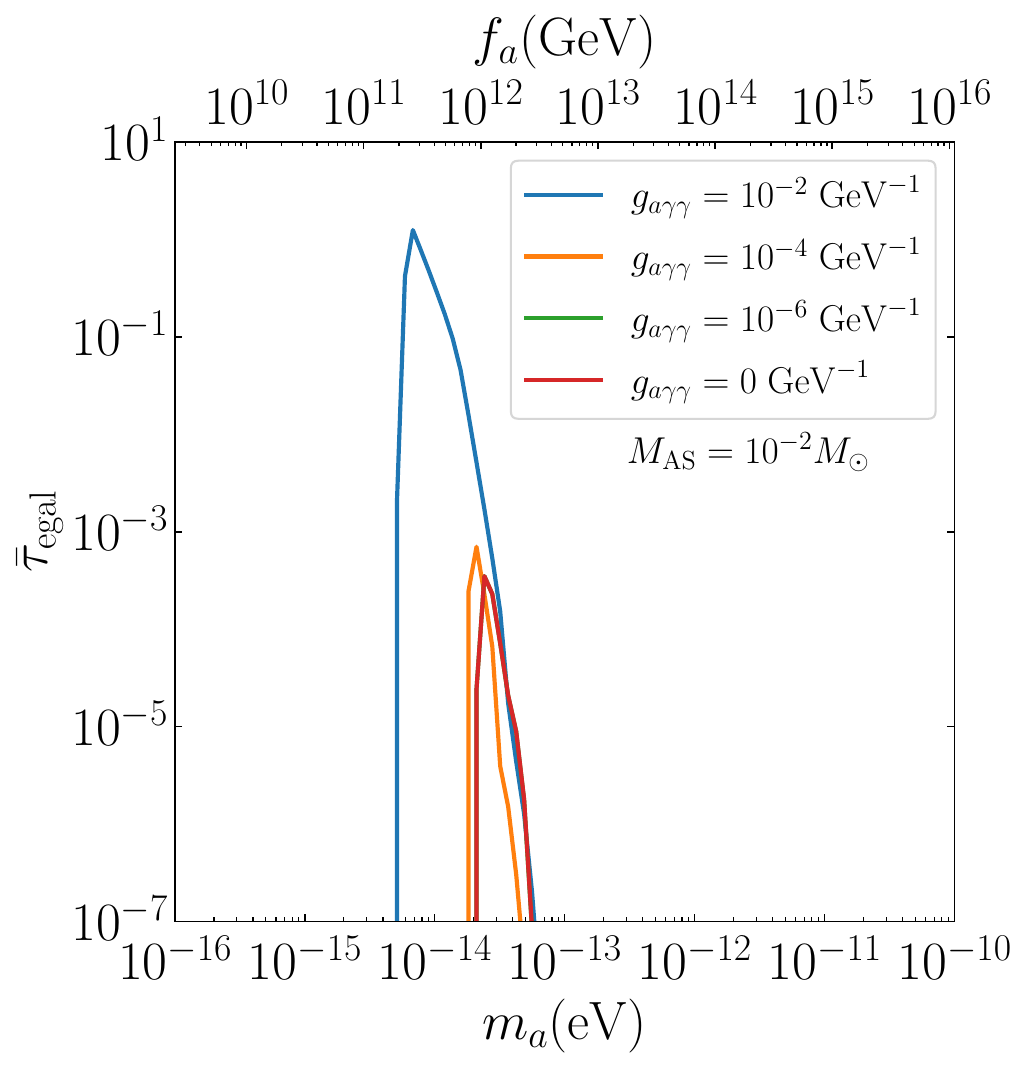}&\includegraphics[scale=0.27]{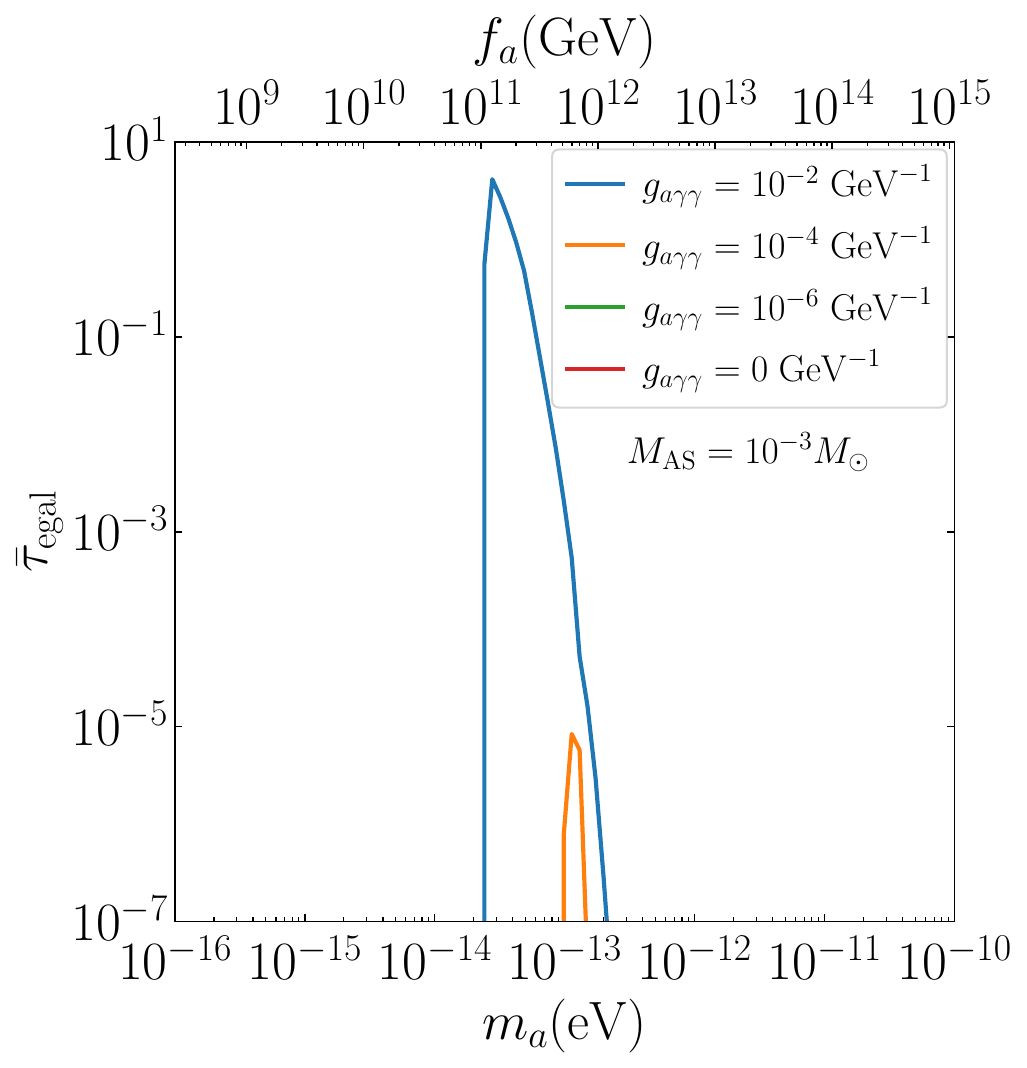}\\
    (a) & (b) & (c)\\
    \includegraphics[scale=0.27]{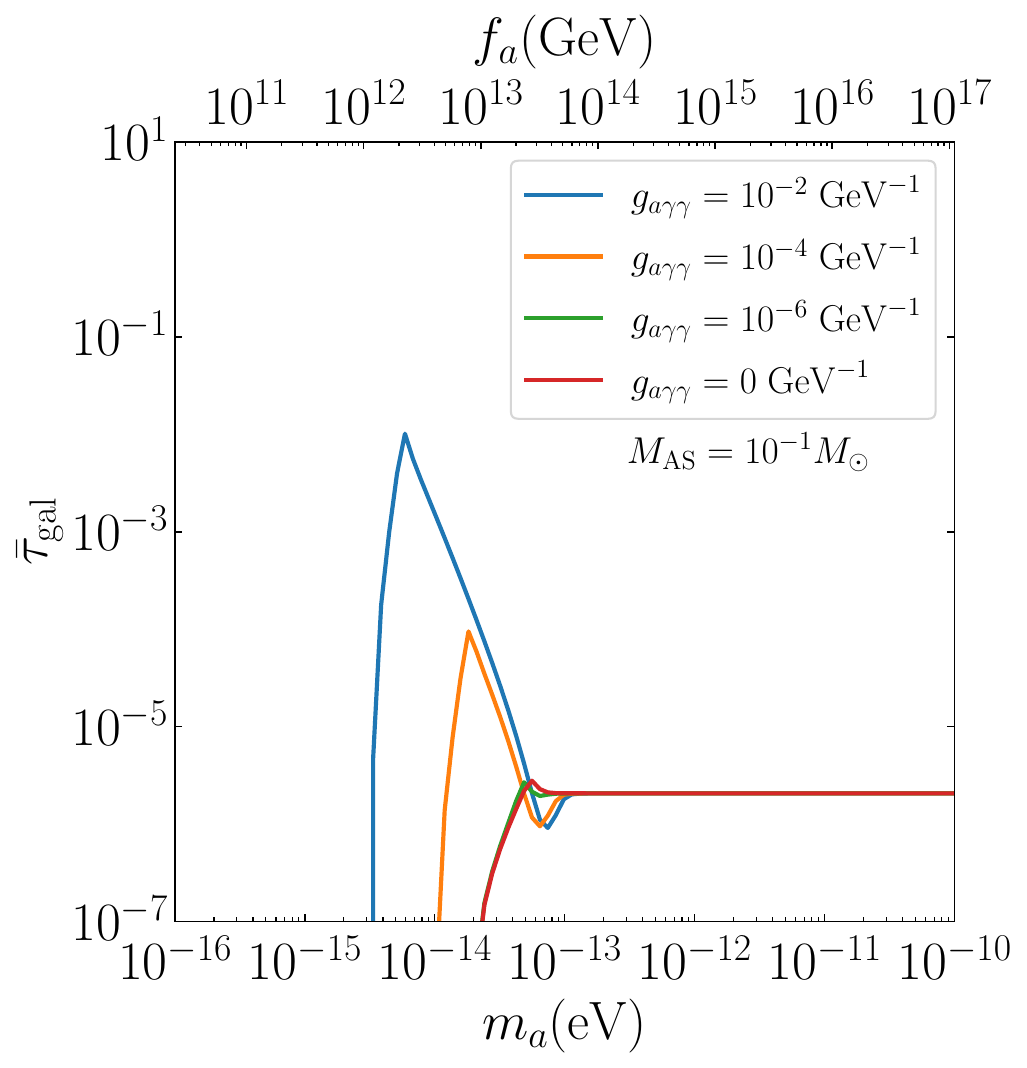}&\includegraphics[scale=0.27]{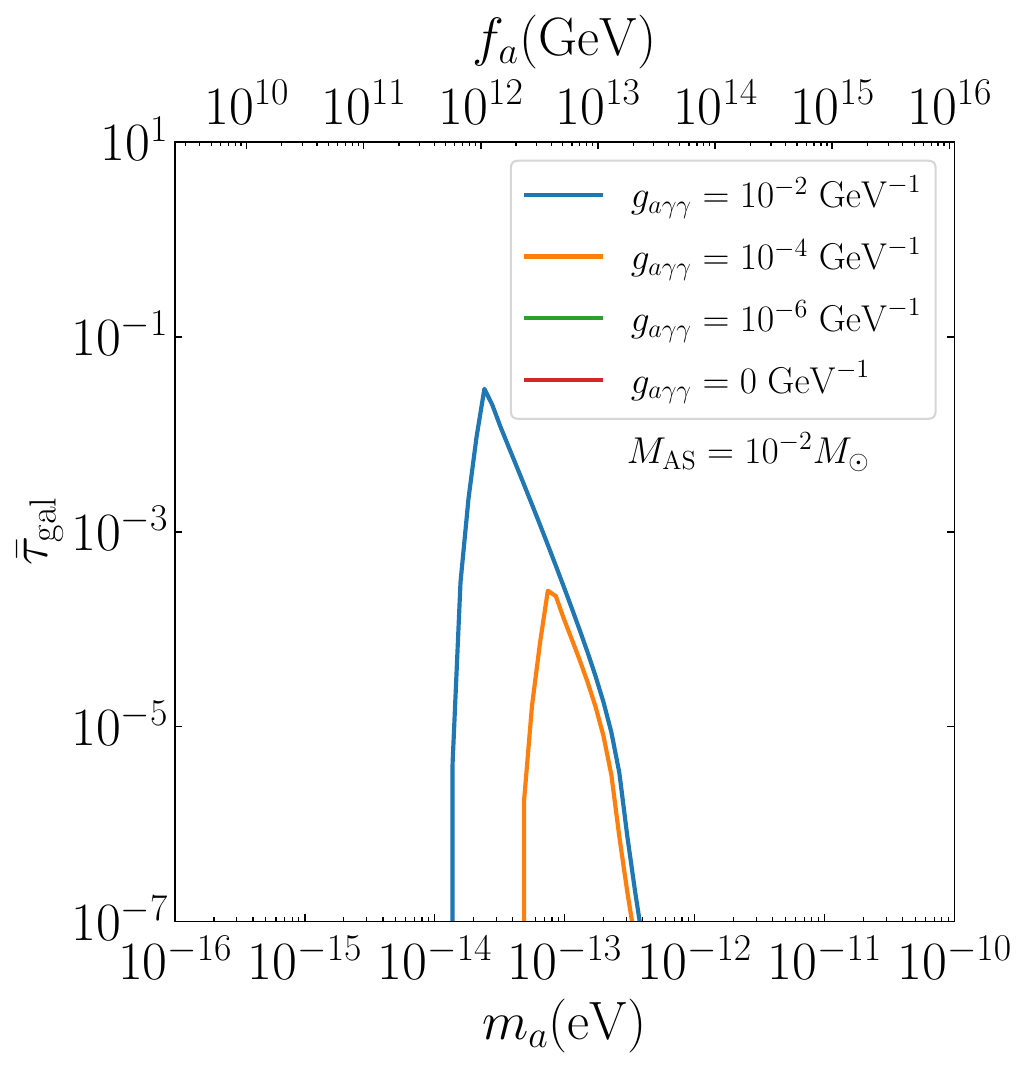}&\includegraphics[scale=0.27]{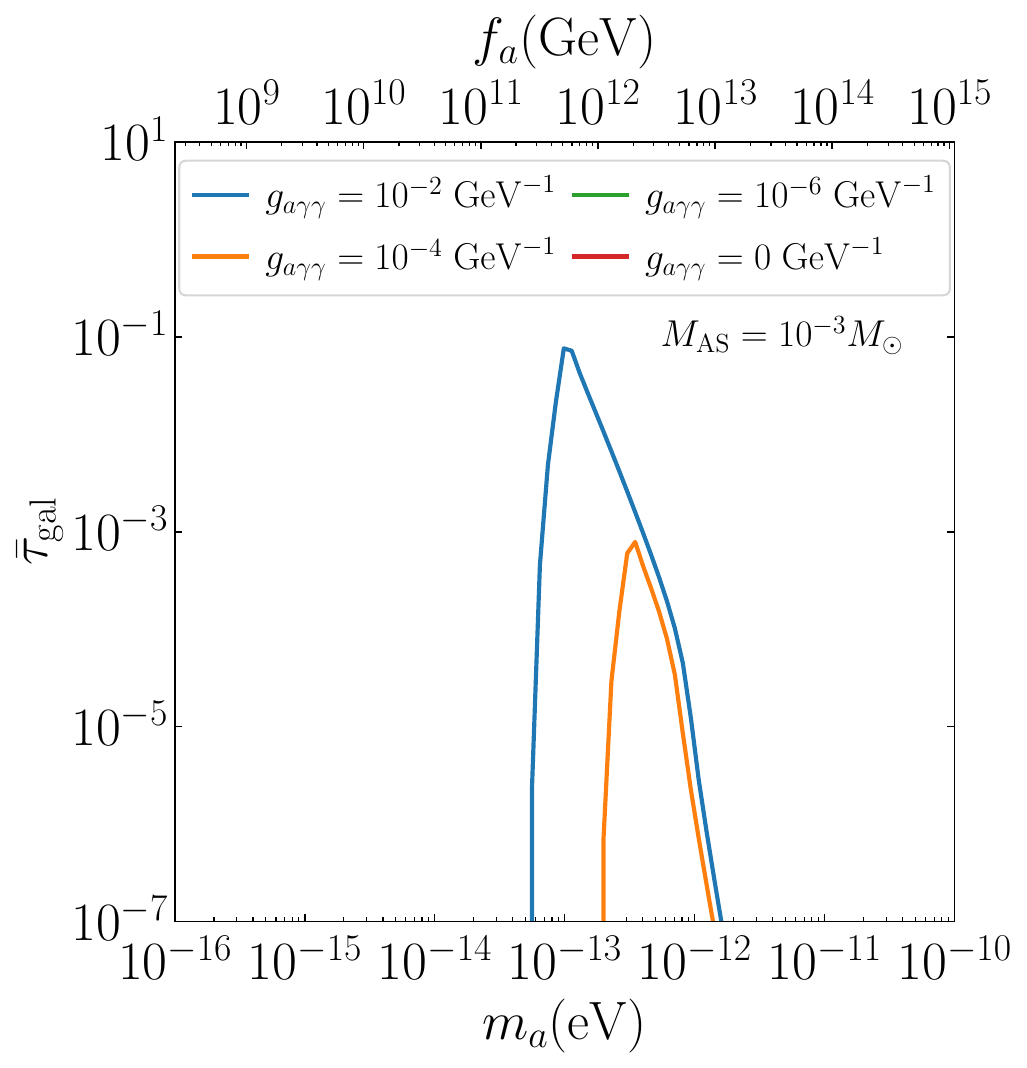}\\
    (d) & (e) & (f)
    \end{tabular}
    \caption{Integrated optical depth for the case of ALP axion stars, for $M_{\rm AS} = 10^{-1} M_\odot$ (left column), $M_{\rm AS} = 10^{-2} M_\odot$ (middle column), and $M_{\rm AS} = 10^{-3} M_\odot$ (right column). Panels a, b, c refer to the extragalactic contribution, while Panels d, e, f refer to the Galactic contribution. Here we take $f_{\rm AS} = 1$. In Panel a, the curves for $g_{a\gamma\gamma} = 0, 10^{-6}$, and $\unit[10^{-4}]{GeV^{-1}}$ almost overlap.}
    \label{fig:ALP_IOD}
\end{figure}

\begin{figure}[t]
    \centering
    \begin{tabular}{ccc}
    \includegraphics[scale=0.3]{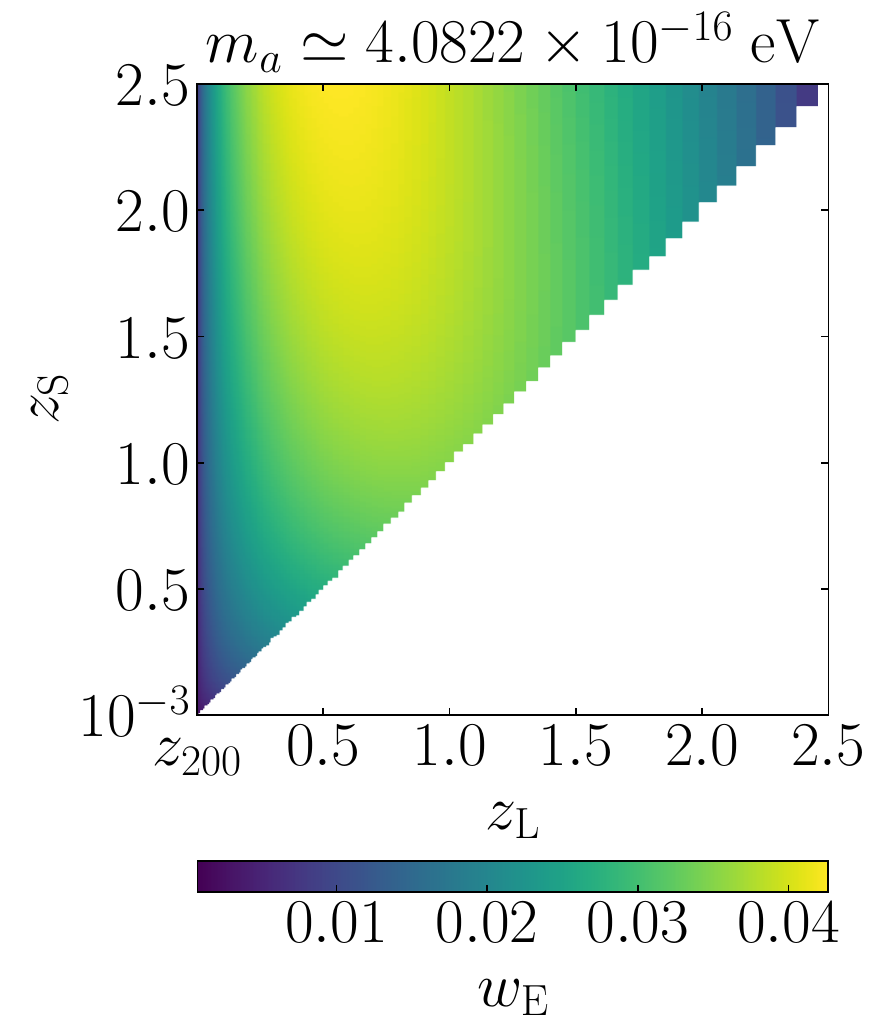}&\includegraphics[scale=0.3]{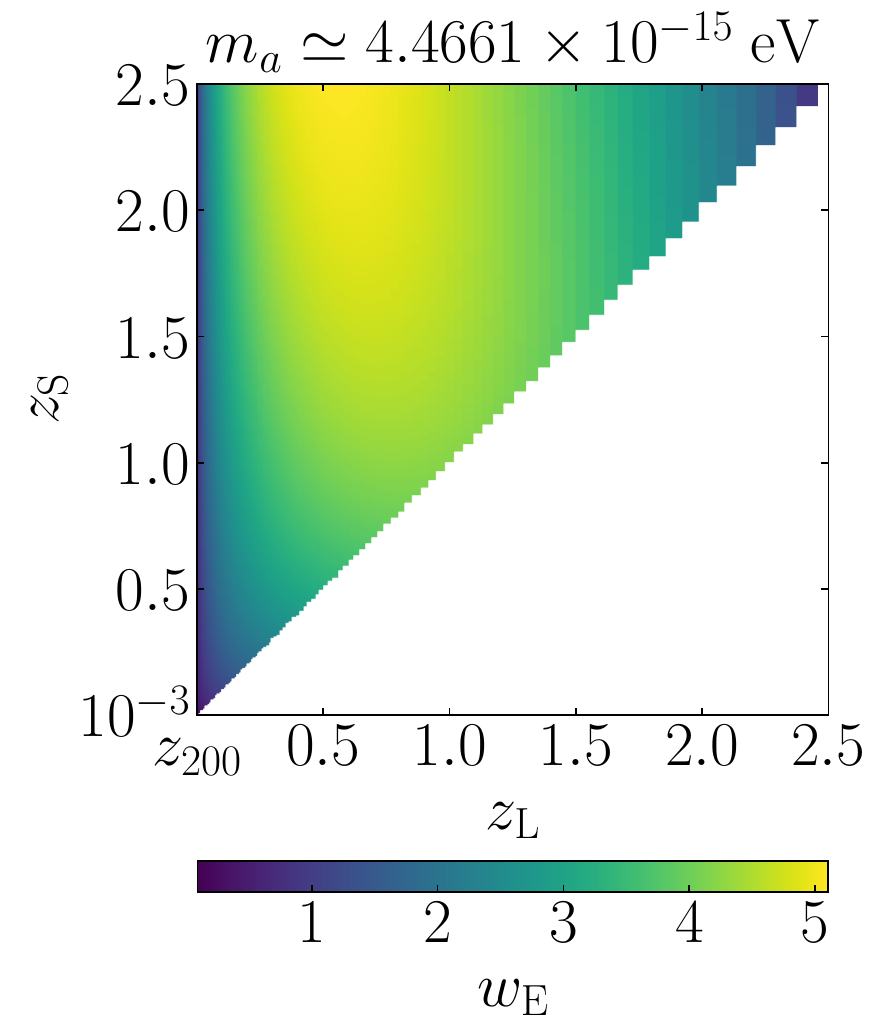}&\includegraphics[scale=0.3]{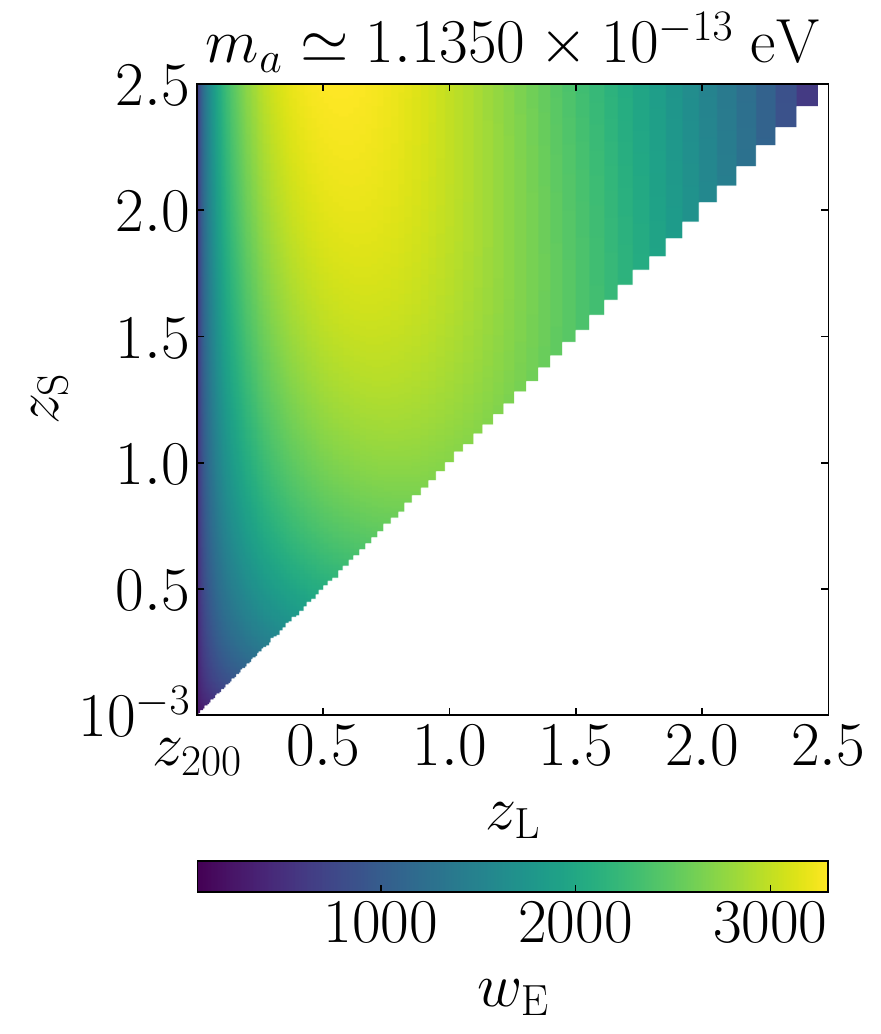}\\
    (a)&(b)&(c)
    \end{tabular}
    \caption{The corresponding distribution of $w_{\rm E}$ of Fig. \ref{fig:ALP_IOD} (a), Panel a on $z_{\rm L}$-$z_{\rm S}$ plane for $M_{\rm AS}=10^{-1}M_{\odot}$ with $m_{a} \simeq 4.0822 \times 10^{-16}$ eV (Panel a), $m_{a} \simeq 4.4661 \times 10^{-15}$ eV (Panel b) and $m_{a} \simeq 1.1350 \times 10^{-13}$ eV (Panel c). For Panel a, since $w_{\rm E}$ is much less than 1, the lens equation becomes $\tilde{\beta} \approx \tilde{\theta}$, hence no $\tilde{\beta}_{c}$ and we get $\bar{\tau}=0$. For Panel c, $w_{\rm E}$ is much greater than 1, so the lens equation is reduced to the case of pointlike lens.}
    \label{fig:ALPwESample}
\end{figure}

We present the plots of the integrated optical depth for the ALP case in Fig. \ref{fig:ALP_IOD}, where the axion star mass in each column is fixed to the following values: $M_{\rm AS} = 10^{-1} M_\odot, 10^{-2} M_\odot$, and $10^{-3} M_\odot$. We separate the extragalactic contribution in Panels a, b, and c from the Galactic contribution shown in Panels d, e, and f. The results for the integrated optical depth are shown for axion-photon coupling values $g_{a\gamma\gamma} = 10^{-2}, 10^{-4},$ and $\unit[10^{-6}]{GeV^{-1}}$, as well as in the case where the axion-induced effect is turned off, $i.e.$ $g_{a\gamma\gamma}=\unit[0]{GeV^{-1}}$. Compared with the pointlike case shown in Fig. \ref{fig:QCDAxionCase}, we observe some similarities with the pointlike case, particularly for larger values of $m_a$ where the integrated optical depth stays at a fixed value ($\bar{\tau}_{\rm egal} \simeq 10^{-2}$ and $\bar{\tau}_{\rm gal} \simeq 2 \times 10^{-6}$), as seen in Panels a and d of Fig. \ref{fig:ALP_IOD}. For fixed $g_{a\gamma\gamma}$, the level of the peak decreases with $M_{\rm AS}$, simply due to the fact that the lensing cross section is proportional to the axion star mass. We also notice a decrease in $\bar{\tau}$ for low $m_a$, as in the pointlike case. In the ALP case, this is due to $w_{\rm E}$ going below the threshold, given in Eq. (\ref{wEThreshold}), which ensures the presence of multiple images. Note that
\begin{eqnarray}
\label{ALPwE}    w_{\rm E}^2 \simeq 0.3597 h^{-1} \left(\frac{M_{\rm AS}}{M_\odot}\right)^3\left(\frac{m_a}{\unit[2 \times 10^{-16}]{eV}}\right)^4\left(D_{\rm S} H_0\right)\left(\frac{D_{\rm L}}{D_{\rm S}}\right)\left(1-\frac{D_{\rm L}}{D_{\rm S}}\right),
\end{eqnarray}
and the threshold value can be obtained from
\begin{eqnarray}
   \nonumber \frac{1}{w_{\rm E, \rm th}^2} - 1 &\simeq& 1.73268 \times 10^{-22} \left(\frac{\unit[600]{MHz}}{f_0}\right)^2 \left(\frac{M_{\rm AS}}{10^{-1} M_\odot}\right)^2\\
   \label{ALPwEThreshold} &\times&\left(\frac{m_a}{\unit[2\times 10^{-16}]{eV}}\right)^4 \left(\frac{g_{a\gamma\gamma}}{\unit[10^{-2}]{GeV^{-1}}}\right)^2\left(\frac{0.3}{\vert \gamma \vert}\right)^{\frac{1}{2}}.
\end{eqnarray}
A more careful assessment is shown in Fig. \ref{fig:ALPwESample}, where Panels a, b, and c show scans of $w_{\rm E}$ over source-lens redshift space for $M_{\rm AS} = 10^{-1} M_\odot$, for the following selection of axion masses: $m_a \simeq \unit[4.08 \times 10^{-16}]{eV}$ (Panel a), $\unit[4.47 \times 10^{-15}]{eV}$ (Panel b), and $\unit[1.14 \times 10^{-13}]{eV}$ (Panel c), which correspond to the vertical dashed lines in Panel a of Fig. \ref{fig:ALP_IOD}. Based on the estimate in Eq. (\ref{ALPwEThreshold}), and taking $g_{a\gamma\gamma} = \unit[10^{-2}]{GeV^{-1}}$, the threshold is $w_{\rm E, \rm th} \simeq 1$ for all the axion mass parameters depicted in Panels a, b, and c. However, only Panel a of Fig. \ref{fig:ALPwESample} demonstrates the scenario where the $w_{\rm E}$ over the scanned region of lens and source redshifts is below threshold; thus, this case does not lead to any lensing and the optical depth is zero.

\begin{figure}[t!]
    \centering
    \includegraphics[scale=0.42]{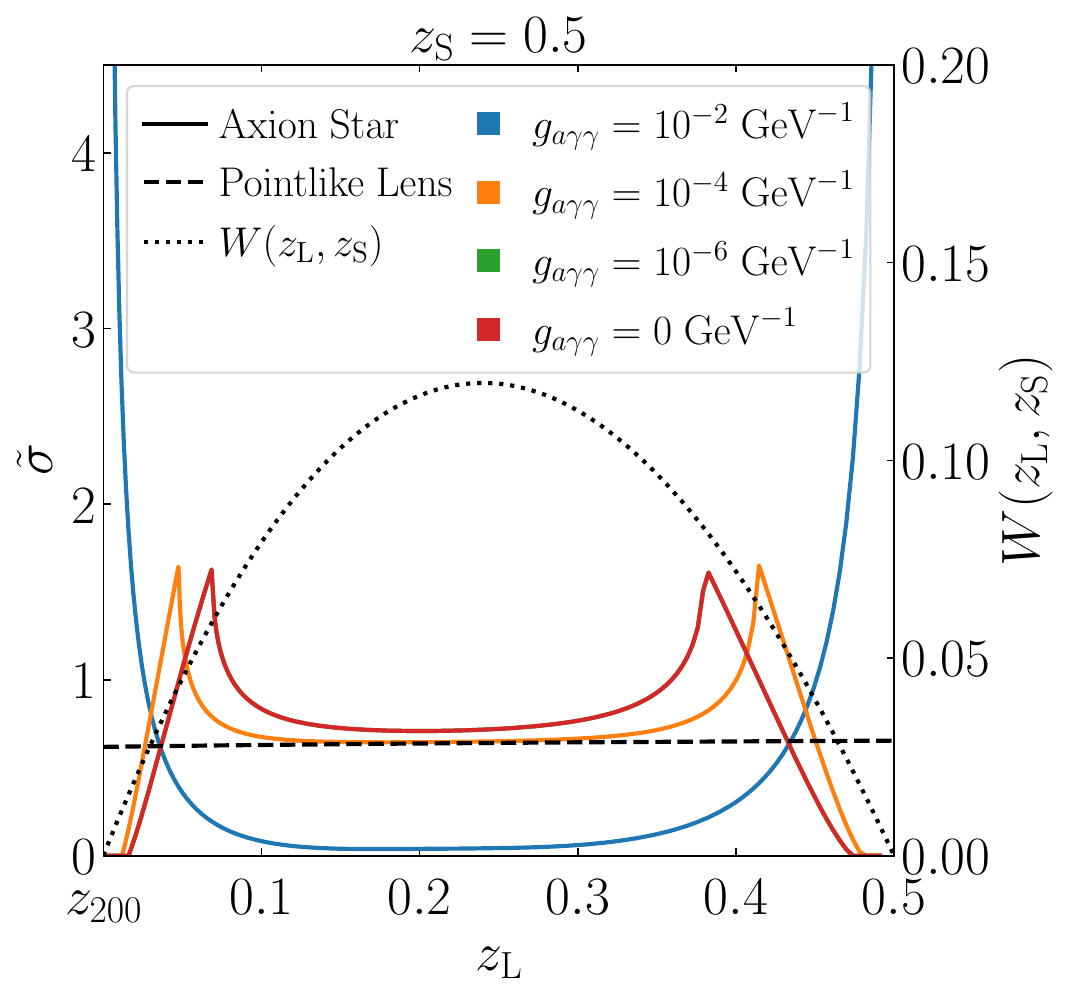}
    \includegraphics[scale=0.42]{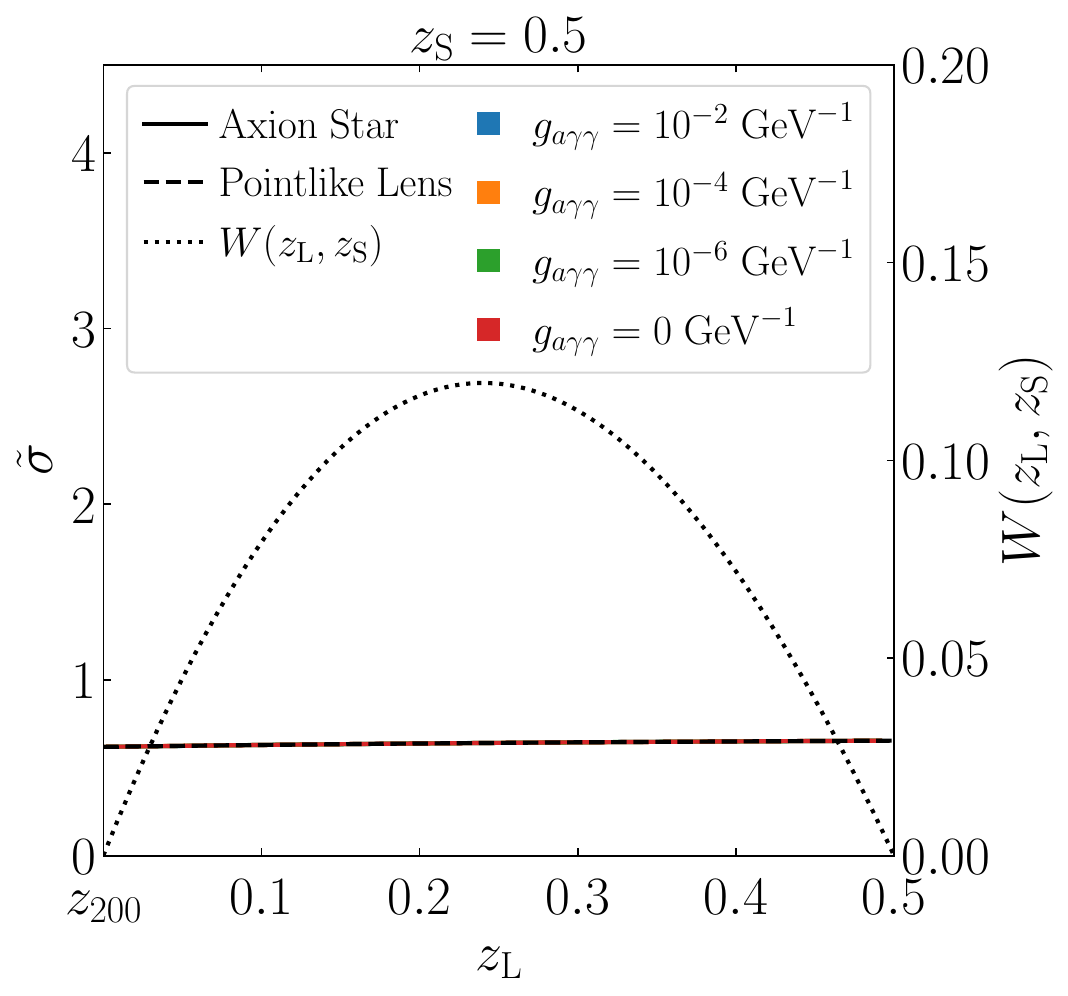}  
       \caption{The dimensionless cross section for (left) $m_a \simeq 4.4661 \times 10^{-15}$ eV and (right) $m_a \simeq 1.1350 \times 10^{-13}$ eV and $z_{\rm S}=0.5$, $M_{\rm AS}=10^{-1}M_{\odot}$. In the left panel, cases of $g_{a\gamma\gamma}=0$ and $10^{-6}$ GeV$^{-1}$ almost overlap, while in the right panel, all cases converge to the pointlike limit. See text for more details.}
       \label{fig:ExplainDip}
\end{figure}

One can observe some peculiar features in the integrated optical depth, both in the case of nonzero $g_{a\gamma\gamma}$, as well as in the case of lensing solely through gravitational interactions. This is an indication that these features can be attributed to the novel lensing effects. In particular, looking at Panel a of Fig. \ref{fig:ALP_IOD} we observe a dip in $\bar{\tau}$ for $m_a \simeq \unit[4.461 \times 10^{-15}]{eV}$. We can understand this behavior by splitting the extragalactic contribution to the optical depth as a product of the dimensionless cross section $\tilde{\sigma}$, and a window function $W(z_{\rm L}, z_{\rm S})$ which depends solely on source and lens redshift positions, \textit{i.e.}
\begin{eqnarray}
    \tau_{\rm egal}(z_{\rm L}, z_{\rm S}) = \frac{3}{2}\Omega_{\rm m}~\tilde{\sigma}(\beta_c)~W(z_{\rm L}, z_{\rm S}),\quad W(z_{\rm L}, z_{\rm S}) \equiv \frac{D_{\rm LS}D_{\rm L}}{D_{\rm S}}\frac{H_0^2}{H(z_{\rm L})}(1+z_{\rm L})^2.
\end{eqnarray}
In Fig. \ref{fig:ExplainDip} we show plots of the dimensionless cross section versus lens redshift position, for two axion mass parameters $m_a \simeq \unit[4.4661 \times 10^{-15}]{eV}$, $\unit[1.1350 \times 10^{-13}]{eV}$, which correspond to two of the vertical dashed lines in Panel a of Fig. \ref{fig:ALP_IOD}. We plot the dimensionless cross section for $g_{a\gamma\gamma} = 10^{-2}, 10^{-4}, 10^{-6}$, and $\unit[0]{GeV^{-1}}$. In both panels we included the window function (dotted curve), for an FRB source at $z_{\rm S} = 0.5$, which determines which lens redshift positions will eventually be relevant in the calculation of the integrated optical depth. We also show the dimensionless cross section in the pointlike case (dashed curve), to highlight the impact of the finite size effect on the cross section and optical depth. The right panel of Fig. \ref{fig:ExplainDip}, corresponding to $m_a \simeq \unit[1.1350 \times 10^{-13}]{eV}$ which falls in the pointlike case according to Panel a of Fig. \ref{fig:ALP_IOD}, shows that the dimensionless cross section for all the featured $g_{a\gamma\gamma}$ values, coincides with the pointlike limit. On the other hand, the left panel, corresponding to $m_a \simeq \unit[4.4661 \times 10^{-15}]{eV}$, exhibits a double peak feature in the dimensionless scattering cross section, for all the featured $g_{a\gamma\gamma}$ values. In particular, increasing $g_{a\gamma\gamma}$ leads to the increase in both the width of the valley between the peaks, and the heights of the peaks. However, for larger $g_{a\gamma\gamma}$, the peaks in the dimensionless cross section lie beyond the lens redshift positions where the window function is dominant. Furthermore, the height of the valley also decreases with larger $g_{a\gamma\gamma}$. Both behaviors in the dimensionless optical depth provide explanations to the dip in $\bar{\tau}$, in Panel a of Fig. \ref{fig:ALP_IOD}.
\begin{figure}[t]
    \centering
   \includegraphics[scale=0.4]{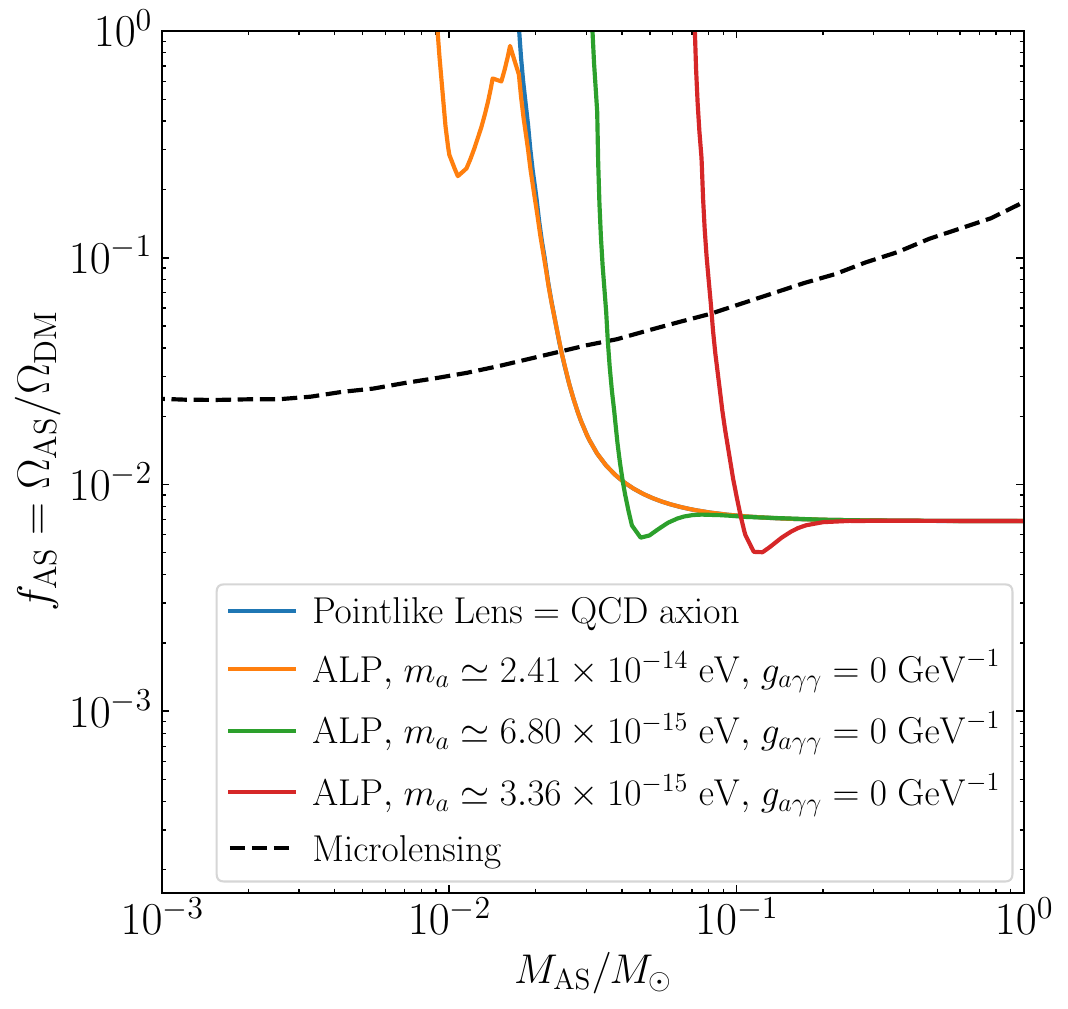} \includegraphics[scale=0.4]{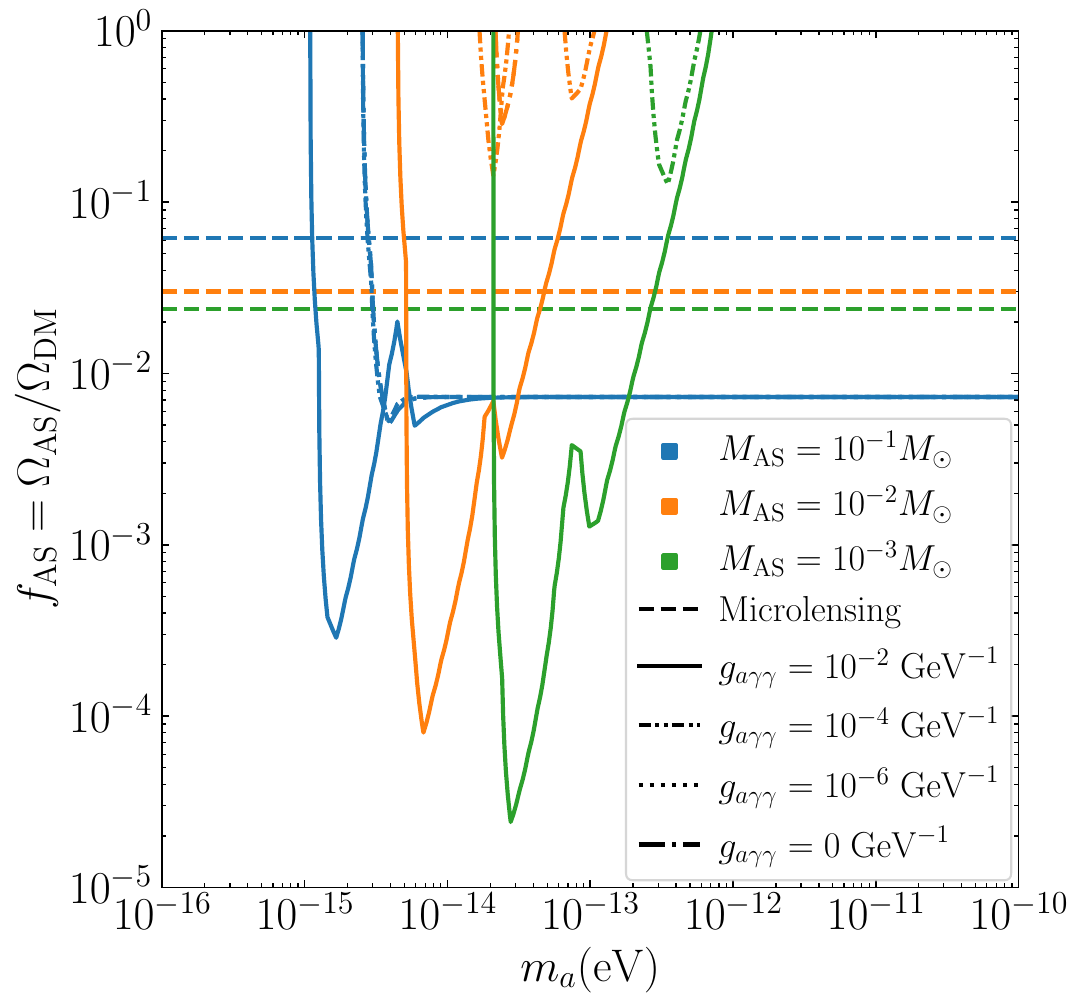}
    \caption{Sensitivity curves with $N_{\rm obs}=10^4$ and $\Delta t_{\min}=1\,\mu{\rm s}$ in the $f_{\rm AS}$-$M_{\rm AS}/M_{\odot}$ plane and $f_{\rm AS}$-$m_a$ plane, for the case of axion stars made of QCD axions and ALPs of fixed $m_a$ (left panel) and ALPs (right panel).}
    \label{fig:ConstraintfAS}
\end{figure}
\begin{figure}[t!]
    \centering
    \includegraphics[scale=0.4]{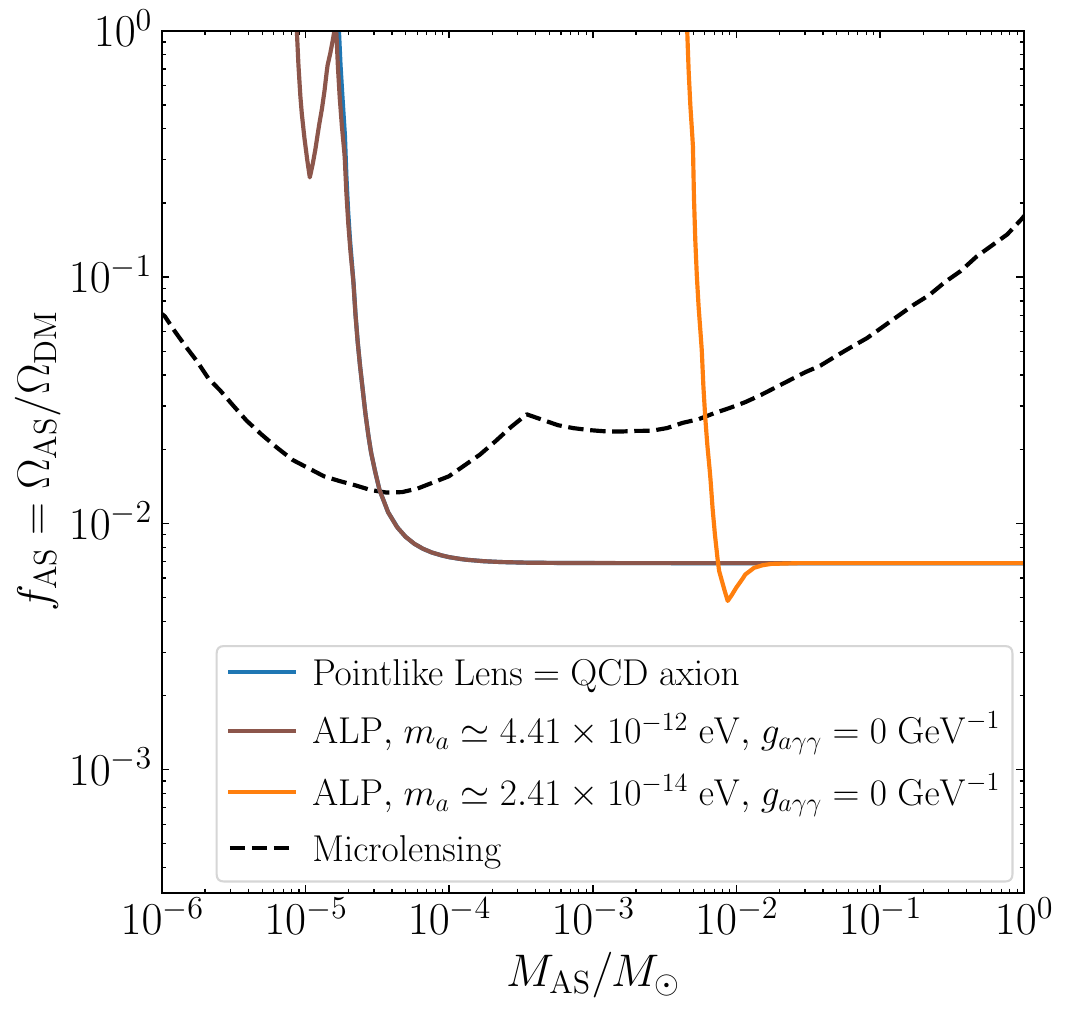}
    \includegraphics[scale=0.4]{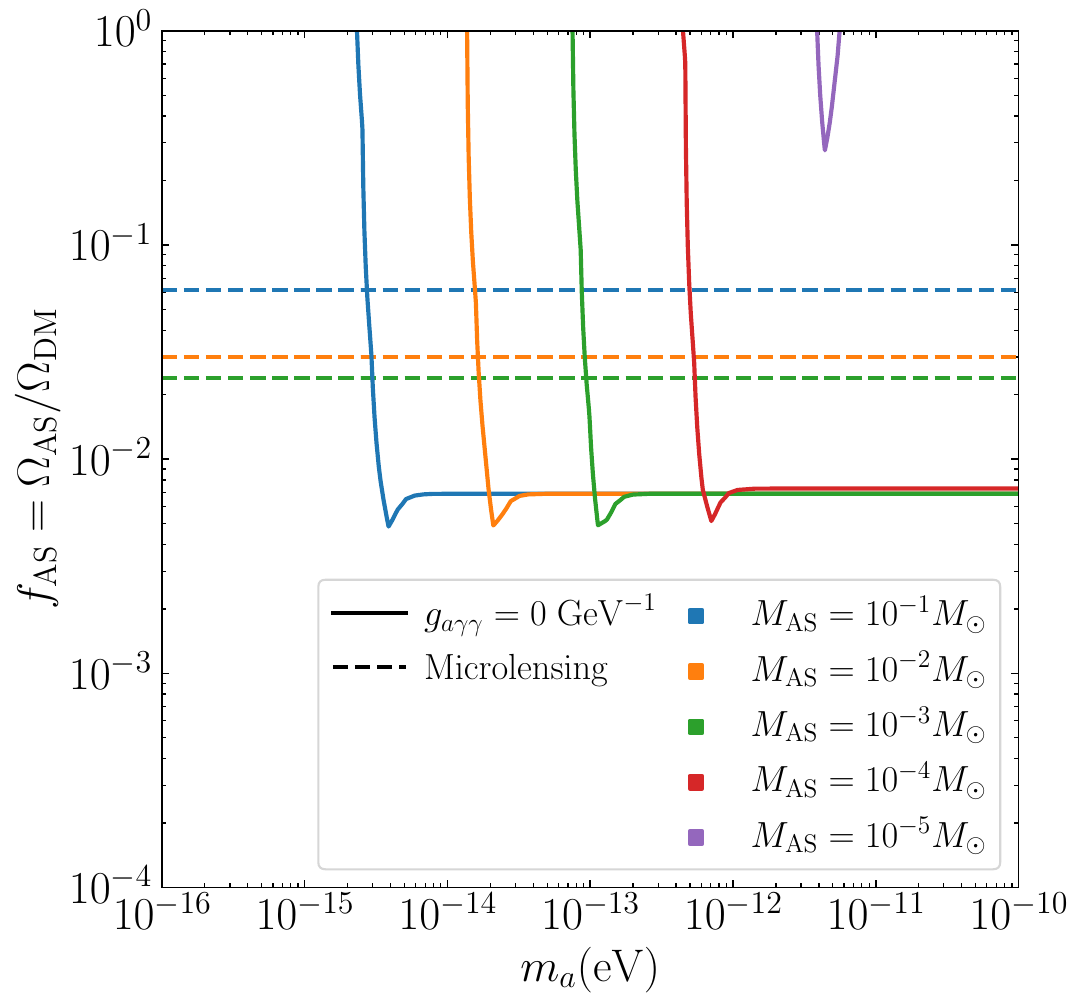}  
       \caption{Same as Fig.~\ref{fig:ConstraintfAS}, but assuming $\Delta t_{\min}=1\,{\rm ns}$ while fixing $g_{a\gamma\gamma}=0 \, {\rm GeV}^{-1}$.}
    \label{fig:ConstraintfAS_future_1}
\end{figure}
\begin{figure}[t!]
    \centering
    \includegraphics[scale=0.4]{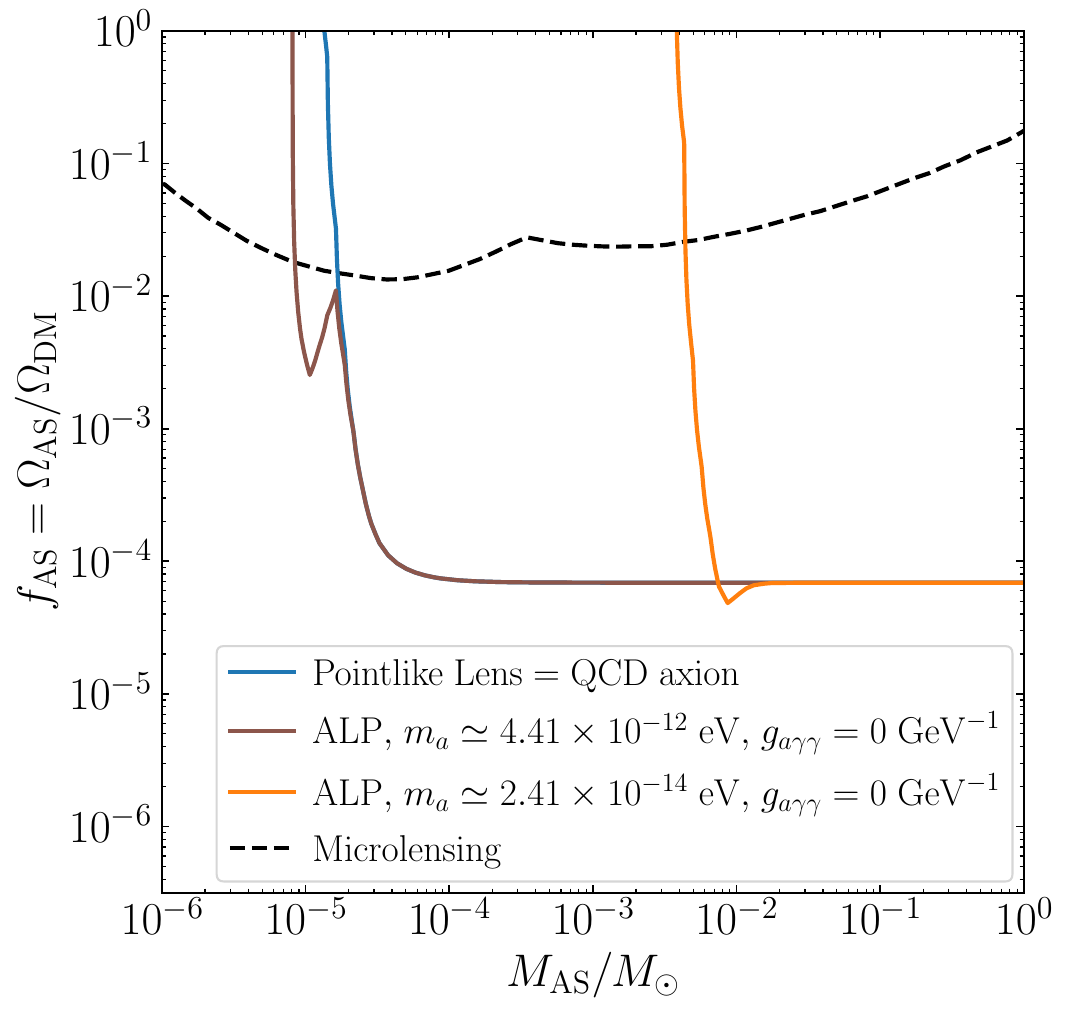}
    \includegraphics[scale=0.4]{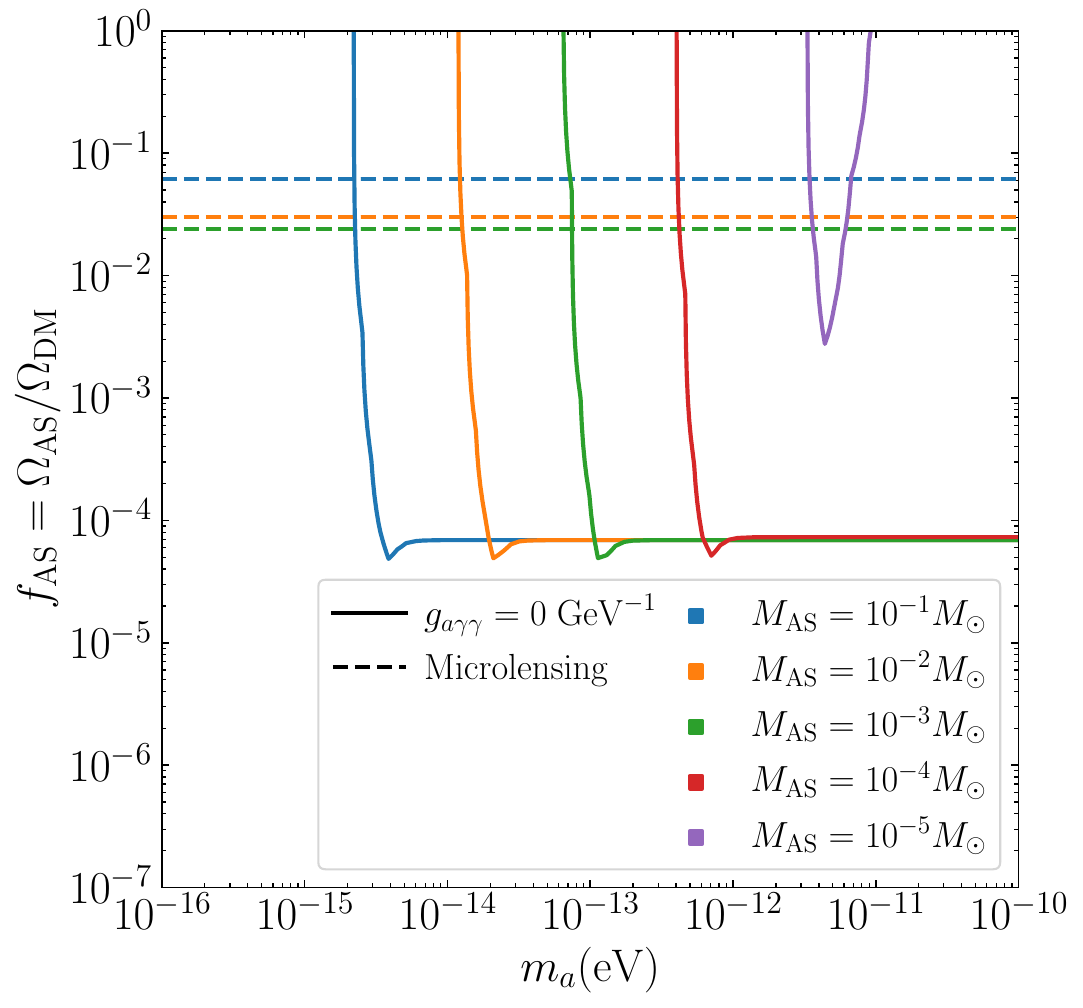}  
       \caption{Same as Fig.~\ref{fig:ConstraintfAS}, but assuming $\Delta t_{\min}=1\,{\rm ns}$ and $N_{\rm obs}=10^6$ while fixing $g_{a\gamma\gamma}=0 \, {\rm GeV}^{-1}$.}
    \label{fig:ConstraintfAS_future_2}
\end{figure}

\subsection{Sensitivity forecasts}
\label{sec:exclusion}
In Fig. \ref{fig:ConstraintfAS} we finally present our sensitivity forecasts in the $f_{\rm AS}$-$M_{\rm AS}/M_\odot$ plane (left panel), and in the $f_{\rm AS}$-$m_a$ plane (right panel). These constitute the main results of this work. The region lying above the plotted curves corresponds to a predicted number of lensing events $N_{\rm lensed} > 1$. From Eq. (\ref{NLens}) and taking the optimistic scenario of $N_{\rm obs} = 10^4$ observed FRBs, we find that the exclusion criterion implies that
\begin{eqnarray}
    \bar{\tau} > -\ln\left(1 - \frac{N_{\rm lensed}}{N_{\rm obs}}\right) \simeq 10^{-4}.
\end{eqnarray}
The curves shown in the left panel of Fig. \ref{fig:ConstraintfAS}, feature results for the following: QCD axion case, where the relationship between $m_a$ and $M_{\rm AS}$ is fixed; and in the ALP case where one can choose a fixed axion mass $m_a$ and vary $M_{\rm AS}$ by changing the decay constant $f_a$. We see that the sensitivity in the QCD axion case agrees with earlier results in the literature (\textit{e.g.} \cite{Ho:2023feo}), and the novel effects from finite-size and axion-induced lensing effects are barely noticeable because the axion star is effectively pointlike.

On the other hand, when we take the case of ALPs and fix the axion mass $m_a$, we find some additional features in the sensitivity curves, compared to the QCD axion case, which we mainly attribute to the finite size lensing effect. In particular, applying the same lensing criteria for the QCD axion case and for the ALP case, we find improved sensitivity at $M_{\rm AS} = 10^{-2} M_\odot$, for an axion mass $m_a \simeq \unit[2.41 \times 10^{-14}]{eV}$. For comparison, we also show the constraint curve for boson stars obtained by using microlensing reported in \cite{Croon:2020ouk}. On the other hand, the right panel which shows the corresponding constraint curve for the ALP case, exhibits the novel effects, most notably in the case where $g_{a\gamma\gamma} = \unit[10^{-2}]{GeV^{-1}}$. Note that each color corresponds to a different axion star mass, and we limited our scans to $M_{\rm AS} = 10^{-1} M_\odot, 10^{-2} M_\odot$, and $10^{-3} M_\odot$. The dips in the exclusion curves shift to heavier axion masses in the case of lighter axion stars, mainly due to the fact that we want $w_{\rm E}$ to be fixed to a small value to observe the novel effects, and $w_{\rm E} \propto \sqrt{M_{\rm AS}}/R \propto M_{\rm AS}^{3/2} m_a^2$, which follows from Eq. (\ref{ALPwE}). 

As a side remark, earlier work by \cite{Chang:2024fol} concluded, based on \cite{Croon:2020ouk}, that microlensing searches are insensitive to axion stars within $10^{-14} M_\odot$ to $10^{-11} M_\odot$. From our results, we find that this mass range is also beyond the reach of FRB lensing searches, and thus falls beyond our mass range of interest. On the other hand, reference~\cite{Chang:2024fol} also pointed out that axion stars in the ALP scenarios they considered fall in the range $0.1 M_\odot \lesssim M_{\rm AS} \lesssim 10^8 M_\odot$, where microlensing searches lose sensitivity. Indeed, the microlensing constraint (black dashed line) in the left panel of Fig. \ref{fig:ConstraintfAS} shows a loss in sensitivity from $0.1 M_\odot$ to $1 M_\odot$. This is due to the suppression of the microlensing event rate by the lens mass \cite{Griest:1990vu}. In contrast, FRB lensing, an observational technique that is distinct from microlensing, does not suffer from this mass suppression, at least in the range $0.1 M_\odot$ to $1 M_\odot$; this can be understood by looking at the estimate of the optical depth in Eq. (\ref{TauEstimatePointLike}). In the limit of heavier masses, other works~\cite{Oguri:2022fir} and \cite{Kalita:2023eeq} both demonstrated a loss of sensitivity in the fraction of pointlike objects, but providing different explanations for such behavior. For the ALP case, the axion stars in this limit are effectively pointlike, so we can use Eqs. (\ref{PointLikeDt}) and (\ref{PointLikeRf}) to estimate the time delay and magnification ratio, respectively. One finds that $R_f \rightarrow 1$. Meanwhile $\Delta t$ is proportional to $M_{\rm AS}$, such that the time delay may exceed the maximum observation time of the telescope, and hence there is no lensing event.

\subsection{Improvements in timing resolution and FRB count}
We provide the sensitivity projections on $f_{\rm AS}$ versus $M_{\rm AS}/M_\odot$, and on $m_a$, where we anticipate improvements in the timing resolution. For illustration, we consider $\Delta t_{\min} = \unit[1]{ns}$. A decrease in $\Delta t_{\min}$ accommodates much lower source angular positions, leading to an increase in the dimensionless cross section. Then keeping the optical depth to be the same leads to opening up the viable parameter range, for smaller values of $M_{\rm AS}$. For the QCD axion case, we see an improvement in the sensitivity reach in $f_{\rm AS}$ for axion star masses up to $M_{\rm AS} \simeq 4 \times 10^{-4} M_\odot$. In the ALP case, the $g_{a\gamma\gamma} = 0 \; {\rm GeV}^{-1}$ curves, \textit{i.e.} pure gravity induced lensing with finite size effects, in the right panels of Figs.~\ref{fig:ConstraintfAS} and \ref{fig:ConstraintfAS_future_1}, show that contours corresponding to axion star masses of $10^{-3}M_\odot$ to $10^{-5} M_\odot$ appear. As in the previous case where we considered $\Delta t_{\min} = \unit[1]{\mu s}$, the dip features in the sensitivity curves appear as a result of the impact of the finite size effect in lensing. In the right panel of Fig.~\ref{fig:ConstraintfAS_future_1}, the loss in sensitivity below certain values of $m_a$, for contours of constant $M_{\rm AS} = 10^{-1} M_\odot$ to $10^{-4}M_\odot$, is still attributed to configurations having $w_{\rm E}$ values going below the threshold of ensuring multiple images given in Eq.~(\ref{ALPwE}). The contour $M_{\rm AS} = 10^{-5}M_\odot$ shows a different qualitative behavior, where we only see a single dip feature in the sensitivity limit. The scaling $w_{\rm E} \propto M_{\rm AS}^{3/2}m_a^2$, suggests that the dip feature appears at larger $m_a$, for lighter axion star masses. From Fig. \ref{fig:ConstraintfAS_future_1}, we observe that the dip occurs at $m_a \simeq 4.41\times 10^{-12}$ eV in the right panel, and corresponds to the the dip at $M_{\rm AS}=10^{-5}M_\odot$ in the left panel. In Fig. \ref{fig:ConstraintfAS_future_2} we also show sensitivity projections in the highly optimistic case of $N_{\rm obs} = 10^6$ observed FRBs, which simply amounts to an overall rescaling of the sensitivity curves to lower axion star fractions.
\section{Conclusions}
\label{sec:conclusions}
It has been pointed out in earlier literature that extended objects, in the form of axion stars, can be created out of axion-like particles. In this work, we studied the lensing of fast radio bursts (FRBs), by these axion stars. This astrophysical signal may be used to infer the properties of the constituent axions that make up these compact objects. 

We assumed that the axion stars are spherically symmetric. By using the nonrelativistic effective field theory framework for axions, and assuming a Gaussian profile for the axion field configuration in the axion star, we established a connection between: the macroscopic variables characterizing the axion star, namely the axion star mass $M_{\rm AS}$ and radius $R$; and the fundamental axion parameters, such as the axion mass $m_a$ and decay constant $f_a$. In studying the lensing of FRBs by axion stars, we accounted for novel effects that arise from the finite size of the lensing object, as well as the additional light bending induced by the axion-photon coupling $g_{a\gamma\gamma}$. Assuming that the plasma effect can be neglected, the latter effect occurs at order $g_{a\gamma\gamma}^2$. Furthermore, the effect induced by the axion-photon coupling is more pronounced at, or below the radio frequency band. Thus we have chosen FRBs as the target signal, which can be observed by radio telescopes such as CHIME. 

We developed a formalism to calculate the time delay and magnification ratio, which are relevant in identifying the presence of a lensing signal. For a given observer-axion star lens-FRB source configuration, we identified two main quantities that control the finite size and axion-photon induced effects, respectively: $w_{\rm E}$, which is the size of the Einstein radius relative to $R$; and $A$, which is proportional to $g_{a\gamma\gamma}^2$. We note that the axion-photon induced effect will only appear if the finite size effect is present. For this work, the lensing criteria we used are defined by the presence of a time delay above the CHIME threshold of $\unit[1]{\mu s}$, and a magnification ratio below 5, between the two brightest lensing images. These requirements define a lensing cross section, which can be used to obtain the optical depth of radio signals, passing through an extragalactic and Galactic population of axion stars. In both populations, we assumed a common energy density fraction $f_{\rm AS}$ of nonbaryonic matter residing in axion stars. We have also accounted for the redshift distribution of the FRB sources, which are mostly extragalactic, to obtain the integrated optical depth $\bar{\tau}$, which eventually determines the fraction of FRBs that undergo lensing. Assuming the optimistic case of $10^4$ observed FRBs, the lack of observed FRB lensing signals can be used to set forecasts on the viable axion star and axion parameters.

In the scenario where axion stars are made up of QCD axions, the finite-size parameter $w_{\rm E}$ is significantly large within the redshift range we are interested in. The axion star is effectively a pointlike object and leads to an exponential suppression in the lens equation. Because of this exponential suppression, there are no notable signatures attributed to the finite-size effect and the axion-induced lensing effect in the integrated optical depth curves, for axion stars located in extragalactic space and the Milky Way. We found that a radio telescope facility like CHIME constrains the axion star fraction $f_{\rm AS}$ at a level above $7 \times 10^{-3}$, for a range of axion star masses $M_{\rm AS}$ above $5 \times 10^{-2} M_\odot$. The sensitivity in $f_{\rm AS}$ is lost for lower axion star masses because the time delay generally goes below the 1-$\mu$s threshold. 

For axion stars composed of ALPs, the axion parameters $m_a$ and $f_a$ can be chosen independently. The novel lensing effects appear, since it is possible to find situations where the axion star effectively becomes an extended object so that $w_{\rm E} \sim O(1)$, and $A \gg 1$ given a sizable $g_{a\gamma\gamma}$. For axion stars within the mass range of $10^{-3} M_\odot$ to $10^{-1} M_\odot$, we identified the target range of $m_a$ within $\unit[10^{-15}]{eV}$ to $\unit[10^{-10}]{eV}$, which corresponds to $f_a$ between $\unit[10^{11}]{GeV}$ and $\unit[10^{17}]{GeV}$. With respect to the projected sensitivity for the pointlike case, we find situations where the inclusion of the finite size effect may allow for increased sensitivity in $f_{\rm AS}$ for axion star masses about $10^{-2} M_\odot$. Increasing $g_{a\gamma\gamma}$ leads to more stringent sensitivities on $f_{\rm AS}$, up to a level of $\sim 10^{-4}$. For $g_{a\gamma\gamma} = \unit[10^{-4}]{GeV^{-1}}$, we open up the possibility of excluding $f_{\rm AS} \lesssim 1$ for axion stars with mass $10^{-3} M_\odot$, with a corresponding axion mass range within $\sim \unit[2-6\times 10^{-13}]{eV}$; for $g_{a\gamma\gamma} = \unit[10^{-2}]{GeV^{-1}}$, this axion mass range widens to $\unit[2\times 10^{-14}]{eV}-\unit[6\times 10^{-13}]{eV}$. We emphasize that while the axion-photon induced lensing effects manifest for $g_{a\gamma\gamma}$ couplings that are already ruled out by more robust experimental probes, we are still able to see novel lensing features for negligible $g_{a\gamma\gamma}$ that is mainly attributed to the finite size of the axion star.

As we enter into the era of radio astronomy, where new facilities are being constructed, we expect to obtain better timing resolution for the arriving radio signals. As a reference scenario, a timing resolution at the nanosecond level translates to the possibility of probing axion stars at lower masses to $M_{\rm AS} \simeq 10^{-4} M_\odot$ for the pointlike case, and up to $M_{\rm AS} \simeq 10^{-5} M_\odot$ for the ALP case where novel features from the finite size effect can manifest. Meanwhile, the expected increase in the number of observed FRB events leads to an improvement in probing the axion star fraction $f_{\rm AS}$. As for probing the axion-photon induced lensing effect, the resulting optical depth for the pure gravity case is indistinguishable for smaller $g_{a\gamma\gamma}$ couplings. This may still pose a challenge for future generation telescope facilities, which warrants a better proposal to probe $g_{a\gamma\gamma}$ using axion stars.
\section{Acknowledgments}
We acknowledge the kind support of the National Science and Technology Council of Taiwan R.O.C. (formerly the Ministry of Science and Technology), with grant number NSTC 111-2811-M-007-018-MY2. P.Y.T. acknowledges support from the Physics Division of the National Center for Theoretical Sciences of Taiwan R.O.C. with grant NSTC 114-2124-M-002-003. K.Y.C.  is supported in part by NSTC with Grant No. 111B3002I4 and Grant No. 113J0073I4 and by the Ph.D. scholarship from the Ministry of Education of Taiwan R.O.C. The authors would like to thank Daniel Harsono, Alvina Yee Lian On, Martin Spinrath, and Yiming Zhong for fruitful discussions and constructive comments.

\appendix
\section{Gaussian profile}
We collect some results for certain quantities, specific to the case of the Gaussian profile
\begin{eqnarray}
a_0 =  \sqrt{\frac{2N}{\pi^{3/2}m_a R^3}}&,&\quad f(r/R) = \exp\left(-\frac{r^2}{2R^2}\right)
\end{eqnarray}
The form factor and Abel transform are
\begin{eqnarray}
F(r/R; m_a R) &=& \left(r^2/R^2 - m_a^2 R^2 - 1\right)\exp\left(-r^2/R^2\right),\\
\mathcal{F}(\xi_\perp/R; m_a R) &=& \frac{\sqrt{\pi}}{2}\exp\left(-\xi_\perp^2/R^2\right)\left(2\xi_\perp^2/R^2 - 2m_a^2 R^2 - 1\right).
\end{eqnarray}
The volume density, lens plane surface density, and projected mass on the lens plane are respectively given by
\begin{eqnarray}
    \rho &=& \frac{m_a N}{\pi^{3/2} R^3}\exp\left(-\frac{r^2}{R^2}\right),\quad \Sigma(s) = \frac{m_a N}{\pi R^2}\exp\left(-\frac{s^2}{R^2}\right),\\
    \mathcal{M}(s) &=& m_a N \left[1 - \exp\left(-\frac{s^2}{R^2}\right)\right].
\end{eqnarray}
The lensing potential is $\Psi = \Psi_{\rm axion}+\Psi_{\rm gravity}$, where
\begin{eqnarray}
    \Psi_{\rm axion}(\vec{\theta}) &=& \frac{D_{\rm LS}}{D_{\rm L}  D_{\rm S} }\frac{g_{a\gamma\gamma}^2 N}{16\pi R~(\kappa_0 R)^2(m_a R)}\left(\frac{2D_{\rm L} ^2 \theta^2}{R^2}-2m_a^2 R^2 + 1\right)\exp\left(-\frac{D_{\rm L} ^2 \theta^2}{R^2}\right),\\
    \Psi_{\rm gravity}(\vec{\theta}) &=& \frac{2D_{\rm LS}}{D_{\rm L}  D_{\rm S} } G_N m_a N \left[\Gamma(0,~D_{\rm L} ^2/R^2 \theta^2)+2\ln\vert\vec{\theta}\vert\right],
\end{eqnarray}
and $\Gamma(0,z)$ is the (upper) incomplete Gamma function. The derivatives are
\begin{eqnarray}
    \nonumber \Psi_{\rm axion}'(\theta) &=& \frac{D_{\rm LS}D_{\rm L} }{D_{\rm S} }\frac{(g_{a\gamma\gamma}/R)^2 N}{8\pi R~(\kappa_0 R)^2(m_a R)}\exp\left(-\frac{D_{\rm L} ^2\theta^2}{R^2}\right)\\
    &\times&\theta\left(1 + 2m_a^2 R^2 - \frac{2D_{\rm L} ^2 \theta^2}{R^2}\right),\\
    \Psi_{\rm gravity}'(\theta) &=& \frac{4D_{\rm LS}}{D_{\rm L}  D_{\rm S} } G_N m_a N \left[\frac{1 - \exp\left(-D_{\rm L} ^2/R^2~\theta^2\right)}{\theta}\right],
\end{eqnarray}
and 
\begin{eqnarray}
  \nonumber  \Psi_{\rm axion}''(\theta) &=& \frac{D_{\rm LS}D_{\rm L} }{D_{\rm S} }\frac{(g_{a\gamma\gamma}/R)^2 N}{8\pi R~(\kappa_0 R)^2(m_a R)}~\exp\left(-\frac{D_{\rm L} ^2\theta^2}{R^2}\right)\\
    &\times&\left(2m_a^2 R^2 - \frac{8D_{\rm L} ^2 \theta^2}{R^2}+\frac{4D_{\rm L} ^4\theta^4}{R^4}+1 - 4 D_{\rm L} ^2 m_a^2 \theta^2\right)\\
    \Psi_{\rm gravity}''(\theta) &=& \frac{4D_{\rm LS}}{D_{\rm L}  D_{\rm S} } G_N m_a N \left[\frac{2D_{\rm L} ^2}{R^2}\exp\left(-\frac{D_{\rm L} ^2 \theta^2}{R^2}\right)-\frac{1 - \exp\left(-D_{\rm L} ^2/R^2~\theta^2\right)}{\theta^2}\right].
\end{eqnarray}

\bibliographystyle{JHEP.bst}
\bibliography{refs.bib}

\end{document}